\documentclass[12pt]{ucsddissertation}
\usepackage[NoDate]{currvita}
\usepackage{array}
\usepackage{tabularx}
\usepackage{booktabs}
\usepackage{ragged2e}
\usepackage{microtype}
\usepackage[breaklinks=true,pdfborder={0 0 0}]{hyperref}
\usepackage{graphicx}
\usepackage{bibentry}
\usepackage[square,numbers]{natbib}
\usepackage{float}    
\usepackage{multirow}
\usepackage{enumitem}
\usepackage{color,soul}
\usepackage{todonotes}
\usepackage{longtable}
\usepackage{arydshln}
\usepackage{amsmath}
\AtBeginDocument{%
	\settowidth\cvlabelwidth{\cvlabelfont 0000--0000}%
}

\usepackage{trace}
\usepackage[english]{babel}
\usepackage{blindtext}
\title{\textbf{Improving Human-Autonomous Vehicle Interaction in Complex Systems}} %Developing Situational Awareness For Human-Autonomous Vehicle Systems %Developing Situational Awareness For Joint Action With Autonomous Vehicles %A Complex Systems Approach To Understanding and Improving Human-Autonomous Vehicle Interaction
\author{Robert A. Kaufman}
\degree{Cognitive Science}{Doctor of Philosophy}
\cochair{Professor David Kirsh}
\cochair{Professor Nadir Weibel}
\committee{Professor David Danks}
\committee{Professor Jim Hollan}
\degreeyear{2025}

\begin{document}
% Begin with frontmatter and so forth
\frontmatter
\maketitle
\makecopyright
\makesignature
% Optional
\begin{dedication}
\setsinglespacing
\raggedright % It would be better to use \RaggedRight from ragged2e
\parindent0pt\parskip\baselineskip
To those with minds fractured and reformed: there is strength and beauty in new shapes.
\end{dedication}

\begin{epigraph}
\vskip0pt plus.5fil
\setsinglespacing
\vfil
\noindent “Of course it is happening inside your head...\\
but why on earth should that mean that it is not real?”

\vskip\baselineskip
\hskip0pt plus1fil\textit{Albus Dumbledore}\hskip0pt plus4fil\null

\vfil
\end{epigraph}

% Next comes the table of contents, list of figures, list of tables,
% etc. If you have code listings, you can use \listoflistings (or
% \lstlistoflistings) to have it be produced here as well. Same with
% \listofalgorithms.
\tableofcontents
\listoffigures
\listoftables

% Preface
% \begin{preface}
% Almost nothing is said in the manual about the preface. There is no
% indication about how it is to be typeset. Given that, one is forced to
% simply typeset it and hope it is accepted. It is, however, optional
% and may be omitted.
% \end{preface}

% Your fancy acks here. Keep in mind you need to ack each paper you
% use. See the examples here. In addition, each chapter ack needs to
% be repeated at the end of the relevant chapter.
\begin{acknowledgements}
It would not have been possible for me to complete my PhD without the support of numerous individuals. I benefitted from working with two great advisors, Nadir Weibel and David Kirsh, whose mentorship showed a fine balance between patience and unwavering dedication to rigor. Nadir: thank you for opening your lab to me and demonstrating to me the power of open-mindedness, interdisciplinary thinking, and excitement for cutting-edge research. David: thank you for teaching me how to ask the right questions and embrace complex problems with curiosity and systematic thinking. I am grateful to my PhD committee as well, Jim Hollan and David Danks, for invaluable feedback that helped bring new meaning and depth to the work.

I was surrounded at each step of my PhD with an amazing set of collaborators and research assistants. I would like to thank Emi Lee, Aaron Broukhim, Manas Bedmutha, Eve Kimani, and Jean Costa for their direct contributions to the research in this dissertation. I would also like to thank Saumitra Sapre and Rohan Bhide for their amazing work on our lab's own custom driving simulator, without which the study in Chapter 3 would not have been possible. 

I would like to thank Michael Haupt, with whom I co-founded the Cognitive Media Lab, and Alejandro Jinich, with whom I co-founded the Discovery Way Foundation, for supporting my academic and non-academic pursuits in equal measure.

I am inspired by present and past mentors, who -- for reasons I have yet to fully comprehend -- took a chance on me even when I had nothing yet to show for myself. Chris Simons at Ohio State, you gave me the freedom to pursue my first ever independent research project, from which I fell in love with science. Sara Lazar, you opened your lab to me at Massachusetts General Hospital and showed me how to lead with compassion and mindfulness.

To Alexina, thank you for helping me keep my plates spinning and know when to put them down. 

To my friends, it is impossible to describe how much your love and support mean to me. The laughter and adventures we share bring meaning to my life and gave me the energy to push through the challenges that this PhD presented. Thank you for accepting me as I am and standing by me as I grow. In no particular order: Aaron, Megan, Andrew H, Eric, Alex, Max, Nicky, Noah, Andrew B, Sam, Ailie, Chris, Tricia, Amy, Steff, Tanya, Matt, Cody, Jenna, Jose, Cat, and Ashley.

To my partner Ariana (a.k.a. Archie): you are a constant source of joy and comfort in my life. It is a privilege to share a barn, laugh, and adventure with you each and every day.

To my brother, Michael, thank you for constant laughter and always being there for me -- you inspire me with your creativity, hilarity, and zest for wooden clocks. I am grateful to call you my brother.

And finally, to my parents, Marcy and Dale -- your unending support and unconditional love mean the world to me. I am grateful every day for how you taught me to engage with the world. Thank you for never giving up on me, even when I broke the garage.
\newline

Chapter 1, in part, contains reprinted materials as they appear in \textit{Developing Situational Awareness for Joint Action With Autonomous Vehicles} (ArXiv~2024). Kaufman, Robert; Kirsh, David; Weibel, Nadir. The dissertation author was the primary investigator and author of this paper.

Chapter 2, in full, is a reprint of the material as it appears in \textit{Effects of Multimodal Explanations for Autonomous Driving on Driving Performance, Cognitive Load, Expertise, Confidence, and Trust}
(Nature: Scientific Reports~2024). Kaufman, Robert; Costa, Jean; Kimani, Everlyne. The dissertation author was the primary investigator and author of this paper.

Chapter 3, in full, has been accepted for publication using the title \textit{What Did My Car Say? Impact of Autonomous Vehicle Explanation Errors and Driving Context On Comfort, Reliance, Satisfaction, and Driving Confidence}, in the Proceedings of the 2025 CHI Conference on Human Factors in Computing Systems (CHI ’25). Kaufman, Robert; Broukhim, Aaron; Kirsh, David; Weibel, Nadir. The dissertation author was the primary investigator and author of this paper.

Chapter 4, in full, has been accepted for publication using the title \textit{Predicting Trust In Autonomous Vehicles: Modeling Young Adult Psychosocial Traits, Risk-Benefit Attitudes, And Driving Factors With Machine Learning.}, in the Proceedings of the 2025 CHI Conference on Human Factors in Computing Systems (CHI ’25). Kaufman, Robert, Lee, Emi, Bedmutha, Manas, Kirsh, David, Weibel, Nadir. The dissertation author was the primary investigator and author of this paper.
\end{acknowledgements}

% Stupid vita goes next
\begin{vita} %includes current phd!!
\noindent
\begin{cv}{}
\begin{cvlist}{}
\item[2013--2017] Bachelor of Science in Neuroscience and Psychology, Ohio State University
\item[2019--2021] Master of Science in Cognitive Science, University of California San Diego
\item[2021--2025] Doctor of Philosophy in Cognitive Science, University of California San Diego
\end{cvlist}
\end{cv}

% This puts in the PUBLICATIONS header. Note that it appears inside
% the vita environment. It is optional.
\publications %%ADD REFERENCES TO BIBLIOGRAPHY FOR EACH? like \cite[ref]

\noindent \textbf{Kaufman, R. A.}, Broukhim, A., Kirsh, D., Weibel, N. (2025). What Did My Car Say? Impact of Autonomous Vehicle Explanation Errors and Driving Context On Comfort, Reliance, Satisfaction, and Driving Confidence. \textit{Proceedings of the 2025 CHI Conference on Human Factors in Computing Systems (CHI ’25).}
\\

\noindent \textbf{Kaufman, R. A.}, Lee, E., Bedmutha, M., Kirsh, D., Weibel, N. (2025). Predicting Trust In Autonomous Vehicles: Modeling Young Adult Psychosocial Traits, Risk-Benefit Attitudes, And Driving Factors With Machine Learning. \textit{Proceedings of the 2025 CHI Conference on Human Factors in Computing Systems (CHI ’25).}
\\

\noindent \textbf{Kaufman, R. A.}, Broukhim, A., Haupt, M. (2025). WARNING This Contains Misinformation: The Effect of Cognitive Factors, Beliefs, and Personality on Misinformation Warning Tag Attitudes. \textit{Proceedings of the 2025 ACM Conference on Computer Supported Cooperative Work (CSCW).}
\\

\noindent \textbf{Kaufman, R. A.}, Kirsh, D., Weibel, N. (2024). Developing Situational Awareness for Joint Action With Autonomous Vehicles. \textit{ArXiv.}
\\

\noindent \textbf{Kaufman, R. A.}, Costa, J., Kimani, E. (2024). Effects of Multimodal Explanations for Autonomous Driving on Driving Performance, Cognitive Load, Expertise, Confidence, and Trust. \textit{Nature: Scientific Reports.}
\\

\noindent \textbf{Kaufman, R. A.}, Kirsh, D. (2023). Explainable AI and Visual Reasoning: Insights from Radiology. \textit{2023 Computer-Human Interaction (CHI) Workshop on Human-Centered Perspectives in Explainable AI.}
\\

\noindent \textbf{Kaufman, R.}*, Haupt, M.*, Dow, S. (2022). Who’s In the Crowd Matters: Cognitive Factors and Beliefs Predict Misinformation Assessment Accuracy. \textit{Proceedings of the 2022 ACM Conference on Computer Supported Cooperative Work (CSCW).}
\\

\noindent \textbf{Kaufman, R. A.}, Kirsh, D. (2022). Cognitive Differences in Human and AI Explanation. \textit{In Proceedings of the Annual Meeting of the Cognitive Science Society.}
\\

\noindent Soltani, S., \textbf{Kaufman, R. A.}, Pazzani, M. (2022). User-Centric Enhancements to Explainable AI Algorithms for Image Classification. \textit{In Proceedings of the Annual Meeting of the Cognitive Science Society.}
\\

\noindent Pazzani, M., Soltani, S., \textbf{Kaufman, R.}, Qian, S., Hsiao, A. (2022). Expert-Informed, User-Centric Explanations for Machine Learning. \textit{Thirty-Sixth AAAI Conference on Artificial Intelligence.}
\\

\noindent Sevinc, G., Rusche, J., Wong, B., Datta, T., \textbf{Kaufman, R.}, Gutz, S., ... Lazar, S. W. (2021). Mindfulness Training Improves Cognition and Strengthens Intrinsic Connectivity Between the Hippocampus and Posteromedial Cortex in Healthy Older Adults. \textit{Frontiers in Aging Neuroscience, 565.}
\\

\noindent Binder, F. J., Jones, C. R., \textbf{Kaufman, R. A.}, Lin, N. T., Poole, C. R., Vul, E. (2021). Cognitive cost and information gain trade off in a large-scale number guessing game. \textit{In Proceedings of the Annual Meeting of the Cognitive Science Society (Vol. 43, No. 43).}
\\

\noindent \textbf{Kaufman, R. A.}, Simons, C. (2017). The Effects of Linalool and Peppermint Aroma on Cognitive Performance (Undergraduate Thesis). \textit{The Ohio State University Press.}

% % This puts in the FIELDS OF STUDY. Also inside vita and also
% % optional.
% \fieldsofstudy
% \noindent Major Field: Cognitive Science (Specialization or Focused Studies)
% \vskip\baselineskip
% Studies in Applied Mathematics\par
% Professors Alpha Beta and Gamma Delta
% \vskip\baselineskip
% Studies in Mechanices\par
% Professors Epsilon Zeta and Eta Theta
% \vskip\baselineskip
% Studies in Electromagnetism\par
% Professors Iota Kappa and Lambda Mu
 \end{vita}

% Put your maximum 350 word abstract here.
\begin{dissertationabstract}
Advances in Autonomous Vehicle (AV) technology promise significant individual and societal benefits. However, unresolved questions about how to meet the informational needs of riders hinder real-world adoption. Complicating our ability to satisfy rider needs is the intuition that different people, with different goals, and in different driving contexts, may have different requirements for what constitutes a successful interaction. Unfortunately, most human-AV research and design today treats all people and all situations as if they are the same. As a result, it is crucial that we understand what information an AV should communicate to meet rider needs, and how these communications should change when aspects of the human-AV complex system change, including when communications fail. I argue that understanding the relationships between different aspects of the human-AV system can help us build better AV communications and improve interactions beyond one-size-fits-all approaches. I support this argument using three empirical studies. First, by exploring the novel application of an AI race car driving coach, I identify optimal communication strategies that enhance driving performance, confidence, and trust for learning in even the most extreme driving environments. Findings highlight the need for task-sensitive, modality-appropriate communications tuned to learner cognitive limits and goals. Next, I examine the effect of AV communication errors on rider trust, showing that an error's impact is dependent on the driving context within which it occurs. Results highlight the consequences of deploying faulty communication systems and emphasize the need for context-sensitive communications. Third, I explore individual differences in trust perceptions, using machine learning (ML) to illuminate personal factors predicting young adult trust in AVs. This study highlights the importance of tailoring designs to individual traits and concerns, with implications for personalized design. Together, this dissertation supports the necessity of transparent, adaptable, and personalized AV systems that cater to individual needs, goals, and contextual demands. By considering the complex system within which human-AV interactions occur, we can deliver valuable insights for designers, researchers, and policymakers. This dissertation also provides a concrete domain to study theories of human-machine joint action and situational awareness, and can be used to guide future human-AI interaction research. 
\end{dissertationabstract}

% This is where the main body of your dissertation goes!
\mainmatter

% Optional Introduction
\begin{dissertationintroduction}
\newpage
Complex systems are everywhere. Recognizable by their intricate networks of interdependent parts, understanding how a complex system will behave requires careful unraveling of the relationships between its individual components. How can we predict the weather, prevent diseases from spreading, or determine if the stock market will rise or fall? Studying complex systems can be tricky: when the system’s behavior can’t be pinned to one source, it makes knowing how the system will behave a challenge. 

This dissertation examines an important and timely area of questioning in the field of human-computer interaction -- \textbf{how can we make robust and adaptive autonomous vehicles (AVs) that can work in the real world?} Human-AV interaction is fascinating because the dynamic that exists between a person and a vehicle is emergent from the complex systems within which it is situated, including transportation systems and social networks. This makes human-AV problems challenging to solve. Understanding how best to design human-AV interactions is important, because without satisfying the needs of riders, they won't ride.

Complex systems theory helps us understand that the success of an AV will be determined not just by how well it performs in isolation, but its wider context of use. This includes the diverse goals and traits of the people interacting with the system, as well as the environment they are using the system in. For example, you could have an AV car that drives perfectly, but if people don’t trust it, they won’t ride. In other words -- the systems people use don’t exist in a vacuum, so we shouldn’t design them like they do. 

Concretely, designing successful AVs is dependent on meeting the needs of real (human) riders driving in real (imperfect and often dangerous) driving situations. As a result, rider needs depend not just on the goal of the interaction (like to safely get from point A to B), but also who the rider is, what are the abilities of the AV, and what is the driving context. It’s easy to imagine that grandma may have different needs than her 16-year-old grandson, and interactions may need to change when an AV is driving in a blizzard vs. a sunny afternoon. The major implication is that when aspects of the system change, so do the design requirements that need to be met. 

Unfortunately, most human-AV (and human-AI in general) research and design today treats all people and all situations as if they are the same. One of the reasons is because it’s really difficult to account for the complex web of interacting parts. But -- if we can break off chunks of the system bit by bit and study them, there are a ton of opportunities to make the system better.

In this dissertation, I take a first step towards improving human-AV interaction, applying this complex systems lens along with cognitive theories of joint action and situational awareness to answer three distinct questions. From studies on AV communication techniques for learning to drive a racecar (Chapter 2), an assessment of AV errors and driving context (Chapter 3), and a study on personal traits predicting trust in AVs (Chapter 4) -- I provide both theoretical insights and practical implications that can immediately be used to design better AVs, and understand the consequences of deploying AVs that are imperfect.

These studies fit into a larger research agenda: working in domains ranging from healthcare AI to misinformation mitigation on social media, I have spent my PhD showing how the success of the AI systems we design is dependent who is interacting with the system, why they are interacting with it, and what situation they are immersed in. It is my hope that this dissertation improves more than just human interaction with autonomous vehicles -- by providing a scaffold for future designers and researchers, I hope this work can serve as an inspiration to anyone seeking to improve interactions with ever-more-ubiquitous AI across many domains. \textbf{By accounting for the complex systems within which human-AI interaction occurs, we can empower users to meet their goals and ensure that systems work for many diverse people in many diverse situations.}
\end{dissertationintroduction}

\chapter[Human-AV interaction as a complex system]{Human-AV interaction as a complex system}
\newpage
Advances in Autonomous Vehicle (AV) technology promise huge individual and societal benefits, from increased driving safety to reduced environmental impact~\cite{fagnant2015preparing}. However, unanswered questions about how AV interaction design can support the critical informational needs of people riding with AVs are hindering real-world adoption~\cite{hegner2019automatic, detjen2021increase}. A well documented example is that people often lack of trust in AV decision-making, particularly when the AV’s decision procedures are opaque and not human-understandable~\cite{kohn2021measurement, meyer2022baby, morra2019building, ekman2017creating, frison2019ux, wiegand2019drive}.  

One reason we think that the success of AV technologies will likely depend on designing for the social informational needs of people is because we know that theory of mind is a core part of how we think and move about the world, particularly when we are trying to understand decisions made by other actors~\cite{cuzzolin2020knowing, premack1978does, wang2021towards}. A person seeking to make decisions within a human-AV system necessarily confronts many questions about AV decision-making. Most pressingly for the decisions of the person, questions involve both present and future actions on the road and their predictability: \emph{Why is the AV moving to the left? Does the AV see the pedestrian ahead? What will it do next? Will this always be the case?}

As AI-based systems advance, their increasing scope of decision-making actions shifts them from being tools for human use to agents in a human-AI team~\cite{dafoe2021cooperative}. In the case of autonomous vehicles, current theory proposes that successful designs should conceptualize a human and an AV as team members that must work together to achieve common goals, such as safe transportation. 

The human-AV team does not operate in a vacuum; it's sensitive to the situated, contextual demands of the physical and social environment~\cite{clancey1997situated, kirsh2009problem, ladyman2013complex}. Applying a complex systems \cite{ladyman2013complex} framing to human-AV interaction allows us to understand that the attributes of one component of the system will impact the informational needs of the human-AV team as a whole. These determine the necessary communicative actions that should be taken. For example, different communications may be necessary when driving in dangerous situations~\cite{stanton2001situational, liu2023design} or when a rider may have additional interaction goals like learning from the AV’s driving decisions~\cite{ghai2021explainable, soltani2022user, kaufman2024learning}. It is intuitive to expect that a specific person's traits -- such as their expertise or risk tolerance -- will impact how they perceive and behave towards AVs \cite{ferronato2020examination, ayoub2021modeling}. Likewise, we'd expect the specific attributes of the AV -- including how well it drives or if it communicates with errors -- will impact how it will be utilized \cite{hoff2015trust}. What happens to a person's trust and future communication needs when an AV makes a mistake? In all cases, the attributes of each system component, and their relationships with each other, will dictate how the human and the AV can and should interact.

\begin{figure}[H]
  \centering
  \includegraphics[width=\linewidth]{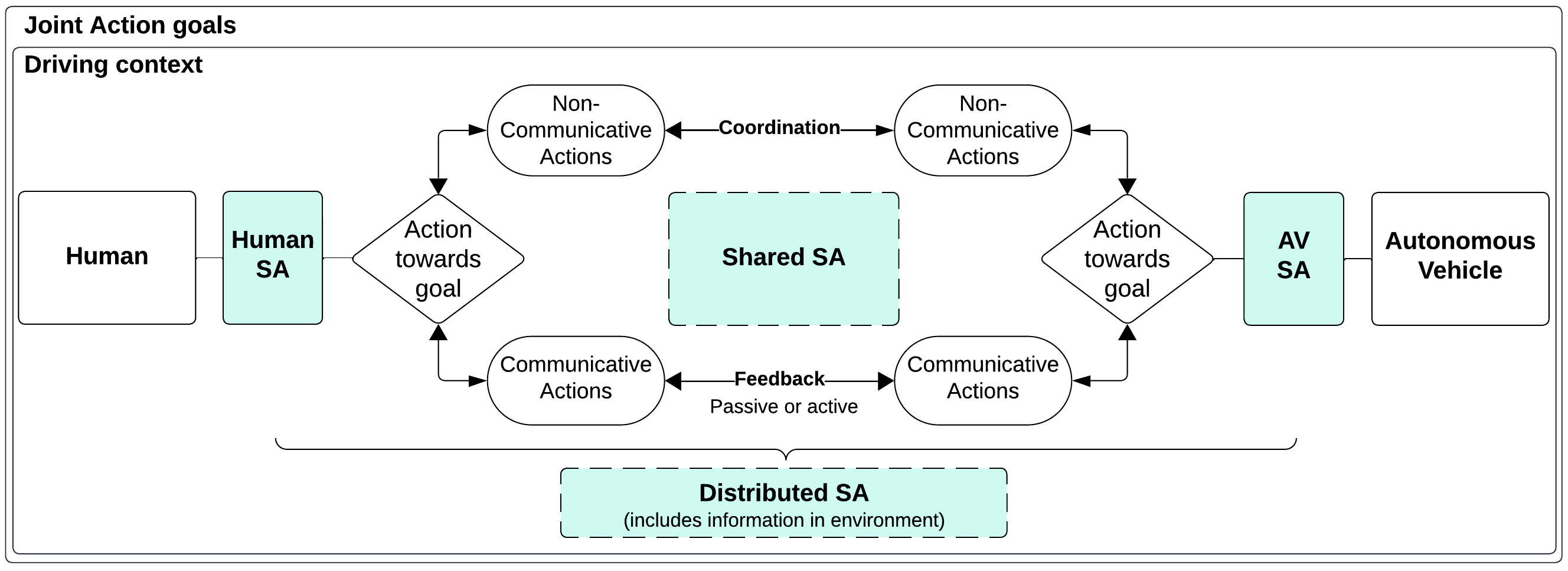}
  \caption[\textbf{Overview of the Human-AV system.}]{\textbf{Overview of the Human-AV system.} Action goals, the driving context, and different traits (human and AV traits) are fundamental parameters determining what actions -- communicative and not -- need to be jointly taken by the team. Actions build, and are iteratively impacted by, the different types of situational awareness held by the system as the team works towards goal success.}
  \label{SA_framework}
\end{figure}

This dissertation posits that by understanding and accounting for the dynamic interplay between the components of the shared situation, specifically the driving context, interaction goals, human traits, and AV traits --  a human-AV team can successfully act to achieve their goals. I specifically focus on goals which involve satisfying information needs by building the team's individual and collective situational awareness. This primarily involves building what the AV knows, so that it can communicate to a person what \emph{they} need to know, in the right way, and at the right time. See Figure \ref{SA_framework} for a system overview. \textbf{For human-AV communication design, the implication is that AV communications should be tailored based on these four factors of the complex system, and be sensitive to when components of the system change.} 

Across three studies, this dissertation examines different components of the human-AV system with the goal of improving the design of future systems. Importantly, I shift away from viewing the human and AV as separate agents, instead highlighting how a dynamic approach to human-AV interaction can provide new insights into how designers and researchers can better support human-AV team success.

AVs of the (not so distant) future may play an expanded role in facilitating more than just safe transportation: managing social, emotional, educational, and cultural experiences for their passengers. For example, an AV could take the scenic route through a city, comment on notable historical centers, and adjust its communication style to avoid waking a sleepy travel companion. AVs could teach driving skills, soothe nervous passengers, and entertain children on a road trip. There are near-endless opportunities. It is likely that there will be multiple AI systems working together: some managing the task of driving and analyzing the outside environment, while others will focus on detecting and moderating passenger experiences. \textbf{Without addressing the salient information needs of passengers, however, very few AV functions can be fully realized.}

\section{Situational awareness for joint action success}
A theory which has helped researchers understand many of the informational problems that stem from interacting with an AV is through the lens of situational awareness (SA), where people want to be able to perceive, comprehend, and project into the future what is happening in a given driving situation~\cite{endsley1995toward}. A lack of knowledge in how an AV will behave or why it does so prevents a person interacting with the AV from achieving situational awareness. This is because they do not have a means to access the information required to develop understanding or predict future behaviors that the AV might take~\cite{capallera2022human}.

\vspace{1em}
\noindent
Lacking situational awareness:
\begin{itemize}[topsep=0pt, nolistsep]
\item Impedes a person from being able to develop trust, a reliance strategy, or expertise from learning, limiting the usefulness of the technology~\cite{vorm2022integrating, marangunic2015technology}.
\item Removes a person’s agency, forcing them to rely on blind faith and abandon their desire for understanding~\cite{tulli2024explainable}.
\end{itemize}

\vspace{1em}
Models of SA tend to focus on either the needs of the human(s) or the AV in the system, independently of one another. However, achieving necessary situational awareness in human-AV partnership requires bidirectional communication~\cite{chen2018situation}. Some situational awareness needs to be held by the human and AV independently, while other SA must be shared or distributed between them~\cite{salas1995situation, stanton2006distributed} (Figure \ref{SA_framework}). 

Individual situational awareness becomes shared when there is a \textit{common} representation of a situation, where each team member can understand and predict a particular aspect of their environment in synchrony \cite{salas1995situation, nofi2000defining, saner2009measuring}. \textit{Crucially, this may include an understanding of the state, needs, or actions of the other team member.} For instance, when an AV shares on a display screen that it “sees” other cars around it (perhaps as a visualization of the scene around the vehicle), it allows a rider to ensure that they share a common representation of the external driving environment. If there was a car in front of the AV that the rider sees, but the AV does not display, a rider may be concerned that there is a gap in the AV’s situational awareness. This may impact the rider's trust and willingness to rely on the AV.

Distributed situational awareness, by contrast, proposes that SA is emergent from the system as a whole, rather than held by a given individual \cite{stanton2006distributed, salmon2022distributed, salmon202013}. To act as a team, some information can be held individually, while other information needs to be shared or held collectively. How much situational awareness is necessary, and who possesses it, is determined by the team's driving goal.

Numerous questions remain on how communication needs should be satisfied based on theories of situational awareness and how to account for the complex sociotechnical system within which the human and AV interact \cite{stanton2006distributed}. \emph{What does each team member need to know about each other, and the broader driving context? How do these informational needs change -- and how should communication strategies change -- when components of the system change?} 

In human-AI teaming, a person and an AV would be bound by principles of joint action similar to those of coordinated human-to-human activities~\cite{sebanz2006joint}. Joint actions are those which involve two or more people or agents working together in coordination to achieve a common goal. For example, a rider and an AV need to act jointly, in coordination, to achieve the goal of safe transportation: the AV drives, while the rider wears their seatbelt and does not jump out of the car into traffic. To keep the rider comfortable or get them to ride in the first place, the AV will need to know what and how to communicate to meet the rider's informational needs.

Joint actions are behaviors oriented towards achieving action goals~\cite{sebanz2006joint}. These include higher-level goals like safe transportation or learning as well as lower-level subgoals that are necessary preconditions for achieving higher-level goals, like trust, error-free driving, or sharing the right information. The hierarchy of high-level goals and their constituent subgoals may be best viewed as a multi-level checklist, where high-level goals have criteria to meet, and those criteria, themselves, have their own criteria to meet. Mathematically, this can be written as a multi-level success function, where humans and AVs act in coordination to achieve action goals by accomplishing their conditional criteria. Some conditional criteria are met by the human, some by the AV, and some jointly through coordinated behavior.
%\vspace{1em}

%\vspace{1em}

\begin{figure}[H]
  \centering
  \includegraphics[width=.9\linewidth]{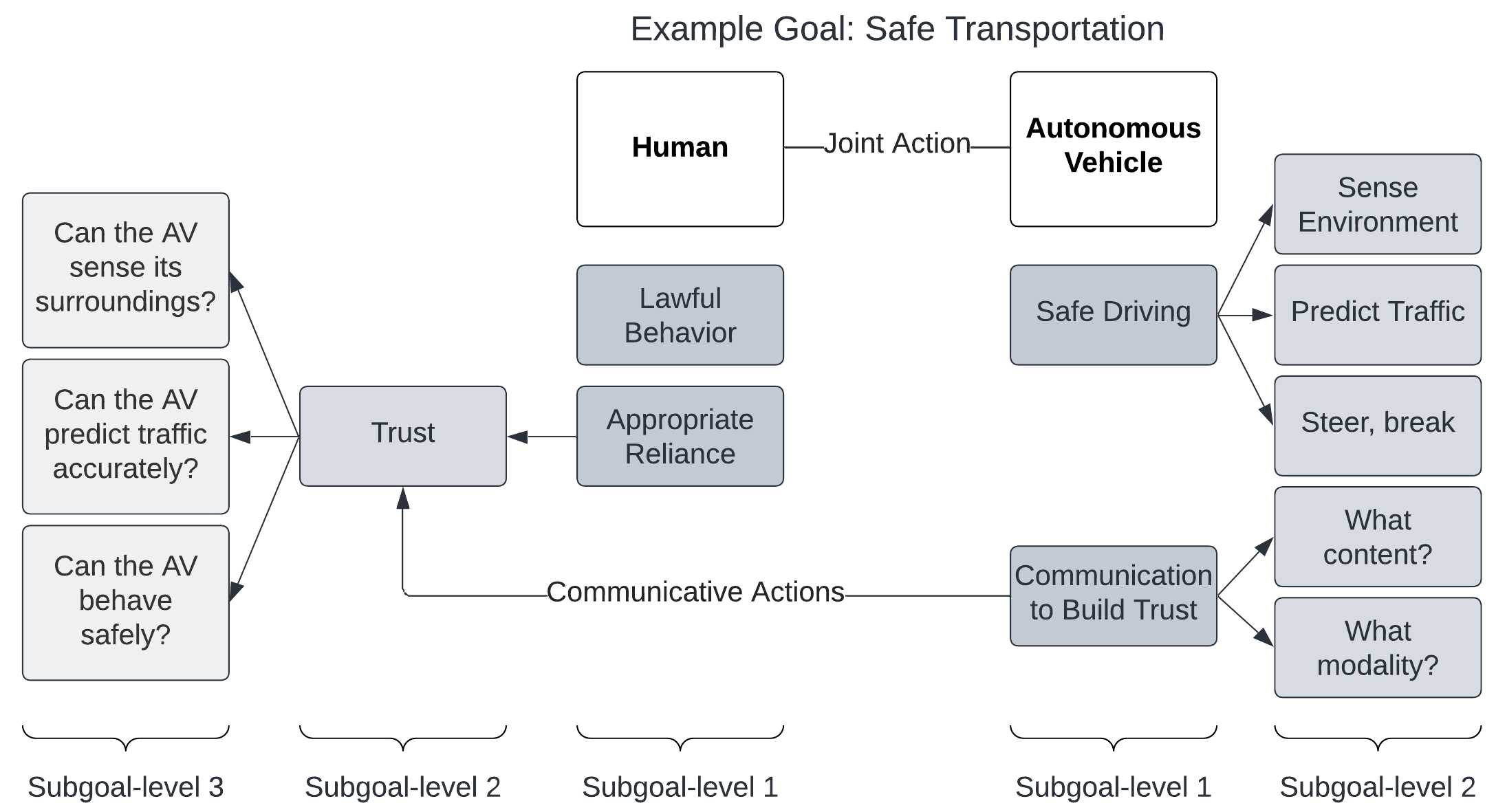}
  \caption{\textbf{Example Success function for the goal of Safe Transportation.}}
  \label{action_goals}
\end{figure}

To successfully achieve the action goal of safe transportation, for example, the subgoal of calibrated reliance must be met amongst others. To achieve calibrated reliance, knowing when and when not to trust an AV is necessary. To know when and when not to trust an AV, a person needs to be aware of specific information -- such as the AV's ability to perform in a particular environment -- that can be used to develop a trust and reliance strategy. This information has its own criteria, based on principles of communication, including criteria on the amount, type, and presentation modality of the information to be conveyed. A simplified example is shown in Figure \ref{action_goals}.

Some subgoals like building trust in an AV’s decisions have received a lot of focus by the research community. The issue of trust spans nearly all human-AV goals, and is perhaps the most pressing issue hindering AV adoption and the adoption of AI systems more generally~\cite{choi2015investigating}. Chapters 3 and 4 of this dissertation are devoted primarily to the study of trust. Goals like learning from an AV’s decisions are severely underexplored by the research community, and is the focus of Chapter 2.

This dissertation focuses on goal criteria -- such as developing trust or learning from an AV -- related to informational needs. This is done by developing the situational awareness of each team member as well as the team as a whole. How can we present information to help a novice learn to drive (Chapter 2)? How do information errors impact trust formation, and how does this change based on the driving context (Chapter 3)? How do personal traits impact trust dispositions, and how can this information be used by an AV to build trust with diverse riders (Chapter 4)? These questions are examples of those which can be asked (and answered) by looking at human-AV teaming as joint action within a complex system.

\textbf{In sum, we have two assumptions: (1) each goal has criteria for success, and (2) criteria for success change when aspects of the system change.}

\section{Communication for situational awareness}
As human cognition and decision-making capacity are necessarily limited, an emphasis on explainability in AI has recently emerged as a key factor supporting successful human-agent interactions. This approach emphasizes that by giving the right information, in the right way, at the right time -- an AV can help a person achieve many of their interaction goals, thereby increasing the overall situational awareness that an individual has when driving with a vehicle. This concept is a central fixture for the field of explainable AI (XAI), which emerged in recent years as a means to generate trust with autonomous decision-makers like AVs~\cite{dzindolet2003role, benbasat2005trust}. DARPA’s XAI program posed explainable AI as a means for humans to gain insight into the “black box” nature of machine learned systems so that they know why a decision has been made, when to trust the decision, why errors have occurred, and in what contexts the system will succeed and fail~\cite{gunning2019darpa}. Explainability may be especially important because automation does not have an understanding of cause and effect~\cite{pearl2018book}. Building off this agenda, XAI systems have been used to implement explanations for decisions in many domains~\cite{das2020opportunities, pazzani2022expert, kaufman2022cognitive, kaufman2023explainable}, including autonomous driving~\cite{koo2015did, atakishiyev2021explainable, capallera2022human}.

Huge knowledge gaps remain on how to provide the right information, in the right way, at the right time to satisfy the informational success criteria for a particular driving goal. To design effective communications, it is important to know both what type of information is most effective for building satisfactory situational awareness, and what modalities of presentation will most effectively allow access to that information. Choosing the right communication strategy is based on satisfying the information criteria necessary to achieve a joint action goal. To make XAI communication \emph{bidirectional}, knowledge gaps remain on what information the AV needs to know itself -- knowing how to convey information and how information requirements may change between different people or contexts is crucial to designing systems that can meet informational goal criteria. Knowing what happens when communications are wrong is equally important. All of these remain unknown in most cases. Augmenting AVs with this understanding is foundational to supporting a human-AV team's success. Without careful consideration of these factors, ill-fitting communications may fail to meet a rider’s needs or even derail progress towards a goal. These knowledge gaps related to informational needs are the major focus of this dissertation.

\section{Connecting the dots}
In summary, I have discussed how sufficient situational awareness (individual, shared and/or distributed) is essential for a human-AV team to meet the success criteria for many action goals. High-level goals, such as safe transportation or learning from an AV’s behavior, have success functions that need to be satisfied based on meeting conditional criteria like establishing trust or transferring appropriate information. Many of these conditional criteria have their own success functions, forming checklist-like hierarchies. Focusing specifically on goals and criteria related to informational needs, communication strategies such as AI explanations can help a rider learn or calibrate trust. Providing information to a person or AV prior to an interaction may also help tailor behavior accordingly.

\vspace{1em}
\noindent \textbf{To achieve the \underline{ultimate vision} of AVs that can meet the informational needs of diverse riders, contexts, and goals -- there is a specific process we can follow}:
\begin{enumerate}[topsep=0pt, nolistsep]
\item First, the major components of the human-AV joint-action system must be mapped as a network, showing how components interact. In our case, Figure \ref{SA_framework} maps human traits, AV traits (including performance/errors), goals, the driving context, and actions (communicative and not). 
\item For a specified goal, such as safe transportation with an AV, a success function can be defined. Establishing and meeting success criteria may be impossible without understanding how relevant factors (e.g. specific traits) within each system component (e.g. human) will impact needs (e.g. informational needs related to situational awareness).
\item In our example, we know that trust is a necessary condition for safe transportation with an AV. So, we can identify relevant factors within each component that impact trust. For example, how do individual traits of a rider, abilities of the AV, or contextual elements of the driving situation impact trust? This is the primary focus of \textbf{Chapter~3}, where I test how AV mistakes and contextual elements impact trust, as well as \textbf{Chapter~4}, where I identify personal traits that influence trust.
    %\item Once relevant factors are identified, we need to understand how \textit{variance} in that factor impacts the condition of interest. For example, high or low difficulty driving environments may impact trust differently. This is discussed in \textbf{Chapter~2}, where we test the impact of driving difficulty and contextual harm on trust.
\item Insights can then be translated into specific informational needs (success criteria to address). For instance, certain context elements (e.g. high harm scenarios) might lower trust, creating an information requirement in those contexts. Information requirements are distilled from study results in the \textbf{“design~implications”} of each chapter.
\item Communication strategies must be designed to meet these needs; certain communication strategies may be more effective than others. This is the primary focus of \textbf{Chapter~2}, where I test how explanation content and presentation modality impact learning. 
\item Finally, pairing various combinations of factors (e.g. a novice in a dangerous driving context with a poor performing AV) with communication strategies is necessary to ensure adaptability when factors change. A deep understanding of the relationships between components is key to realizing personalized and contextually-adaptive AVs.
\end{enumerate}

\vspace{1em}
Countless open questions remain regarding how system factors translate into informational needs, and how to meet those needs. Knowledge gaps remain on how to effectively communicate the right information, the consequences of delivering information with errors, and how broader system factors -- such as human traits and contextual characteristics -- may dictate information requirements. In this dissertation, I take a small but meaningful step toward addressing this vast challenge. This work offers insights at different steps of the process that can be directly applied to the design and study of human-AV interactive systems. I also provide a methodological framework to guide future applied and theoretical research.

\section{Contributions of this dissertation}
\begin{figure}[H]
  \centering
  \includegraphics[width=.9\linewidth]{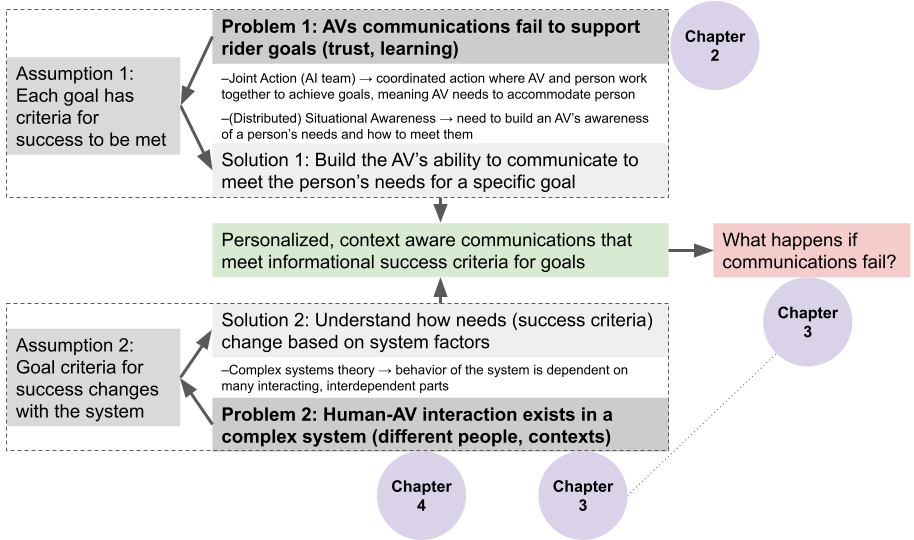}
  \caption[\textbf{Outline of the premises of this dissertation.}]{\textbf{Outline of the premises of this dissertation.} This dissertation is emergent from the intersection of two problems: (1) failure to support goals and (2) complex systems change success criteria needed to meet goals. By bridging these two problem areas, we can improve human-AV interaction. Each chapter in this dissertation looks at different areas of this problem-solution puzzle.}
  \label{thesis_diagram}
\end{figure}

In this dissertation, I examine the complex relationship between humans and autonomous vehicles, focusing on key informational factors influencing trust, understanding, and reliance in this rapidly evolving domain. I argue that accounting for the dynamic interplay between components of the human-AV system can improve interactions by successfully filling gaps in the rider and/or the AV's situational awareness. Figure \ref{thesis_diagram} gives an overview of the premises that relate to this dissertation.

Chapter 2 explores the novel application of an AV race car driving coach, identifying optimal communication strategies that enhance driving performance, confidence, and trust for learning in even the most extreme driving environments. Through an investigation of information types and presentation modalities, I find that AVs can successfully teach performance driving skills, but communications should be task-sensitive, modality-appropriate, and tuned to learner cognitive limits and goals. I provide design implications for AI coaches and human-machine interfaces supporting learning and knowledge transfer more broadly.

Chapter 3 investigates the impact of explanation errors and contextual characteristics on trust in autonomous vehicles, topics not widely explored in previous research. I find that errors negatively impact trust even in cases where the AV drives perfectly, and that an error's impact is dependent on the driving context within which it occurs. Further, certain personal traits may improve system perceptions overall. Results highlight the consequences of deploying faulty communication systems and emphasize the need for contextually adaptive and personalized AV explanations.

Chapter 4 deepens our understanding of trust formation, using machine learning (ML) to predict trust in young adults based on a wide range of personal factors. Results show that risk and benefit perceptions are the most significant predictors of trust in AVs, overshadowing the influence of other personal characteristics. This study highlights the importance of tailoring designs to individual traits and concerns: I discuss how this information can be used by an AV to personalize communications for riders with different trait compositions. The methodology that can be replicated for future research investigating trust in AVs and other AI systems for diverse populations.

Each of the three studies offers unique insights into how information shapes people's perceptions and interactions with AVs. All three independently deepen our theoretical understanding of how to address gaps in situational awareness within complex human-AV systems, and each study provides practical design implications for immediate improvement of state-of-the-art AV communication design. This dissertation ultimately shows how a systems' lens can lead to more successful AV (and AI more generally) design and interaction. By exploring different components of the human-AV interactive system, from foundational trust-building mechanisms to practical design interventions, I aim to demonstrate the value of a multifaceted investigation. As such, I hope this work can guide future researchers and designers in studying and developing better AI.

\section{Acknowledgements}
This introduction contains reprinted materials as they appear in \textit{Developing Situational Awareness for Joint Action With Autonomous Vehicles} (ArXiv~2024). Kaufman, Robert; Kirsh, David; Weibel, Nadir. The dissertation author was the primary investigator and author of this paper. \cite{kaufman2024developing}

\chapter[\underline{Communication Strategies:} Effects of Multimodal Explanations for Autonomous Driving on Driving Performance, Cognitive Load, Expertise, Confidence, and Trust]{\underline{Communication Strategies:} \\ Effects of Multimodal Explanations for Autonomous Driving on Driving \\ Performance, Cognitive Load, \\ Expertise, Confidence, and Trust}
\newpage

\section{Interim Summary and Chapter 2 Overview}
We start by studying communications themselves. If our ultimate vision is for AVs to meet the informational needs of their (diverse) riders in (diverse) contexts, this is a little like starting at the end. However, knowing how to satisfy information needs for a specific goal brings a clear and concrete understanding of what the end result may look like. In this chapter, I discuss how to optimize communications for the goal of learning from an AV, exploring techniques to help learners improve their race car driving. This chapter gives clear design insights for communications teaching a novice to drive and highlights other considerations that may be important, including trust and cognitive traits.

\textbf{Chapter 2 Overview} -- \textit{Advances in autonomous driving provide an opportunity for AI-assisted driving instruction that directly addresses the critical need for human driving improvement. How should an AI instructor convey information to promote learning? In a pre-post experiment (n = 41), we tested the impact of an AI Coach’s explanatory communications modeled after performance driving expert instructions. Participants were divided into four (4) groups to assess two (2) dimensions of the AI coach’s explanations: information type (‘what’ and ‘why’-type explanations) and presentation modality (auditory and visual). We compare how different explanatory techniques impact driving performance, cognitive load, confidence, expertise, and trust via observational learning. Through interview, we delineate participant learning processes. Results show AI coaching can effectively teach performance driving skills to novices. We find the type and modality of information influences performance outcomes. Differences in how successfully participants learned are attributed to how information directs attention, mitigates uncertainty, and influences overload experienced by participants. Results suggest efficient, modality-appropriate explanations should be opted for when designing effective HMI communications that can instruct without overwhelming. Further, results support the need to align communications with human learning and cognitive processes. We provide eight design implications for future autonomous vehicle HMI and AI coach design.}
\newpage

\section{Introduction}
Recent years have seen vast improvements in autonomous driving technology, with on-road vehicle testing currently being conducted in major cities around the world. The proposed benefits of autonomous vehicles (AVs) include increases in driving safety and efficiency while reducing driving infractions, traffic, and passenger stress \cite{fagnant2015preparing}. Though AVs may have a bright future, real-world deployment is hindered by a number of technological, infrastructural, and human-interaction roadblocks that are likely to take decades to solve. Putting it succinctly -- human-AV interaction exists within complex systems, making human-AV problems challenging to address. Meanwhile, the National Highway Traffic Safety Administration (NHTSA) estimates that in the United States alone, there are over 6 million car crashes each year, resulting in over 2.5 million injuries, 40,000 deaths, and \$340 billion in damages \cite{NHTSA2021, blincoe2022economic}. Of these, it is estimated that 94\% of vehicle crashes are due to human error \cite{singh2015critical}. Therefore, there is a large and pressing demand for technologies that may improve human driving ability. 

Understanding how to optimize AV communications to meet specific goals and needs of human riders remains a critical and open challenge. In this study, we seek to test a novel use of autonomous vehicle technology: can AVs teach humans to be better drivers?

We propose that augmenting driver learning by observing an AI driving coach may help address the glaring need for human driving improvement. This study explores a specific goal within the broader framework of this dissertation: meeting informational needs for enhancing human learning through AV communication. By investigating how explanatory techniques influence performance, this work contributes to the larger dissertation argument of how AV systems can be designed to satisfy informational criteria for success for specific goals in specific driving contexts. 

To test this concept, we conducted a mixed-methods driving simulator experiment in which participants observed one of four AI coaches providing different types of explanatory instructions. We evaluate how the AI coaching sessions impact learning outcomes such as participant-driving performance, cognitive load, confidence, expertise, and trust.

We leverage the domain of performance or “racetrack” driving to test study objectives. Performance driving is more challenging but very related to everyday driving, and it allows driving skills and knowledge to be built and tested more objectively. The goal of performance driving is to drive as fast as possible on a given track, maximizing the vehicle’s limits while minimizing the distance needed to travel \cite{braghin2008race}. Many performance driving skills directly translate to real-world driving contexts -- such as improvements to vehicle handling and situational awareness \cite{van2017differences, mckerral2021identifying} -- and thus, it is an appropriate proxy to study driving skill learning. 

Testing the potential of an AI driving coach in the context of performance driving has several major benefits over everyday driving. First, performance driving has specifically defined parameters for success, so we can objectively measure the effectiveness of our AI coach in a controlled environment. Next, it is a driving task many people are unfamiliar with, and thus we are able to test the potential of an AI driving coach on true novices. This helps maintain a consistent knowledge and skill baseline for our study sample, providing further consistency. This is important, as this dissertation's thesis suggests that the efficacy of communications may differ based on personal traits, like expertise level. In the context of this dissertation, leveraging a performance driving context can serve as a case study for how AVs can tailor communication strategies to support complex, goal-oriented tasks. By testing our AI coach on learning an extreme and challenging driving task, we hope to derive insights that can be generalized to even the most intense real-world driving situations, where user learning and situational awareness are essential for effective human-AV interaction.

Though novel in the realm of autonomous driving, learning from AI, particularly by means of AI explanations, is not a new concept \cite{carbonell1970ai}. Learning from AI is a rapidly expanding area of interest, particularly given the proliferation of AI-based large language model tools such as ChatGPT \cite{baidoo2023education, mozer2019artificial}. In domains such as medicine, explainable AI-based systems – for example, image classification systems for radiology – have been shown to help physicians learn new techniques to identify pathologies \cite{irvin2019chexpert, hosny2018artificial, duong2019artificial}. These systems present justifications in the form of `explanations' which provide a rationale for a system's decisions. Specific cases where human-interpretable AI explanations of decisions are produced have been singled out as especially helpful for learning and improvement \cite{holzinger2019causability, wang2019designing, soltani2022user}, particularly in cases where AI explanations are based on the explanations of human experts \cite{pazzani2022expert, kaufman2022cognitive}. AI has the ability to advance educational techniques in many other domains, from second language learning \cite{ruan2021englishbot} to programming \cite{becker2023programming}. Measuring the impact of interacting with AI systems on learning outcomes can be a challenge, and we can differentiate between explicitly testing learning via knowledge tests and operationalizing learning more functionally via outcomes like task performance \cite{zheng2023effectiveness}. For the present study, we measure learning in both ways. The main learning outcomes assess changes in driving performance before and after exposure to the AI coach; secondarily, we assess knowledge via quiz. In this way, we are able to measure the learning impact of the AI Coach directly from multiple perspectives.

In the context of autonomous driving, we propose that observing an AI driving coach may be an effective way to transfer knowledge and teach critical driving skills. How can success criteria for learning be effectively met by an AV? There have been a number of studies on human-machine interfaces (HMIs) for AVs, which often take the form of Heads-Up-Displays (HUDs) aimed at building transparency, accountability, and trust \cite{currano2021little, omeiza2021explanations}. Though highly relevant, these studies have majorly focused on non-learning areas of the driver experience, and none to our knowledge have tested the impact of HMI elements on driver learning specifically. For example, in the large corpus of work on human-AV trust formation, Morra et al. found that showing more informative displays increased participant willingness to test a real AV \cite{morra2019building}, Ruijten et al. found that displays mimicking human explanations increases trust \cite{ruijten2018enhancing}, and Koo et al. found that including why-type information in  display improves trust \cite{koo2015did}. Other HMI studies have emphasized the complexity of trust formation, proposing holistic and multifactor approaches to designing trustworthy interfaces \cite{kaufman2024developing, ekman2017creating, frison2019ux}. Beyond trust, \citet{schartmuller2019text} assessed display modality differences in driver comprehension and multitasking, finding that certain interfaces could reduce workload better than others. HMI experiments aimed at building situational awareness with AVs have shown improved awareness with display elements that allow a person to comprehend of why or why-not (contrastive) a vehicle is taking an action \cite{wiegand2019drive, omeiza2021not, omeiza2021towards}. Though these prior studies support vehicle HMIs as a promising avenue for coaching applications, how these may apply in instructional applications is currently unexplored. The work presented here seeks to contribute to the knowledge gap that exists between HMI design research and learning communities focused on optimal coaching techniques.

We focus on the role an AI Coach can play in improving a person’s driving ability -- using explanatory communications modeled after the instructions of human driving experts. As the most common in-car method for introductory performance driving instruction is to begin with observational learning (i.e., a passenger observing an expert driver), our explanatory communications are framed within this observational learning context. Using a state-of-the-art, full-motion driving simulator, we explore both if an AI coach’s instructions can improve a novice’s driving performance more than observation alone and, if so, what explanatory methods may be best.

Inspired by explanations given by expert driving instructors, we explore the role that HMIs may play in performance driving instruction. To design an effective HMI, it is important to determine both the type of \textit{informational content} that should be conveyed, as well as the \textit{presentation modality} of the information. This is especially important in the context of HMIs for safety-critical and cognitively demanding tasks like driving \cite{liu2023design}. Studies have shown mixed results on the impact of modality of information presentation. Some studies claim visual techniques should be preferred \cite{schartmuller2019text, chang2016don}, while others suggest auditory feedback as a better strategy \cite{jeon2009enhanced, locken2017towards}. Impact of the type and content of information presented also shows mixed results in terms of driver performance and preference, including differences between ‘what is happening’ or ‘why it is happening’ information \cite{koo2015did, omeiza2021not}. These explanation attributes have not been explored in the domain of AI coaching for driving instruction. 

To address these gaps of knowledge, this study uses a mixed-methods approach to assess study outcomes. Participants were randomized into one of four AI coaching groups, differing by the type and presentation modality of the information they presented. Before and after observation, measures were taken to compare changes in participant driving performance, cognitive load, confidence, expertise, and trust. Each of these factors have been highlighted as important for the development and adoption of HMIs and explainable AI systems more generally. Interviews were conducted to contextualize findings within the larger learning context, including building a more general understanding on concerns with AVs and how these may be mitigated. This broader view of AVs can help illuminate additional roadblocks that must be addressed for successful future adoption of AI driving coaches.
\vspace{1em}

\noindent
\textbf{Our research questions are as follows:}
\begin{enumerate}[topsep=0pt, nolistsep]
\item What are the pre-post impacts of observing an AI performance driving coach compared to a pure observation (no explanation) control, including the impacts of explanation information type and presentation modality?
\item What is the process of a novice learning performance driving, and how can an AI Driving Coach facilitate this learning?
\item What concerns do novices have in general about the deployment of AV technology, and how can these further inform future AV HMI design?
\end{enumerate}

\vspace{1em}
\noindent
\textit{Results from this study support the premise of AI coaching as a promising method for driving instruction and lead to important considerations for future HMI design. }
\newline
\newline
\textbf{Specifically, we contribute:} 
\begin{enumerate}[topsep=0pt, noitemsep]
\item A novel assessment of the impact of observing an AI Coach on performance-driving ability, cognitive load, confidence, expertise, and trust.
\item Insight into the impact of information type (‘what’ and ‘what + why’) and information presentation modality (visual and auditory) on the process of learning performance driving.
\item Eight design insights to inform the creation of future human-centered HMIs for driving and AI driving coaches.
\end{enumerate}

Ultimately, this study exemplifies how AI-driven explanations can address specific informational needs within the human-AV system. Building the situational awareness of the AV to include the needs (and how to meet them) of their riders is crucial to promoting successful deployment. These findings align with the dissertation's central argument: by designing AV communication strategies that are sensitive to task demands, user goals, and cognitive processes, we can advance the development of robust and adaptable AV systems.
    
\section{Method}
To address our research questions, participants were divided into one of four experimental conditions differing in the \textit{information type} and information \textit{presentation modality} given by the AI coach. For information type, we test two layers of information: (1) ‘What’ information provides descriptors of where the vehicle should drive; (2) ‘Why’ information explains the rationale behind why that position and movement is optimal, and is meant to build conceptual understanding. Some participants received just `what' information, while others received both `what' and `why' information in combination. With this manipulation, we seek to build upon prior work on the impact of information level on performance, preference, and trust \cite{koo2015did}. For information modality, we manipulated whether ‘what’ explanations are presented auditorily or in the form of a visual racing line projected on the track; These two additional conditions aim to provide clarification on the efficacy of visual and auditory information at conveying information about a vehicle's behavior.

We evaluate changes in driving performance from before observing the AI coaching condition to after observation. One crucial concept of learning performance driving focuses on vehicle positioning, which is a fundamental skill for novices. In performance driving, the optimal position follows a path on the track called the ‘racing line’, which minimizes time cost around corners and allows the driver to move as quickly as possible \cite{xiong2010racing, brayshaw2005quasi}. As the AI coach’s ‘what’ instructions primarily focused on the racing line, the main outcome of our study is how well participants positioned themselves while driving. Other performance measures include lap time, speed, and acceleration; these were addressed only in explanations that included a ‘why’ component. 

Additional secondary outcomes were addressed via interview and questionnaire. These include impacts on trust, knowledge and expertise, driving confidence, and feedback related to how helpful and effective the AI coach was at facilitating the participants’ learning process.

A large corpus of prior work suggests that a major drawback of mid-drive communication is the potential for information overload and high cognitive demand \cite{liu2023design}. We hypothesized that observational learning would provide the opportunity to transfer knowledge to a novice while avoiding cognitive overload, as a participant can pay attention to explanations without task switching between learning and driving decision-making. To further explore this phenomenon, we measure cognitive load for each of our study conditions.

\subsection{Participants}
A total of 41 participants were recruited for the study and completed all study procedures. Participants were all novices in the performance driving domain but were otherwise licensed to drive with at least 5 years of everyday driving experience. This ensured that participants had the baseline motor skills necessary to begin performance driving instruction.

\subsection{Performance Driving Simulation}
\begin{figure}[H]
  \centering
  \includegraphics[width=\linewidth]{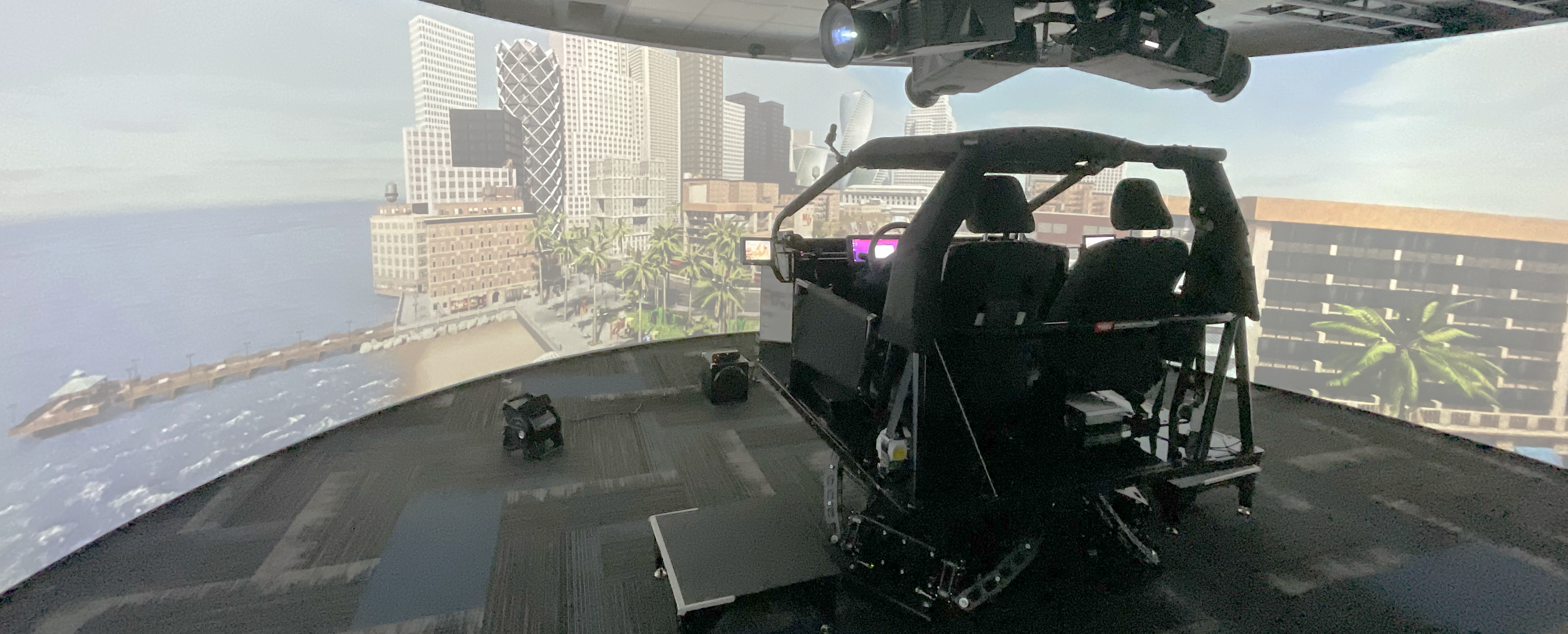}
  \caption{\textbf{Full-motion driving simulator.}}
  \label{sim_pano}
\end{figure}

All driving-related tasks took place in a state-of-the-art, highly immersive, full-motion driving simulator to ensure realism. This included a 270-degree wrap-around screen, a full vehicle cabin with a closely calibrated steering wheel and pedals, sound effects, wind effects, and cabin movement mimicking the feeling of a vehicle moving on the track (Figure \ref{sim_pano}). The simulator has a mounted tablet so that participants can answer survey questions between drives without leaving the simulation, ensuring real-time responses and preserving immersiveness. The specific racing context chosen was a highly accurate simulation of the professional driving course Thunderhill Raceway in Willows, California, USA.

During the AI coaching observation sessions, the vehicle drove autonomously with no user input. The specific model used to control the vehicle was a reinforcement learning agent trained using the DSAC algorithm \cite{ma2020dsac}, with observations and rewards tuned for Thunderhill. The agent's policy was optimized for the physical features and limitations of the vehicle and the geometry of the track \cite{chenlearn}. This model was also used to compute the “ideal” racing line for calculating performance.

\subsection{AI Coach Explanations}
During the AI coach observation sessions, explanatory instructions were provided to all participants except the control group. Auditory explanations were presented via an in-cabin speaker. These were produced using Amazon Polly Text-to-Speech \cite{polly}. Visual explanations were projected directly onto the track in the simulated environment.

We modeled the AI coach’s explanations off instructions given by four real performance driving experts and instructors who all had real-world experience driving and instructing on the real-world Thunderhill Raceway. First, two expert performance drivers were ethnographically observed giving real-world driving instruction at Thunderhill Raceway via several hours of video. Next, this was paired with think-aloud and interview sessions with two additional expert performance drivers as they drove the same simulated Thunderhill track used in the experiment. This qualitative assessment was used to ground and justify the specific explanations given by the AI coach to ensure accuracy. Descriptions given by the experts also helped scaffold the learning procedure, which we have described in detail later in this chapter. 

We found that the most common in-car method for introductory performance driving lessons began with observational learning, specifically when an instructor drives the performance vehicle on the track while providing instruction as the novice observes and listens. This instruction is typically auditory; however, may also include visual explanation via pointing. Thus, our explanatory communications are framed within an observational learning context, and explanations are motivated by these explanatory modalities. In safety-critical domains like driving, observational learning may be a safe and effective method to transfer knowledge to a novice. This study was designed to assess whether observing an AI coach drive and explaining its behavior are effective methods to improve a novice’s driving performance,  among other outcomes.

\subsubsection{Dimension 1: Information Type}
 First, we focus on the type of information presented to a driver: ‘what’ information provides descriptors of where the vehicle should drive, specifically following the ideal racing line; ‘why’ information explains why that position and movement is optimal. These are inspired by the explanations examined in experimental work on explanations of driving behavior by Koo et al. \cite{koo2015did}; discussed in frameworks by Wang et al. \cite{wang2019designing} and Lim et al. \cite{lim2019these}; and explored in cognitive science and philosophy \cite{miller2019explanation}. In our experiment, the ‘what’ explanations are designed to help the participant know where to go on the course in order to follow the racing line, such as “to the left edge of the track.” The ‘why’ explanations are designed to help the participant develop a more generalized understanding of performance-driving concepts, techniques, and decision-making. These include ways to optimize speed, acceleration, and lap time. For example, the vehicle moves “to the left edge of the track \textit{...to decrease the traction needed to get around the curve.}” We separate participants into ‘what’-only and ‘what + why’ conditions and compare these to a \textit{no explanation} control group to assess the effect of information type.

\subsubsection{Dimension 2: Information Modality}
\begin{figure}[htp]
  \centering
  \includegraphics[width=\linewidth]{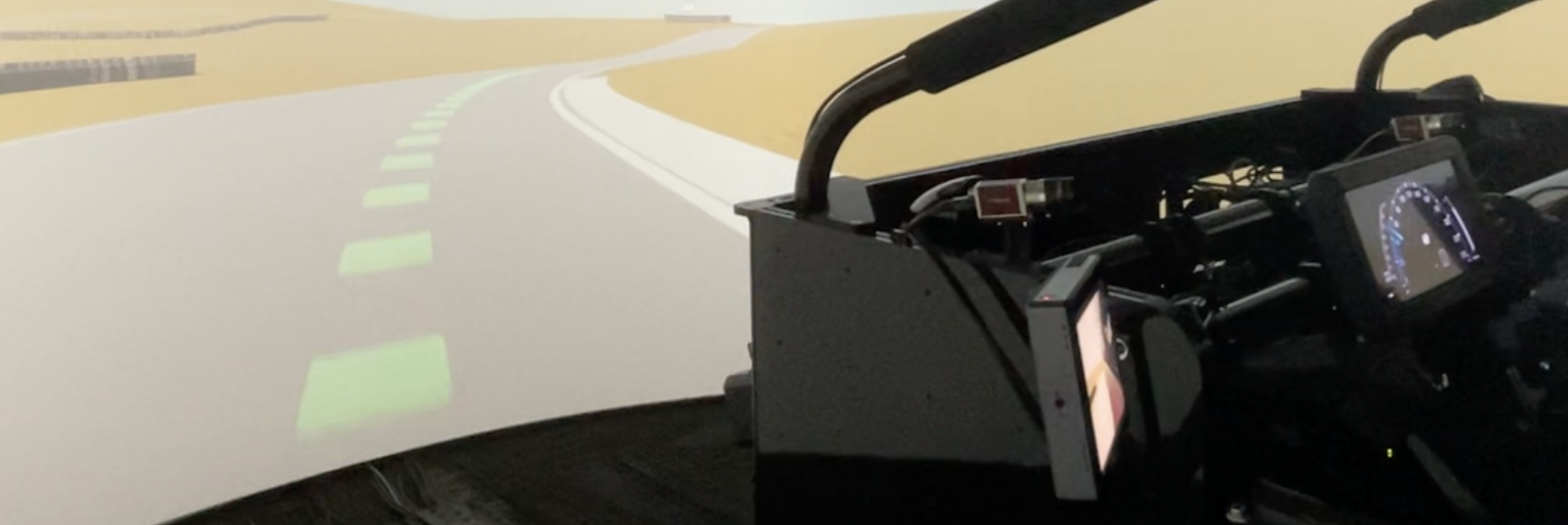}
  \caption[\textbf{Visual `what' racing line projection.}]{\textbf{Visual `what' racing line projection.} The green racing line projected on the track is an example of a visual ‘what’ explanation seen by Group 4.}
  \label{sim_line}
\end{figure}

We are further interested in information presentation modality. To assess the impact of multimodal explanations, two different ‘what + why’ conditions were tested. One received visual ‘what’ explanations, while the other received only auditory cues. This visual ‘what’ explanation shows a visual projection of the racing line (Figure \ref{sim_line}). Auditory ‘what’ explanations were designed to convey the same information given by the visual racing line instead of using verbal cues. Auditory ‘why’ explanation content remained consistent, as the instructions conveyed were deemed too complex for simple visualizations.
%\vspace{2em}

%\vspace{-0.5cm} %brings conditions section higher to fit fig 2 on page
%\vspace{2em}
\subsection{Conditions}
Participants were randomly assigned to one of four conditions (Table \ref{ex_conditions}).

\begin{table}[htp]
\centering
\caption[\textbf{Experimental conditions: Information Content and Modality of Presentaton.}]{\textbf{Experimental conditions: Information Content and Modality of Presentaton.} Visual ‘What’ Example: A visualization of the racing line on the track. Auditory ‘What’ Example: “Move the car to the left edge of the track.” Auditory ‘Why’ Example: “Moving to the left will widen the arc of the next turn, preserving momentum.”}
\begin{tabular}{p{.1\linewidth}|p{.5\linewidth}|p{.1\linewidth}}
\toprule
\textbf{Group} & \textbf{Condition} & \textbf{n} \\
\midrule
\textbf{1} & No explanation - Control & 10  \\
\hline
\textbf{2} & Auditory ‘What’ & 10 \\
\hline
\textbf{3} & \underline{Auditory} ‘What’ + Auditory ‘Why’ & 11 \\
\textbf{4} & \underline{Visual} ‘What’ + Auditory ‘Why’ & 10 \\
\bottomrule
\end{tabular}
\label{ex_conditions}
\end{table}

\subsection{Measures}
\subsubsection{Driving Performance}
\begin{enumerate}[topsep=0pt, nolistsep]
\item \textit{Racing Line Distance}: This is how far participants drove from the ideal racing line, measured in meters. It is calculated as the absolute value of the difference between a participant’s line position and the ideal line position determined by the RL model, averaged across the track. Lower racing line distance means a participant was closer to the ideal.
\item \textit{Maximum speed}: This is the highest speed achieved in miles per hour. Higher speed is indicative of better performance.
\item \textit{Average Acceleration}: This is calculated as the mean change in velocity per 1 second when accelerating forward, measured as meters per second per second. Higher acceleration is indicative of better performance.
\item \textit{Lap time}: The time taken to complete one lap in seconds. Lower time is indicative of better performance.
\end{enumerate}

\subsubsection{Trust} 
\begin{enumerate}[topsep=0pt, nolistsep]
\item \textit{Trust in AV Driving Coach}: This is the average of four, 5-point Likert scale (strongly disagree - strongly agree) questions similar to “I would trust the advice given to me by an AI driving coach.”
\item \textit{General Trust in AVs}: This is the average of four, 5-point Likert scale (strongly disagree - strongly agree) questions similar to “I would trust an autonomous vehicle to drive me around safely.”
\end{enumerate}

\subsubsection{Expertise} 
\begin{enumerate}[topsep=0pt, nolistsep]
\item \textit{Self-Reported Performance Driving Expertise}: This is the average of three, on a 5-point Likert scale (strongly disagree - strongly agree) questions similar to “I understand the concepts behind performance driving.”
\item \textit{Racing Line Knowledge}: This is the number of correct responses given on four true/false questions and one diagram question, designed by the research team to test how well participants understood the racing line concept. It was given after all driving tasks concluded.
\end{enumerate}

\subsubsection{Confidence and Cognitive Load} 
\begin{enumerate}[topsep=0pt, nolistsep]
\item \textit{Performance Driving Confidence}: This is the rating, from 0-10, given in response to the question, “If you were asked to performance drive in real life (without assistance), how confident would you feel?”
\item \textit{Cognitive Load}: This is measured using the widely accepted NASA Task Load Index (TLX) \cite{hart2006nasa}.
\end{enumerate}

\subsection{Procedure}
\begin{figure}[htp]
  \centering
  \includegraphics[width=\linewidth]{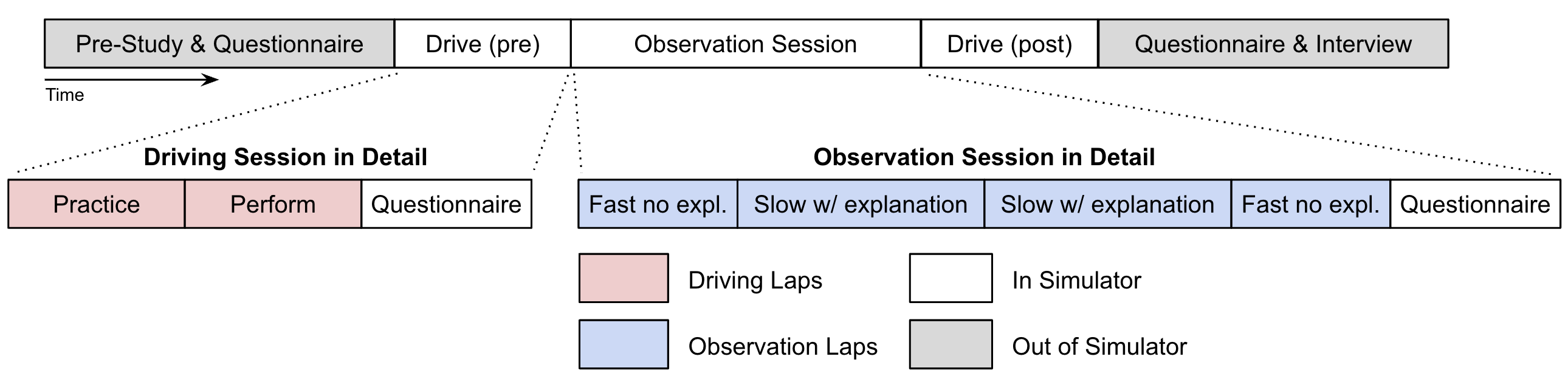}
  \caption[\textbf{Study timeline.}]{\textbf{Study timeline.} The study duration was approximately 1 hour in total and included a questionnaire, participant driving, AI coach observation, and interview.}
  \label{procedure}
\end{figure}

The study was approved for human subjects research by WCG IRB (external) and informed consent was collected from all study participants before procedures began. All study procedures were performed in accordance with relevant guidelines and regulations. Figure \ref{procedure} shows the study procedure.

The study was framed in the context of learning performance driving. Before and after exposure, participants drove two laps. First, a practice to familiarize participants with the track, controls, and vehicle capabilities. Next, a performance where participants were told to drive as fast as possible without going off track or losing control of the vehicle, as if they were going for their best lap time. These instructions align with the standard instructions for racing used in past studies \cite{braghin2008race}.

During the observation, participants observed the AI coach drive four laps around the same track in the driving simulator. During the first and last lap, the RL agent drove at full speed; these did not contain any explanations from the AI coach and remained consistent across all groups. The second and third laps occurred at a slower speed and contained explanations from the AI coach. The same explanations were repeated twice to promote uptake: in all, about 8 minutes of explicit instruction was given to participants. Explanations varied by the condition the participant was assigned. Participants in Group 1 did not receive any verbal or visual instruction during these laps, but still observed the vehicle drive the course. 

Questionnaires in the simulator were given for measures of confidence and cognitive load. Questionnaires outside of the simulator to track trust, and expertise, as well as to get feedback on participants’ experience overall. A 10-15 minute semi-structured interview was conducted at the end, during which participants gave feedback on the AI coach’s instructions, described their own learning procedure, and discussed trust and opinions on autonomous vehicles and coaches.

\section{Results}
\subsection{Analysis Methods}
To assess score differences from before to after observation, we used linear mixed effects (LME) models via the R package ‘lme4’. LME models yield similar results as mixed model ANOVAs, however, they allow for greater flexibility for pre-post experiments while allowing random effects to reduce the probability of a Type 1 error \cite{boisgontier2016anova}. To test the impact of the explanations received by different groups, models were made with fixed effects for Timepoint (pre-post) and Group (1, 2, 3, or 4), with random effects for Subject ID added to control for individual differences. This tests if the pre-post \textit{change} experienced by each experimental group differs from the pre-post \textit{change} of Group 1, the pure-observation control. In some cases, we are also interested in the impact of the observation sessions themselves, regardless of group assignment. This can give insight into the general impact of observing any version of the AI coach. In these cases, the fixed effect for Group is removed, allowing us to test pre-post differences with all groups combined. To avoid skewing results, outliers were removed from racing line analysis when they represented extreme values related to a participant’s vehicle moving far off the track; other racing measures are still valid in these cases.

\subsection{Impact of AI Coaching Information Type and Modality on Driving Performance}
\begin{figure}[htp]
  \centering
  \includegraphics[width=\linewidth]{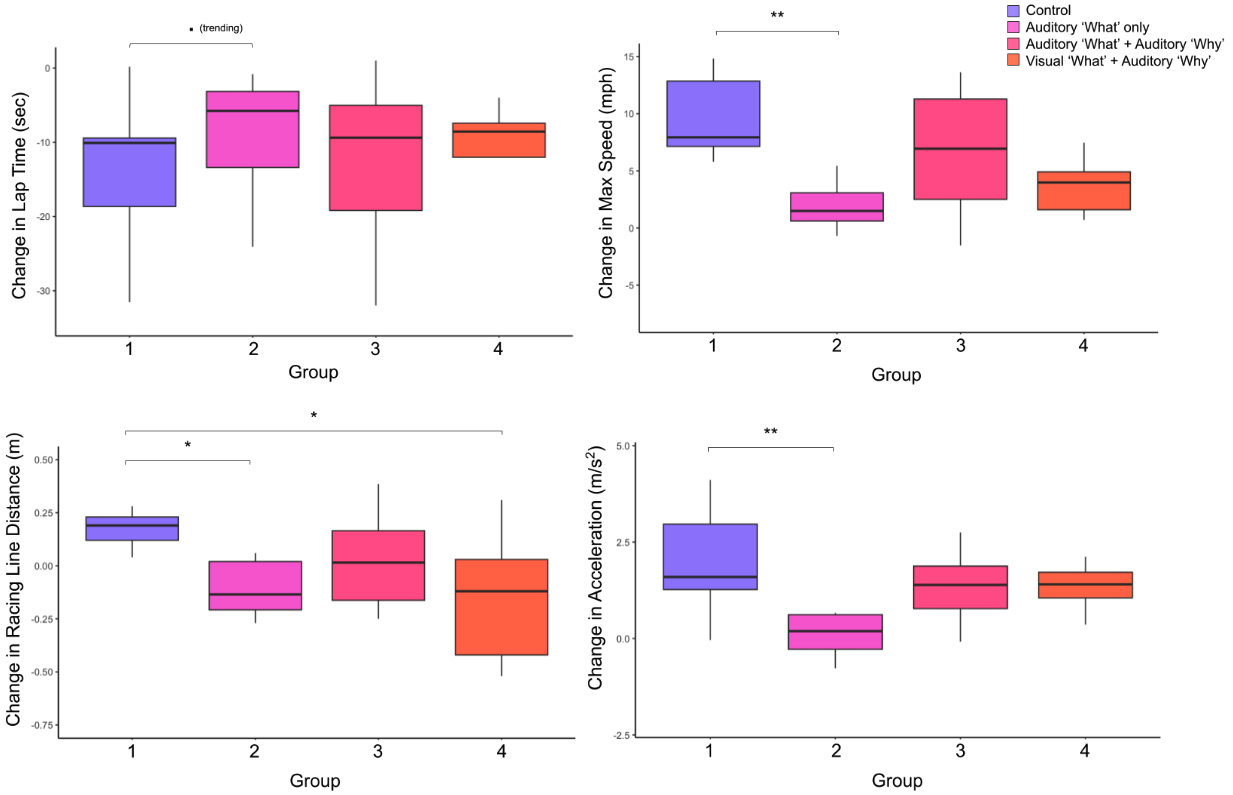}
  \caption{\textbf{Change (\begin{math}\Delta\end{math}) from pre-post observation for driving performance measures by group.}}
  \label{group}
\end{figure}

A summary of driving performance results is included in Table \ref{results_summary} and Figure \ref{group}.

\begin{table}[htp]
\centering

\caption{\textbf{Results summary: Impact of AI Coaching measured via LME.}}
\begin{tabular}{p{.4\linewidth}|p{.3\linewidth}|p{.2\linewidth}}
\toprule
\textbf{\parbox{\linewidth}{\vspace{1pt} Measure \vfill}} & \textbf{\parbox{\linewidth}{\vspace{1pt} \begin{math}\Delta\end{math} Group vs. \\ \begin{math}\Delta\end{math} Group 1 Control \vfill}} & \textbf{\parbox{\linewidth}{\vspace{1pt} \begin{math}\Delta\end{math} Pre-Post \\ (all combined) \vfill}} \\
\midrule
Lap Time & . (trending) Group 2 (+) & *** (-) \\
\hline
Max Speed & ** Group 2 (-) & *** (+) \\
\hline
Racing Line Distance & * Group 2 (-), * Group 4 (-) & NS \\
\hline
Average Acceleration & ** Group 2 (-) & *** (+) \\
\hline
Trust in AV Coach & NS & *** (+) \\
\hline
General Trust in AVs & NS & ** (+) \\
\hline
Self-report Perf. Driving Expertise & NS & *** (+) \\
\hline
Self-report Driving Confidence & NS & *** (+) \\
\bottomrule
\multicolumn{3}{p{.95\linewidth}}{\raggedright \textit{Sig. Codes: ‘***’ p $<$ 0.001 $|$ ‘**’ p $<$ 0.01 $|$ ‘*’ p $<$ 0.05 $|$ ‘. (trending)’ p $<$ 0.1 $|$ Direction of effect in parentheses.}}
\end{tabular}
\label{results_summary}
\end{table}

Distance to the ideal racing line is our primary measure of AI coaching success, as vehicle positioning was the primary focus of the training. This is a measure unlikely to improve from practice alone due to its lack of intuitiveness for a novice driver. Looking between groups, we find significant differences in racing line distance. Specifically, differences are found between groups receiving explicit instruction (2, 3, 4) compared to the control group (1) who did not receive explicit instruction. All groups with explicit instruction saw favorable pre-post change (i.e. were closer to the ideal racing line), while the control group itself got worse pre-post. For Group 2 (\begin{math}\beta\end{math} = -0.3, t(35) = -2.3, p $<$ .05) and Group 4 (\begin{math}\beta\end{math} = -0.3, t(35) = -2.0, p $<$ .05), differences were significant compared to Group 1 (Figure 4). These results imply that explicit instruction on the racing line was helpful for learning beyond simple observation and that both information content type and information modality play a role in how effective an AI coach will be.

Among other driving performance measures, Group 2 saw the most differences compared to Group 1. Group 2 improved less on lap time (\begin{math}\beta\end{math} = 10.3, t(37) = 1.8, p = 0.08), max speed (\begin{math}\beta\end{math} = -9.0, t(37) = -3.4, p $<$ .01), and acceleration (\begin{math}\beta\end{math} = -1.6, t(37) = -2.7, p $<$ .01) compared to the pre-post improvement of the control (Figure 4). Group 3 and 4 did not see any significant pre-post differences compared to the control by these measures.

Combining across all groups allows us to assess pre-post changes that stem from simply observing any version of the AI coach. When participants are lumped together, results show improvement from pre- to post-exposure for several aspects of driving performance, including faster lap times (\begin{math}\beta\end{math} = -13.8, t(40) = -6.8, p $<$ .001), greater average acceleration (\begin{math}\beta\end{math} = 1.3, t(40) = 5.9, p $<$ .001), and greater max speed (\begin{math}\beta\end{math} = 5.7, t(40) = 5.5, p $<$ .001). Interestingly, we did not find an all-groups-combined benefit of the AI coach on a participant’s average distance to the ideal racing line, further emphasizing the importance of the specific explanation received. These results as a whole demonstrate a clear positive trend on the impact of observing the AI driving coach in helping improve these aspects of driving performance. They also introduce nuance in the impact of different types and modalities of explanations on performance results.

%\vspace{1em}
\subsection{Impact of AI Coaching Information Type and Modality on AV Trust, Self-perceived Confidence, and Expertise}
As a whole, impressions of the AI coach were very positive across all groups: 90\% of participants agreed the AI coach was helpful, 93\% agreed it helped them understand performance driving better, and 85\% agreed it helped improve their driving.

\begin{figure}[htp]
  \centering
  \includegraphics[width=\linewidth]{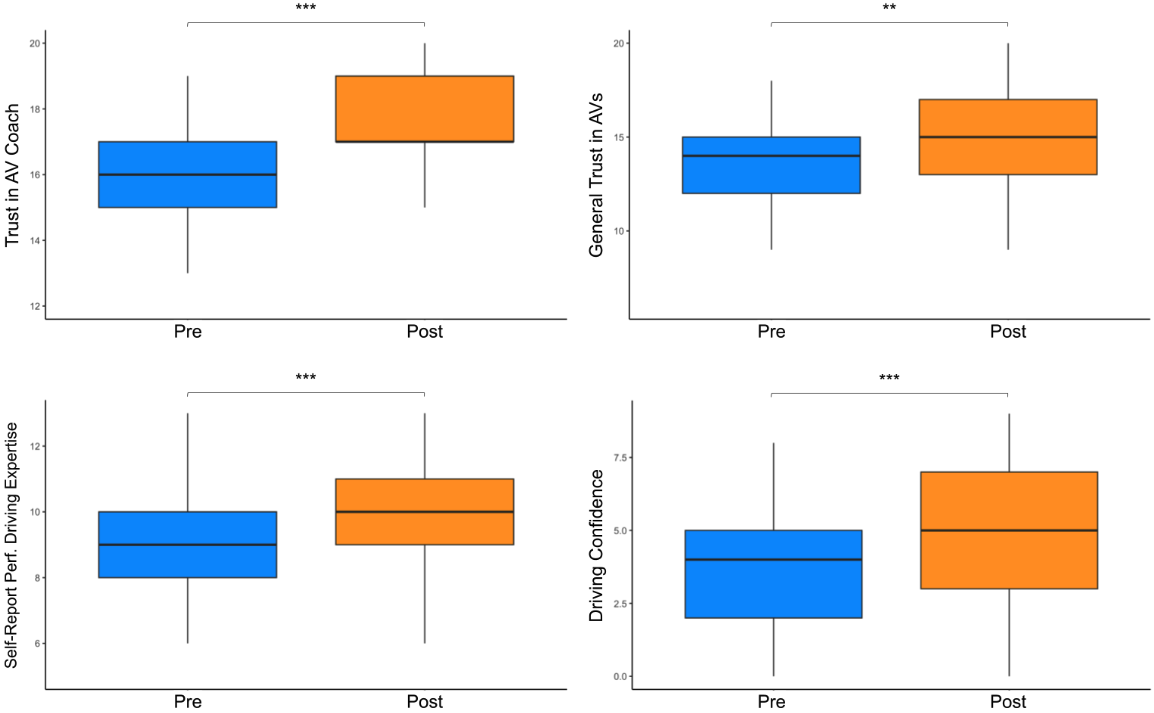}
  \caption{\textbf{Pre-post scores for self-report measures, all groups combined.}}
  \label{pre-post}
\end{figure}

Observing the AI driving coach was subjectively impactful for all groups with regard to trust, confidence, and self-reported expertise. We found all-groups-combined pre-post increases in perceived trust in driving coach (\begin{math}\beta\end{math} = 1.5, t(40) = 5.6, p $<$ .001) and in autonomous vehicles in general (\begin{math}\beta\end{math} = 0.9, t(40) = 3.3, p $<$ .01). Participants had higher self-report performance driving expertise (\begin{math}\beta\end{math} = 1.0, t(40) = 3.7, p $<$ .001) and confidence in their performance driving skills (\begin{math}\beta\end{math} = 1.4, t(40) = 6.6, p $<$ .001). We did not find significant differences in these measures between groups, however. See Table \ref{results_summary} and Figure \ref{pre-post} for a summary.

Scores on the 5-question racing line knowledge quiz – which was only given post-observation – showed that Group 4 trended towards being higher than Groups 1 (p = 0.06) and 2 (p = .10), but did not reach statistical significance. These differences were assessed using ANOVA with a Tukey HSD post-hoc test.

We assessed if any of these measures – trust, confidence, or expertise – were able to predict a participant’s pre-post change in racing line distance. This was done by categorizing participants into groups based on score percentile. We found participants in the top 33\% percent of trust in coach (\begin{math}\beta\end{math} = 0.4, t(36) = 3.1, p $<$ .01) and trust in AVs in general (\begin{math}\beta\end{math} = 0.3, t(35) = 2.2, p $<$ .05) moved closer to the racing line compared to those in the bottom 33\% of each trust measure respectively. Confidence, self-report expertise, and expertise measured via knowledge quiz did not appear to impact change in racing line distance.

%\vspace{-1em}
\subsection{Impact of AI Coaching Information Type and Modality on Cognitive Load}
We are interested in understanding the impact of each type of AI coach on cognitive load. Looking specifically at cognitive load directly after the observation sessions, we find that Group 3 had the highest cognitive load (mean = 14.2, se = 5.7), followed by Group 2 (mean = 14.0, se = 3.2), Group 1 (mean = 13.1, se = 7.0), and Group 4 (mean = 11.9, se = 5.4). Though not statistically significant from each other via ANOVA, these group scores follow the same trend reported in the interview phase and thus may be meaningful for explaining differences observed between groups.

We investigated if cognitive load before or after observation impacted pre-post change in racing line distance. We found individuals in the lowest 33\% of cognitive load pre-observation improved less with respect to the racing line compared to those in the highest 33\% (\begin{math}\beta\end{math} = 0.29, t(37) = 2.5, p $<$ .05). We also found those in the middle 33\% of cognitive load post-observation got closer to the racing line compared to those in the highest 33\% for the cognitive load (\begin{math}\beta\end{math} = -0.28, t(36) = -2.5, p $<$ .05). These imply that cognitive load impacts a participant’s ability to learn and implement racing line knowledge.

\subsection{Interview Insights}
Qualitative analysis of the post-observation interview revealed insights into the learning processes of participants, how the HMI helped or hindered these processes, and brought a new understanding of participant dispositions towards trusting autonomous vehicles. Key themes emerged from the interviews that may inform future HMI design for AV Coaches.

\subsubsection{Information Content, Presentation, and Modality}
\textbf{All groups wanted explanations of additional types.} For participants in Group 1 – who received no explanations – explicit ‘what’ and ‘why’-type information, such as those given to other groups, were desired by nearly all participants. Though many noticed patterns, participants experienced a large amount of uncertainty around what the AI Coach was trying to teach them. One participant explicitly stated, “at first … I didn’t get what the [AI coach] was teaching me.” Participants in Group 2 – who received auditory ‘what’ information only – wanted a further rationale for the `what' instructions they received (i.e. they wanted `why' information). One participant commented that they “didn’t know why the vehicle [moved in a particular way], so [they] couldn’t generalize.” Participants in Groups 3 and 4 expressed a desire for additional visual explanations to help them with the timing and magnitude of accelerating and braking inputs. Finally, a few members of Group 4 suggested that real-time or post-drive feedback would also be helpful for their improvement.

\textbf{Too much auditory information caused a feeling of being overwhelmed and a desire for more efficiency.} Participants in Group 3 – who received auditory `what' and auditory ‘why’ information – chiefly expressed a desire for \textit{more efficiency} in the information presented to them. This was the result of Group 3’s tendency to be overwhelmed with the amount of information presented. “Substantial”, “overwhelming”, and “a lot” were all used to describe the amount of information presented. In essence, participants still wanted comprehensive instruction covering a wide range of racing topics, but they wanted this instruction to be easier to consume and delivered in a less burdensome way. The rationale for this request was that it was too difficult to attend to and comprehend all of the auditory ‘why’ information while also receiving auditory `what' information (though the two never directly overlapped). 

\textbf{No explanation and auditory ‘what’ explanations introduced uncertainty.} In addition to uncertainty about what the coach was trying to teach them, Group 1 participants were unsure of exactly how to position themselves on the track or why they should do so. “It was hard to know what the important parts are [with observation alone],” noted another participant. This introduced additional challenges to the learning process. Participants in Groups 2 and 3 expressed difficulty with the preciseness of auditory ‘what’ instructions, as they had to internally map auditory visuospatial cues such as “move to the left edge” to precise visual positions on the track with inherent uncertainty. These participants wanted higher specificity in the position information they received in order to reduce the amount of guesswork.

\textbf{Visual `what' information was preferred or desired near-universally.} Many participants in Groups 2 and 3 explicitly requested a visual racing line. For example, one participant expressed a desire for a “visual line showing … details such as how much to go left, etc.” This was primarily as a means to overcome the uncertainty of auditory ‘what’ explanations. Group 4 – who received visual `what' and auditory `why' information – appeared far more satisfied with their explanations than the rest. They expressed that the visual projection of the racing line was helpful, and in contrast to Group 3, they did not report any issues with overwhelm nor a desire for more efficiency. One participant commented, “the path on the track was ... very helpful. [It] helped me feel more comfortable and confident”.

\subsubsection{Trust in Autonomous Vehicles}
For nearly all 41 participants, a lack of trust in autonomous vehicles was a key issue hindering their willingness to adopt AV technology. Aligning with prior research on AV trust, the most common trust concern for our participants was over the AV's ability to perform safely and reliably \cite{choi2015investigating}. There was a variance between participants: while some only expressed concerns for extreme or rare circumstances, others felt less confident in AVs for any circumstance involving the potential for unexpected occurrences (such as animals), pedestrians, or other drivers. This implies that trust in AVs may be contextually dependent and subject to individual differences. Other concerns included a lack of trust in the companies building the AVs, concerns over AVs running under-tested beta software, legal liability, losing the fun of driving, and AVs taking jobs from humans.

To help alleviate many of these concerns, participants reported that trust could be built through repeated, positive experiences with AVs. Many also suggested that AI explainability would help them feel more comfortable. Specifically, validation that an AI “perceives” what is around it, knowing how an AV will behave in specific driving situations, and understanding the rationale for decisions. Participants wanted to know \textit{when} an AV can be relied upon and to maintain the ability to regain control from an AV if they are feeling uncertain or uncomfortable.

\section{Discussion}
The findings in this chapter demonstrate that AV systems can successfully facilitate human learning by tailoring communication strategies to task-specific goals and individual user needs. The demonstrated ability of the AI coach to enhance performance driving highlights a broader potential for AV systems to serve as collaborative partners in achieving diverse goals.

Succinctly, this study aimed to assess the viability of an AI driving coach for performance driving instruction, as well as provide insight into how HMIs for AI driving coaching can be improved in the future. Using a mixed-methods approach (n = 41), we find that an AI driving coach may be a successful method to improve driving performance. By enabling users to better understand and predict their environment through tailored explanations, the AI coach contributes to the development of shared situational awareness, a key requirement for successful human-AV teaming. These results suggest that AV systems can not only perform driving tasks, but also dynamically adjust their communication to build the collective situational awareness required for joint action. As such, the AI coach serves as a model for how autonomous systems might support users in not only task execution but also skill acquisition and cognitive growth.

The findings have practical design implications for the development of human-machine interfaces (HMIs) in autonomous vehicles. These design implications are a way to meet success criteria for the goal of learning in a performance racing context for this given population. The results underscore the importance of modality-appropriate and task-sensitive communication strategies for effective HMI design. These insights extend beyond performance driving to inform broader principles for designing AV systems capable of adapting to diverse user goals and contexts. Specifically, the results highlight the need for HMIs that mitigate cognitive overload, direct user attention, and enhance trust, particularly in high-stakes or unfamiliar scenarios. By considering how explanations direct attention, reduce uncertainty, and alleviate overload, designers can better align communication strategies with user learning and cognitive processes.

While this study focused on a controlled performance driving context, future research should explore the generalizability of these findings to everyday driving scenarios and diverse user populations. Investigating how communication strategies can adapt to varying user traits and environmental conditions remains a critical next step in realizing the vision of adaptable and personalized AV systems. This future work will help solidify the broader applicability of the communication strategies tested here, ensuring they can address a wider range of user needs and situational demands. We discuss key findings and their implications for future HMI design in these contexts.

\textbf{AI Coaching is a viable method for performance-driving instruction.} Both quantitative and qualitative results of this study support the viability of an AI coach for performance-driving instruction. The primary area of focus for our AI coach was to provide instruction on racing line positioning. For racing line distance, we observed that Group 1 (control) got worse pre-post, while Groups 2 (auditory ‘what’ only), 3 (auditory ‘what’ and auditory ‘why’) and 4 (visual ‘what’ and auditory ‘why’) improved. The benefits of explicit instruction on racing line distance are unsurprising, as positioning on the racing line may be counterintuitive for novices. For instance, the racing line often follows the edge of the track, whereas everyday driving generally requires sticking to the middle of a lane. We also found significant overall pre-post-observation differences in several areas of performance driving, including faster lap times, max speeds, and acceleration rates before and after observation. Survey and interview data support AI coaching as an instructional method: most participants found the coach helpful, and overall boosts were found in participant expertise, trust, and confidence.

Proving nuance to our findings, Groups 2 and 4 were the most effective at improving racing line distance; however, Group 2 saw reductions in other areas – such as speed and acceleration – while Group 4 did not. This suggests that the specific instructions chosen greatly impact an AI coach’s effectiveness. During the interview, each group expressed concerns and desires related to the way the type and modality of information impacted their learning. While these differed across groups, concerns and desires were generally shared by members of the same group. The fact that we see significant differences for Groups 2 and 4 but not for Group 3 further underscores the importance of carefully selecting the information conveyed by an AI coach. 

We attribute group differences to how information directed attention, mitigated uncertainty, and influenced overload experienced by participants. These, in turn, affected how successfully participants could go about the learning processes.

\textbf{The type of information received by participants played a role in directing attention.} This aligns with prior work in explainable AI suggesting that a feature’s presence (or absence) is meaningful and thus relevant for directing focus \cite{miller2019explanation, soltani2022user}. For our study, ‘what’ information aimed to teach participants how to adhere to the racing line. Group 1 – who did not receive this information – suffered in this regard. Group 2 only received ‘what’ information, and thus, we suspect that they focused their attention more on adhering to the racing line and less on optimizing speed, acceleration, and lap time– topics discussed in the ‘why’ explanations. Thus, Group 2 saw less improvement in these other areas. By contrast, Groups 3 and 4 received ‘what’ and ‘why’ information, and thus focused their attention on several areas of performance driving simultaneously. Consequently, Group 4 also got closer to the racing line compared to the control group without seeing sacrifice in other measures like Group 2. We would have expected Group 3 to also have gotten closer to the racing line, as they also received the same types of information as Group 4. We suspect, however, that Group 3’s improvement was hindered by two consequences of information modality: uncertainty and information overload.

\textbf{The uncertainty of auditory and visual ‘what’ explanations impacted the ease of processing.} For individuals in Groups 2 and 3, the uncertainty of auditory ‘what’ information presented to them was a major point of concern. To integrate ‘what’ information effectively, participants needed to transform auditory visuospatial cues such as “move to the left edge” into precise positions on the track. This extra step of transformation, which involves both creating a spatial representation for the linguistic cue and translating it into visual working memory \cite{magliano2016relative}, was not required with the visually presented racing line. Auditory ‘what’ cues were included in this study in alignment with Wickens’ Multiple Resource Theory, which suggests that auditory information should be more efficiently incorporated during visually heavy tasks, such as driving, due to their different channels of processing \cite{wickens2020processing}. In our case, however, we believe that presenting information auditorily made ‘what’ information more difficult to integrate due to the extra step of processing. For example, the instruction “move to the left edge of the track” lacks details specifying exactly \textit{where} the participant should be aiming, and requires them to visually map the auditory spatial cue to a visual position on the track in front of them. This is supported by the modality appropriateness hypothesis of multisensory integration, which suggests that the effectiveness of information integration is dependent on the context of the task \cite{welch1980immediate}. For a visual context of finding the racing line, visual cues were more efficient. Though we had expected the lack of explanatory preciseness for auditory explanations to be supplemented by observing the movements of the vehicle itself, participants expressed that the uncertainty of the instruction created too much “guesswork”. As a result, our results support that visuospatial ‘what’-type information should be presented visually – such as via a racing line projection – as opposed to auditorily. 

\textbf{The modality and type of information affected the cognitive burden and overwhelm experienced by participants.} Decades' worth of prior research shows that increases in epistemic uncertainty or ambiguity can increase the cognitive burden of information processing \cite{kirschner2002cognitive, enke2023cognitive}. Integrating these theories into our observations suggests that Groups 2 and 3 both experienced higher cognitive burdens processing the more uncertain auditory ‘what’ information than Group 4 did processing more precise visual ‘what’ information. For Group 2, who had no other information to pay attention to, this increase in demand may have been frustrating but did not impact their ability to stay on the racing line. For Group 3, we believe the increase in cognitive demand required to deal with uncertainty, combined with the increase in demand from receiving additional auditory ‘why’ information, was sufficient enough to overload Group 3 and prevent them from improving. In other words, there is a combinatorial effect of the burden of dealing with uncertainty and the burden of adding additional information to learn. Indeed, though not significantly different from the other groups, participants in Group 3 had the highest cognitive load and complained the most about being overwhelmed in their interviews. Group 4, who received visual ‘what’ and auditory ‘why’ information, had the lowest cognitive load of all groups. This aligns with prior work suggesting multimodal interfaces may promote more efficient processing of multiple sources of information \cite{turk2014multimodal}, provided that each is integrated in modality-appropriate ways. Our theory is further supported by the Yerkes-Dodson law, where too much cognitive burden is a detriment to performance \cite{yerkes1908relation}. We also found that Group 4’s knowledge of the racing line trended towards being significantly higher than both Group 1 and Group 2, while Group 3 showed no difference between these non-why groups. This implies that the visual ‘what’ explanations and their lower cognitive burden may have helped Group 4’s ability to take in auditory ‘why’ information compared to Group 3. Taken as a whole, these findings suggest that information modality and type influenced the amount of cognitive load and sensation of overwhelm for participants in each group, which affected performance and preference in turn. They also suggest that – in this particular instance – the uncertainty of information impacted the burden more than the sheer amount of information.

The implication of these results is that there is a nuanced relationship between information type and information modality for AI coaching HMI design. HMIs designed for AI coaching should aim to find a balance between comprehensive coverage of relevant topics and communication efficiency. This can help learners gain sufficient knowledge, prioritize attention, and avoid being overwhelmed. Context-based modality-appropriateness is essential for transferring information efficiently.

\subsection{Designing HMIs to Support the Learning Process}
The results of this study have clear implications for the future design of AI coaching and autonomous vehicle HMIs more generally. By augmenting the situational awareness of the AV to include an understanding of how best to meet the needs of their riders, human-AV teams can successfully meet their goals. We briefly summarize eight design considerations.

\textbf{Designing for the Learning Process.} It is important to design for the specific learning process of the learners being taught. Combining participant descriptions with insights from expert performance driving instructors allows us to delineate the process by which participants learned to drive in the driving simulator (Table \ref{process}). Prior work suggests that AI explanations that align with the learning or reasoning processes of their users may be more effective than those that are not \cite{wang2019designing, kaufman2023explainable}. Our study supports these past findings. By directing attention and supporting stages in the learning process differently -- such as presenting information with more or less uncertainty or ambiguity -- we found differences in the performance, knowledge, and preferences of our participants. Some participants expressed a desire for additional support with stages related to learning to brake and accelerate. Findings imply that an AI coach, or HMI more generally, needs to be tuned specifically for the task at hand and that careful consideration should be placed on the process stages of learning that task. We delineated the learning processes of novices seeking to learn performance driving. Though not explored in this study, this presents the unique opportunity to study the effect of specific HMI designs on their effect of learning specific learning stages in future work.

\begin{table}[htbp]
\centering
\caption[\textbf{Participant learning process phases.}]{\textbf{Participant learning process phases.} Though presented linearly, each sequential phase builds on the previous phases, which iteratively impacts the earlier phases in turn. Individual differences may exist.}
\begingroup % Start a new group
\renewcommand{\arraystretch}{1}
\begin{tabular}{p{.03\linewidth}p{.13\linewidth}|p{.75\linewidth}}
\toprule
\textit{\textbf{\#}} & \textit{\textbf{Stage}} & \textbf{Description} \\
\midrule
\textit{1} & \textit{Simulator calibration} & In this phase, participants begin to calibrate the physical inputs on the simulator’s steering wheel and pedals to the movement of the on-screen vehicle. \\
\hline
\textit{2} & \textit{{Vehicle \newline limits}} & Next, participants begin to learn what types of movements are possible from the vehicle. This often involves experimental speeding up, slowing down, and testing the grip of the tires, among other maneuvers. \\
\hline
\textit{3} & \textit{{Track \newline layout}} & Early in the learning process, this phase involves learning the different shapes the track may take. Later on, it may involve memorizing specific turns. \\
\hline
\textit{4} & \textit{{Vehicle \newline positioning}} & Combining phases 2 and 3 brings insight into how to position a vehicle to optimally move around turns. For example, moving to a specific edge. Finding this ideal racing line can be challenging and unintuitive for a novice driver, and was the specific target for the ‘what’ explanations presented. \\
\hline
\textit{5} & \textit{Handling in position} & In this phase, participants build their handling skills by iteratively combining their understanding of vehicle positioning with their knowledge of the vehicle’s limits. This is a mediary step between phase 4 with phase 6, and becomes increasingly important as the limits of speed and acceleration are tested. This was the first target for the ‘why’ explanations presented. \\
\hline
\textit{6} & \textit{{Speed via \newline Acceleration and \newline Braking}} & In an ideal learning process, speed management should be learned in conjunction with positioning and handling. Once a participant understands where to be positioned, they can safely add and subtract speed to find an optimal balance. Adding speed without first understanding positioning can result in unsafe maneuvers, less ideal positioning, and– consequently– slower lap times. This was the second target for the ‘why’ explanations presented. \\
\hline
\textit{7} & \textit{Iteration} & As drivers learn, they are able to build on their knowledge, iterate, and enhance specific motor skills, maneuvers, and racing strategies. \\
\bottomrule
\end{tabular}
\endgroup
\label{process}
\end{table}

\textbf{Directing Attention.} We find that – based on the type of information presented–  attention may be directed in different ways. Omitting certain information, such as ‘why’ information in our study, may have caused participants to overly focus on the racing line to the detriment of other aspects of performance driving. The implication is that the specific information an AI coach presents should be mindfully chosen to ensure attention is being directed appropriately. In many cases, temporal ordering and prioritization of attention can be specifically designed for using techniques like scaffolding, organizing information hierarchically, and employing progressive disclosure so that less prominent details are deprioritized until they are needed. 

\textbf{Balancing Thoroughness with Efficiency.} Carefully balancing information thoroughness with efficiency of presentation is crucial for useful HMIs. Increasing the amount of information conveyed to a novice may be helpful in transferring sufficient information, however, our evidence suggests that efforts will be futile if not done efficiently. As a result, careful selection of the type and amount of information presented is an important consideration for ease of processing. Avoiding overcomplexity and electing for easy to process information can help reduce the possibility of cognitive overload. 

\textbf{Modality-Appropriateness and Minimizing Uncertainty.} Our findings have clear implications on the importance of presenting modality-appropriate information that maximizes ease of processing. For driving instruction and HMIs for driving tasks, visual explanations are ideal for visual aspects of performance, such as showing where on the road to drive. Auditory may be more appropriate for information that is complex or not directly tied to a visual task, such as ‘why’-type explanations. In both cases, information needs to be efficiently presented with emphasis on the precise details necessary for task execution. Presenting thorough, modality-appropriate information can minimize epistemic uncertainty and ambiguity.

\textbf{Trust as a Barrier.} Trust was a major concern for our participants. Participants who trusted the AI coach more showed better performance for non-intuitive aspects of performance driving, like following the racing line. As such, the effectiveness of an AI coach may be dependent on how trustworthy it is. The implications are clear and supported by a plethora of past research: without trust, HMIs will fail. According to our participants, trust can be built through repeated positive exposure, helpfulness, and explainability validating the AV’s ability and reliability. Though coaching is a novel application from explainable AIs studied in prior work, similar principles for building trustworthy AI systems may apply. For the case of learning specifically, explanations that align with the learning and reasoning processes of the learners themselves may help build trust, as these give the learner the agency to cross-examine the information they are receiving in-context \cite{wang2019designing, kaufman2023explainable}.

\textbf{Personalizing Interactions.} Participants differed in their performance, preferences, and trust levels. Just as it is in human-to-human instruction, AI coaching will work best if personalized to the individual needs of the learner \cite{wintersberger2016towards}. This aligns with the complex systems perspective of this dissertation. Attributes for personalization suggested from the study presented here may include ability, expertise, confidence, trust, preference, and cognitive load. For ability and expertise, an AI coach can add or withhold details and alter the complexity of information offered at a given time, among other techniques. For preference, individuals may want the type or modality of information offered to be changed based on their preferred learning style \cite{pashler2008learning}. For trust and confidence, the types of supporting information given to a participant can be varied to address their specific trust concerns. These may include explanations of how the AI system will behave, how it was trained, or even reassuring comments on its ability. Designing for cognitive load may prove a larger challenge, but if an AV can detect when a participant is feeling overwhelmed through biometrics or self-report \cite{radhakrishnan2022physiological}, it can modulate its communication style and behavior in commensurate ways.

\textbf{Contextual Flexibility.} Particularly in discussions about trust, it was clear that certain contexts may require more or different information to mitigate concern than others. In this way, it is clear that pragmatics matter. This aligns with prior work on the impact of contextual factors on explainable AI usefulness \cite{liao2022connecting}, as well as the complex systems perspective of this dissertation. For vehicle HMI design, this means that explanations may need to be altered based on the perceived riskiness or complexity of the driving scenario. In other cases, a vehicle may even give up control to a passenger, or vice versa. For AI coaching, instructional explanations should be grounded in the context within which they should be applied, giving the learner a broader understanding of when certain lessons should be applied and when they should not be.

\textbf{Observation vs. Interactivity.} Our study highlighted the potential usefulness of observational AI coaching. We did not compare our observational methods to more interactive methods, such as question-answering \cite{liao2020questioning}. Some participants requested real-time feedback and a means to interact with the information they received. These may be potentially viable future directions to explore in the context of AI coaching.

\subsection{Limitations and Future Work}
This study was not without limitations. While these are limitations for a study that is designed to be a first step towards AI-driven vehicle coaching, this study still serves as a confirmation of the viability of AI coaching for driving and as a means to drive future work.

While between-group comparisons are the central focus of this research, all-groups-combined results allow us to assess the potential role AI driving coaches play in driving instruction more generally. Both group and combined comparisons, as well as qualitative results from interviews and surveys, support the viability of AI coaching as an instructional method. Combined results should be approached with caution, however, as it is not possible with the current study design to separate the pre-post observation changes from practice effects. Though changes from pre-post observation are far greater in magnitude than we would expect from practice alone, formally testing the impact of practice against the impact of training is left for future study.

We note that during real-world instruction, novices may ask questions to the instructor. Though the present study does not take into account this interaction, the explanations included were designed to cover the most common questions received by our expert drivers. Interactivity and other non-observational methods would be excellent future research directions to explore.

Compared to real-world instruction, our AI coaching sessions were quite short. We would expect larger differences and greater insight after exposure to longer AI coaching sessions and/or sessions that form a series.

Finally, though sufficient to obtain statistical significance and identify several data trends, this study had a relatively small sample size. With a larger sample, we would be able to evaluate interactions between variables such as cognitive load and trust directly on the pre-post differences in driving performance observed by the groups. We would expect to see clearer impacts of AI coaching with a larger study sample.

\section{Conclusion}
In this study, we tested a novel use of AV technology -- if AVs can teach humans to be better drivers. Results from the pre-post observation study reported here support the conclusion that observing an AI driving coach is a promising method to teach novices performance driving. Breaking participants into groups allowed us to determine how information type (‘what’ and ‘why’) and information modality (auditory and visual) influenced outcomes. We saw differences in how information directed attention, mitigated uncertainty, and influenced overload experienced by participants. These differences affected how successfully participants were able to learn. 

This chapter provides foundational evidence that AV systems can effectively enhance human learning through tailored communication strategies, contributing to the broader goal of designing AV systems that dynamically meet user needs and situational demands. Results suggest that explanations for AI driving coaching and vehicle HMI design should seek to strike a balance between comprehensive coverage of relevant topics – such as how to follow the racing line and meet the limits of speed – with information complexity. When designed properly, explanations can direct attention to appropriate details efficiently, supporting the learning process while avoiding learner overwhelm. Context-based, modality-appropriate explanations should be opted for, especially when they mitigate information uncertainty. 

By demonstrating the potential of AI-driven coaching, this work lays a framework for advancing human-AV interaction beyond traditional paradigms. This work also exemplifies how communication strategies designed to support situational awareness can directly address user goals, such as learning or trust-building. We conclude that communications must be designed to align with the needs and learning processes of the learner and present 8 design considerations to inform future HMI design that should generalize to driving in many different contexts, including everyday driving. These include specific suggestions for how to direct attention and choose the modality of an explanation, as well as more general implications on the need for personalized, trustworthy, context-based HMIs.

By focusing on the dynamic interaction between users and autonomous systems, the study underscores the importance of designing AV systems that are sensitive to individual user needs and capable of adapting to specific task requirements. Building on these insights, the next chapter examines how AV explanation errors impact trust and reliance, highlighting the consequences of communication failures and their dependence on driving context.

%\bibliography{coach/sample} %%MOVED TO main_references.bib SO ALL ARE IN THE BOTTOM OF THE DOCUMENT

\section{Acknowledgements}
The authors would like to express gratitude to the Human Interactive Driving and Human-Centered AI groups at Toyota Research Institute. A special thanks to Nadir Weibel, Allison Morgan, Andrew Best, Paul Tylkin, Tiffany Chen, Hiro Yasuda, Hanh Nguyen, and Steven Goldine for their individual contributions, feedback, and support. Chapter 2, in full, is a reprint of the material as it appears in \textit{Effects of Multimodal Explanations for Autonomous Driving on Driving Performance, Cognitive Load, Expertise, Confidence, and Trust}
(Nature: Scientific Reports~2024). Kaufman, Robert; Costa, Jean; Kimani, Everlyne. The dissertation author was the primary investigator and author of this paper. \cite{kaufman2024effects}

\chapter[\underline{Errors and Context:} What Did My Car Say? Impact of Autonomous Vehicle Explanation Errors and Driving Context On Comfort, Reliance, Satisfaction, and Driving Confidence]{\underline{Errors and Context:} \\ What Did My Car Say? Impact of Autonomous Vehicle Explanation \\ Errors and Driving Context On \\ Comfort, Reliance, Satisfaction, and Driving Confidence}
\newpage

\section{Interim Summary and Chapter 3 Overview}
In Chapter 2, I presented a study that tested the feasibility of an autonomous vehicle (AV) coach, empirically investigating how a human and an AV can collaborate as a team -- through optimized communication strategies -- to achieve the goal of human learning from the AV. Using a realistic driving simulator, we found that both the type of information and the modality through which it is presented can significantly influence how effectively a human learns from an AV. This study provided valuable insights into key cognitive aspects of explanation design, such as using information to direct attention and transfer knowledge efficiently, aligning explanations with individual learning processes, and preventing cognitive overload while mitigating uncertainty.

Stepping out to the broader goals of this dissertation, this study offered a concrete domain to examine how informational needs can be met based on the broader dynamics of the human-AV system. In this case, there were clear criteria for success that needed to be met for the rider to achieve the goal of learning, and enabling the AV to meet those needs allowed the team to reach their goals. How to meet these needs was based not just on learning itself: the effectiveness of communications depended on certain human traits -- such as cognitive load, learning processes, and attentional capabilities -- which influence the communicative strategies an AV teammate should employ to help a person achieve their learning goals. Thus, by enabling the AV to communicate using these methods, they will be better able to support their riders.

Looking into other areas of the human-AV system may allow us to tailor these types of communications more closely. Although this study did not test the impact of expertise levels, we would expect that the information required for effective learning would differ if participants had prior experience, such as basic racing instruction. One human factor that did prove important was \textit{trust}: participants who trusted the AI coach showed better learning outcomes and improved performance in mastering key, unintuitive aspects of racing. This implies that trust may be a key sub-goal criteria that needs to be achieved to make higher-level goals like learning possible. In Chapters 3 and 4, we shift our focus specifically to trust, exploring in greater detail how individual traits influence informational needs and perceptions of AVs.

In Chapter 3, we begin to orient our lens towards the importance of the driving context. In Chapter 2, the driving context was consistently challenging and intense, although simplified by the absence of other vehicles competing for space on the road. Considering the influence of the wider human-AV system, how did this context shape the effectiveness of the communications? Certainly, orientation towards the goal of learning played a role: it is reasonable to expect that different informational content or presentation modalities would be necessary if the goal shifted from learning from the AV to relying on it solely for safe navigation around the track. Similarly, in less dangerous or fast-paced environments, or in more familiar driving scenarios, the communicative actions required to meet informational needs would likely differ. In Chapter 3, we shift the goal of the interaction to safe and trustworthy reliance (traveling from point A to point B) and explore the effects of varying driving contexts, examining how these changes influence human perceptions of and behaviors toward AVs.

Finally, a key AV trait, the vehicle's performance, was held constant in Chapter 2. The driving and instructions were always correct and based on human performance driving experts. What if the accuracy of the AV explanations changed? One could hypothesize that if certain instructions were incorrect, unclear, or contradictory, learning and trust would be affected. In Chapter 3, we examine what happens when explanation inaccuracies are introduced on people's perceptions of AVs -- in this case, in the more common context of general commuting.

\textbf{Chapter 3 Overview} -- \textit{Explanations for autonomous vehicle (AV) decisions may build trust, however, explanations can contain errors. In a simulated driving study (n = 232), we tested how AV explanation errors, driving context characteristics (perceived harm and driving difficulty), and personal traits (prior trust and expertise) affected a passenger's comfort in relying on an AV, preference for control, confidence in the AV's ability, and explanation satisfaction. Errors negatively affected all outcomes. Surprisingly, despite identical driving, explanation errors reduced ratings of the AV's driving ability. Severity and potential harm amplified the negative impact of errors. Contextual harm and driving difficulty directly impacted outcome ratings and influenced the relationship between errors and outcomes. Prior trust and expertise were positively associated with outcome ratings. Results emphasize the need for accurate, contextually adaptive, and personalized AV explanations to foster trust, reliance, satisfaction, and confidence. We conclude with design, research, and deployment recommendations for trustworthy AV explanation systems.}
\newpage

\section{Introduction}
One of the most important and well-documented problems with autonomous vehicles is a lack of trust in how they make decisions~\cite{kenesei2022trust}. Low trust remains a critical barrier to adoption, reflecting the broader challenge of designing AV systems that meet complex user needs in real-world scenarios. This is not just a problem with autonomous vehicles specifically, but a widespread issue across many types of AI-based systems~\cite{bedue2022can}. System transparency via explainable AI (XAI) has been proposed as a means to mitigate concerns with AI-based systems like AVs, offering users a look “under the hood” of black-box AI models so they understand what the system is doing and why~\cite{gunning2019darpa, miller2019explanation}.

However, although potentially increasing trust, explanation of AI \color{black} behavior and decision-making \color{black} in the real world can contain errors. Prior work has shown that when autonomous vehicles exhibit \emph{driving} errors, passenger trust and willingness to rely on the vehicle can deteriorate quickly~\cite{seet2020differential, kaplan2023trust}. Far less is known about the impact of \emph{explanation} errors\color{black}, such as when an AV miscommunicates what it is doing or why\color{black}~\cite{cabitza2024explanations, kenny2021explaining}. Particularly for safety-critical systems like autonomous vehicles -- that may rely on explanations and other in-vehicle communications to elicit trust comfort, and safe reliance with users -- knowing the consequences of errors is pivotal to safe deployment. \color{black} We hypothesize that explanation errors, like driving errors, will negatively impact perceptions of an AV. \color{black} Understanding these consequences is essential for user reliance because, without this knowledge, AV systems cannot be deployed safely or ethically~\cite{martinho2021ethical}. Further, it allows us to detail how a key system component -- the attributes of the AV itself -- may contribute to how the system is used and how it should be designed.

\color{black} Autonomous vehicles offer a unique context to study explainable AI errors, as vehicle errors may be easier to detect for lay users with basic driving knowledge compared to errors in more specialized domains, such as medical diagnosis \cite{rajpurkar2017chexnet, kaufman2023explainable} or bird identification \cite{pazzani2022expert, soltani2022user}, where domain expertise is required for error detection. This distinct characteristic allows non-experts to evaluate AV explanations with relatively high confidence, providing a unique perspective on how explanation errors might influence user perceptions, judgments, and trust in AI. Prior research in XAI suggests that task complexity \cite{salimzadeh2023missing} and domain expertise \cite{nourani2020role} are crucial factors in how users interpret and respond to AI explanations -- we extend this understanding by examining explanation errors in a high-risk, non-expert domain, contributing new insights into how explanation accuracy affects user trust and reliance. \color{black}

Core to our discussion is the repeated mantra of this dissertation: Human-AV interactions do not exist in a vacuum, and are sensitive to the contextual demands of the external driving environment~\cite{hoff2015trust}. It is important to consider the driving context in which explanation errors occur, as this may significantly influence how people interact with AI-based systems~\cite{lim2009and, schilit1994context}, including AVs~\cite{capallera2022human, de2020designing}. The complexity of a driving situation and perceived risk of harm have been singled out as particular factors of interest~\cite{ha2020effects, kaufman2024developing}, as these might drive a person's explanatory needs and reliance behaviors. \color{black} Though presently underexplored, we hypothesize that contextual factors may impact rider perceptions of an AV, particularly when errors are introduced. \color{black}

Finally, recent work has posited that AV explanations should be tailored to meet the specific needs of the people interacting with the system~\cite{ma2023analysing, kaufman2024developing}. This aligns with our complex systems approach: it is well known that people may interact differently with AI-based systems~\cite{schneider2019personalized} based on the individual characteristics or prior experiences they may have~\cite{ayoub2021modeling, kaufman2024predicting}. Particularly well studied are domain expertise~\cite{araujo2020ai, pazzani2022expert, kaufman2022cognitive} and initial (dispositional) trust~\cite{hoff2015trust}; \color{black} we hypothesize that these will also impact AV perceptions.\color{black}

In the present study, we examine the impact of explanation errors across a variety of realistic, simulated driving scenarios. To deepen our investigation and understand the significance of the \textit{type} of error presented, participants were shown AV explanations at three distinct accuracy levels: \color{black}(1) accurate explanation of the AV's behavior, (2) accurate explanation of \textit{what} the AV is doing but incorrect rationale for \textit{why}, and (3) incorrect explanation of what \textit{and} rationale of why. \color{black} By focusing on `what' and `why' errors, this chapter builds on the findings of Chapter 2. We measured the effect of these errors on four main outcomes: comfort relying on the AV, preference for control, confidence in the AV's driving ability, and explanation satisfaction. We include measures of scenario context -- perceived harm and driving difficulty -- to assess their impact on our driving outcomes, including how they may moderate the relationship between explanation errors and user perceptions. To explore the influence of individual differences, we also included measures of trust and \color{black}AV domain \color{black}expertise to determine if they predict study outcomes.

Our findings reveal that explanation errors, contextual characteristics, and personal traits significantly impact how a person may think, feel, and behave towards AVs. Explanation errors negatively affected all outcomes, with impacts proportional to the magnitude and negative implications of the error. Harm and driving difficulty directly impacted outcomes as well as moderated the relationship between errors and outcomes, though in opposing ways. Overall, harm was generally seen as more important and more negative than difficulty. Participants with higher \color{black}AV domain \color{black}expertise tended to trust AVs more, and these each correlated with more positive outcome ratings in turn.

\noindent
In sum, \color{black} our research questions are as follows:
\begin{itemize}[topsep=0pt, nolistsep]
    \item \textbf{RQ1:} What is the impact of AV explanation errors on participant ratings of an AV, and does the impact differ for errors of different information types (`why' errors vs. `what and why' errors)?
    \item \textbf{RQ2:} What is the impact of context (perceived harm and driving difficulty) on participant ratings of an AV, and do these contextual factors modify the effect of an error?
    \item \textbf{RQ3:} How does participant initial trust and domain expertise impact their ratings of an AV?
    \item \textbf{RQ4:} How can findings inform future design and research of explainable AI and autonomous vehicles?
\end{itemize}
\color{black}
\vspace{1em}
\noindent

Results highlight the critical need for accurate and contextually adaptive explanations for autonomous vehicles to enhance user trust, reliance, satisfaction, and confidence. Recognizing the implications of explanation errors is vital for advancing AV research and guiding design teams to make informed decisions. This understanding also contributes to the wider themes of this dissertation by demonstrating how AV communication strategies must dynamically adapt to situational demands and individual user needs. By addressing these challenges, this study lays the groundwork for developing context-aware designs, personalized explanation interfaces, and establishing ethical or regulatory guidelines to ensure the deployment of safe and trustworthy explainable AI (XAI) systems for autonomous vehicles.

\section{Related Work}
\subsection{AV Trust and Explainability}
\textbf{Human-Centered Explainable AI} -- Continuous development of explainable AI (XAI) has brought major progress in the quest for trust through AI transparency~\cite{miller2019explanation}. Approaches to XAI vary, often limited by the availability of the model~\cite{simonyan2013deep, ribeiro2016should}. Recent work found that -- even when an explanation is given -- trust and engagement with the system may not improve unless the \emph{right information} is given in the \emph{right way}, at the \emph{right time}~\cite{wang2019designing, liao2021human}. Several studies and theory pieces have demonstrated the value of human-interpretable explanations on user understanding and trust ~\cite{holzinger2019causability, soltani2022user}. In particular, explanations that are modeled off those given by human experts have suggested as a means to increase understanding for experts and novices alike~\cite{pazzani2022expert, kaufman2022cognitive}. There have likewise been calls for the need for explanations to be sensitive to particular user characteristics (such as personal traits or experiences) ~\cite{ehsan2020human, ehsan2021explainable, kaufman2023explainable} or context of use (including the specific use environment and goals of a user)~\cite{schilit1994context, kaufman2024developing}. Despite the large amount of research on explanation design, little has been done to understand what happens when these explanations fail. The study we present here is a first step towards filling this knowledge gap, using autonomous vehicles as a specific domain of interest.

\textbf{Interface Modalities} -- Research on in-vehicle interfaces for explanation, such as visual heads-up-displays (HUDs)~\cite{currano2021little, chang2016don, schartmuller2019text}, audio interaction~\cite{mok2015understanding, jeon2009enhanced, locken2017towards}, and even haptic feedback for drivers~\cite{di2020haptic} has shown that there are a variety of ways explanations can support users, each dependent on the use-case and context of interest. The benefits and drawbacks of different modalities often relate to the complex process of transferring sufficient knowledge to accomplish a task (such as to build understanding through system transparency) without creating too much cognitive load~\cite{kaufman2024effects, kim2023and, colley2021effects}. In the present study, we leverage video and audio explanations for our study on the impact of errors, as these are common and efficient methods to transfer information to a user without overwhelm. We present both modalities of information at the same time to increase the accessibility of our study and ensure there is successful transfer of information.

\textbf{Explanation Content} -- Choosing the appropriate content for an AI explanation is crucial for enhancing rider trust and reliance \cite{miller2019explanation}. Recent work as focused on providing a description of \emph{what} a vehicle is doing and \emph{why} a vehicle is doing it, as these enable a user to create a momentary evaluation of the AV's behavior in terms of reliance \cite{hoff2015trust}. For example,~\citet{koo2015did} present `how', `why', and a combination of both in various simulated autonomous driving scenarios to assess driver attitudes and safety performance. They found improved safety when both `how' and `why' were presented to drivers, but a preference for `why' explanations alone. \citet{kaufman2024effects} leveraged auditory and visual explanations to teach humans to be better drivers via an `AI coach', highlighting the additive value of \emph{what} and \emph{why}-type information for transferring knowledge, adding that \textit{too much} information can cause overwhelm. They emphasize the need for explanations to strike a balance between complexity and comprehensiveness. The impact of \emph{what} (similar to Koo's \emph{how}) and \emph{why}-type explanation \emph{errors} -- explored in the present study -- remains unknown.

\textbf{Factors Impacting Trust} -- Knowing what factors may impact a person's trust and reliance decisions is vitally important to designing XAI explanations to support them. Hoff and Bashir's theoretical model of trust in automation highlights the importance of three distinct yet interdependent facets of trust: dispositional (e.g. personality or cultural attitudes), situational (e.g. based on a particular context of use), and learned trust (e.g. based on a present evaluation of system performance) \cite{hoff2015trust}. Additionally,~\citet{kaufman2024developing} developed a framework to understand situational awareness in joint action between humans and AV, a key focus of explainable AI transparency in safety-critical situations like driving. They describe how communications like AV explanations can enable human-AV teamwork to achieve particular goals like safe and trustworthy driving. Factors of interest include external driving conditions, human traits and abilities, and communication preferences and goals -- all of which are crucial in managing driving difficulty and reducing the risk of harm. Using these system-based models as a backdrop, AI explanation designers can form hypotheses on how to build more trustworthy systems. In the present study, we build off these models by investigating specific driving context factors \color{black}(perceived harm and driving difficulty) \color{black} and personal traits \color{black}(domain expertise and prior trust) \color{black}which may impact a person's reliance judgments and, in the case of context, how much an error matters.

\textbf{Knowledge Gap} -- Indeed, explanation interfaces have been implemented in mainstream deployed autonomous vehicles, such as communication interfaces by Tesla~\cite{TeslaModelYManual} and Waymo~\cite{WaymoOne}. Despite these efforts, we still know very little on the impact of AV explanation errors on how a person will trust and interact with an AV. We know even less about how errors may depend on contextual factors like driving difficulty or harm~\cite{capallera2022human, de2020designing, ha2020effects}. With this work, we seek to understand how contextual harm and difficulty of driving may affect study outcomes, as well as contribute to the body of literature on how personal traits predict a person's interactions with an AV.

\subsection{AI Errors}
Widespread integration of AI systems into everyday tasks has demonstrated huge benefits to productivity and optimization in many domains~\cite{fauzi2023analysing}. However, concerns over AI errors remain a major point of contention, particularly as systems proliferate. Examples of errors with real-world consequences include language models' propensity to hallucinate~\cite{xu2024hallucination} and algorithmic bias that favors men over women used in the hiring process~\cite{dastin2018amazon}. In contrast to fields requiring domain-specific knowledge for error detection (e.g., medical diagnosis \cite{rajpurkar2017chexnet, kaufman2023explainable}), AV explanation errors are often intuitively detectable by lay users. This accessibility may amplify the impact of errors on trust and reliance, as errors are not only observed but judged against personal driving ability. \color{black}

\textbf{General AV Errors} -- Autonomous vehicle driving errors have important real-world consequences, which may include physical harm or even fatalities~\cite{favaro2017examining}. Even in cases where AV driving outperforms humans~\cite{schwall2020waymo}, concerns over vehicle errors can be a major hindrance to adoption and use~\cite{choi2015investigating}. \citet{luo2020trust} shows that errors caused by an AV had a more significant negative impact on user trust than external errors, such as those caused by other drivers or road conditions. Declines in trust may be difficult to recover from and have long-lasting effect \cite{seet2020differential}. Explanations are no panacea: trust is difficult to achieve when the system itself performs poorly, even when explanations meant to elicit trust are presented~\cite{kaplan2023trust}. In this paper, we seek to connect prior work showing the dire impact of driving errors \cite{luo2020trust, zhang2022trust} to broader XAI literature investigating user perceptions of AI explanations \cite{lebovitz2019diagnostic, cabitza2024explanations}.

\textbf{Explanation Errors} -- Some prior work has investigated the impact of \emph{explanation} errors for autonomous systems. In a study on “white box” XAI,~\citet{cabitza2024explanations} found that non-expert users tend to not catch explanations errors and believe the system even when explanations were wrong, attributing the phenomena to the Halo Effect found in social psychology where people assumed correctness of the system without verifying accuracy. Over- or under-reliance is a problem as people learn to calibrate their interactions with AI systems~\cite{endsley2018situation}. Conflicting results suggest that explanations may help reduce over-reliance in some cases~\cite{vasconcelos2023explanations}, but increase over-reliance in others~\cite{kenny2021explaining}. Other research has shown that the influence of explanations on reliance may be based on the systems' performance itself~\cite{papenmeier2022s}.

\textbf{Knowledge Gap} -- Though several prior studies have investigated the consequences of accurate AV explanations, the impact of explanation errors on reliance behavior and related outcomes remains unexplored. Of particular interest to us are cases when the autonomous vehicle's driving performs properly, as this allows us to separate the impact of driving performance from the impact of explanation performance. We address this important knowledge gap.

\section{Method}
We conducted an online experiment using realistic, simulated driving scenarios of an AV driving through various environments. These ranged from rural to urban driving and from routine navigational challenges like driving around construction cones to challenging situations like avoiding collisions with erratic drivers. Participants viewed videos of the scenarios and, after each, provided ratings related to trust, reliance, satisfaction, and evaluation of the AV.

To test the impact of explanation errors, participants were shown three versions of each scenario, \color{black}each differing by the accuracy of the information. This within-subjects design helps us control for individual differences and attribute findings to the error conditions and scenarios themselves. \color{black} In all three versions of each scenario, the AV drove identically; only the explanations differed. The AV always drove accurately and lawfully. Participants were randomly presented 9 of 27 possible scenarios, \color{black} resulting in a unique set for each. This approach helped evenly distribute any order effects across participants, reducing systematic bias that could result from a block design, increasing generalizability while preserving a naturalistic flow. \color{black} Each participant provided 27 total video ratings (9 scenarios X 3 accuracy levels). Scenarios were presented in random order; \color{black} all had approximately the same number of total ratings. \color{black} Scenarios are described in Table \ref{scenario_list}.

\begin{table}[htbp] %change to H if i want to force positioning
  \caption{\textbf{Scenario \color{black}Reference Codes and \color{black}Descriptions}}
  %\Description{A list of all of the Scenarios used in the study, including brief descriptions of each.}
  \label{scenario_list}
  \begingroup
  \renewcommand{\arraystretch}{1}
  \begin{tabular}{>{\centering \arraybackslash}p{0.08\linewidth}|p{0.84\linewidth}}
    \toprule
    \textbf{Code} & \textbf{Description}\\
    \midrule
    10s1 & Left turn in busy intersection, navigating around object in road. \\
    10s2 & Slowing to avoid collision between two vehicles ahead of the ego vehicle. \\
    10s4 & Parallel parking between two cars on side of busy road. \\
    10s5 & Slowing to avoid being sideswiped during turn by a vehicle that crosses into ego vehicle's lane. \\
    10s6 & Sudden stop mid-intersection by large lead vehicle in adverse weather. \\
    10s7 & Left turn quickly followed by right merge for quick right turn in urban environment in adverse weather. \\
    10s8 & Ego vehicle slows for pedestrian crossing road in non-crosswalk area. \\
    4s1 & Reversing in parking lot. \\
    4s2 & Overtaking cyclist in suburban environment. \\
    4s3 & Ego vehicle must stop quickly for a pedestrian who jumps into the road. \\
    4s4 & Merging from far right lane to far left lane on highway to avoid emergency vehicles / car accident. \\
    4s5 & Ego vehicle must avoid object that falls off of lead vehicle on highway at night. \\
    4s6 & Ego vehicle hydroplanes on wet road on highway, and needs to maintain control. \\
    4s7 & Ego vehicle needs to pull over for flat tire at high speed. \\
    4s8 & Ego vehicle merges left to enter a busy highway. \\
    5s2 & Ego vehicle makes blind turn in intersection due to obstructed view. \textbf{[removed]} \\
    5s3 & Ego vehicle turning left and avoids a collision with another vehicle who ran a red light (T-bone). \\
    5s4 & Car in parallel lane merges into ego vehicle’s path. \\
    5s5 & Ego vehicle needs to navigate around a stopped cyclist. \textbf{[removed]} \\
    5s6 & Ego vehicle navigates construction zone at night. \\
    7s1 & Hidden stop sign at night. \\
    7s3 & Lead vehicle quickly decelerates (brake check) \\
    7s4 & Ego vehicle crosses the midline to overtake a slow lead vehicle. \\
    7s5 & Ego vehicle needs to slow quickly from high speed for a stopped cyclist around a turn. \textbf{[removed]} \\
    7s6 & Vehicle failure (flat tire) in small parking lot. \\
    xs2 &  Ego vehicle waits for child to cross crosswalk before turning right at stop sign. \\
    xs3 & Ego vehicle stops during right turn to avoid collision with vehicle turning right from incorrect (left) lane. \\
  \bottomrule
\end{tabular}
\endgroup
\end{table}

We evaluated changes across four major outcomes of interest, with two additional descriptive outcomes, making a total of six ratings per scenario video. After each video, participants rated their: (1) comfort relying on the AV, (2) preference to take control, (3) satisfaction with the explanation, and (4) confidence in the AV's driving ability. We hypothesized that ratings may be context-dependent. As such, we also collected two ratings describing the driving context: (5) perceived harm and (6) perceived difficulty of driving in each scenario. These context descriptor variables were collected after the accurate explanation videos only. Outside of the rating task, participants answered questions about their trust in AVs, expertise, and demographics.

\begin{figure}[!b] %change to H if i want to force positioning
  \centering
  \includegraphics[width=\linewidth]{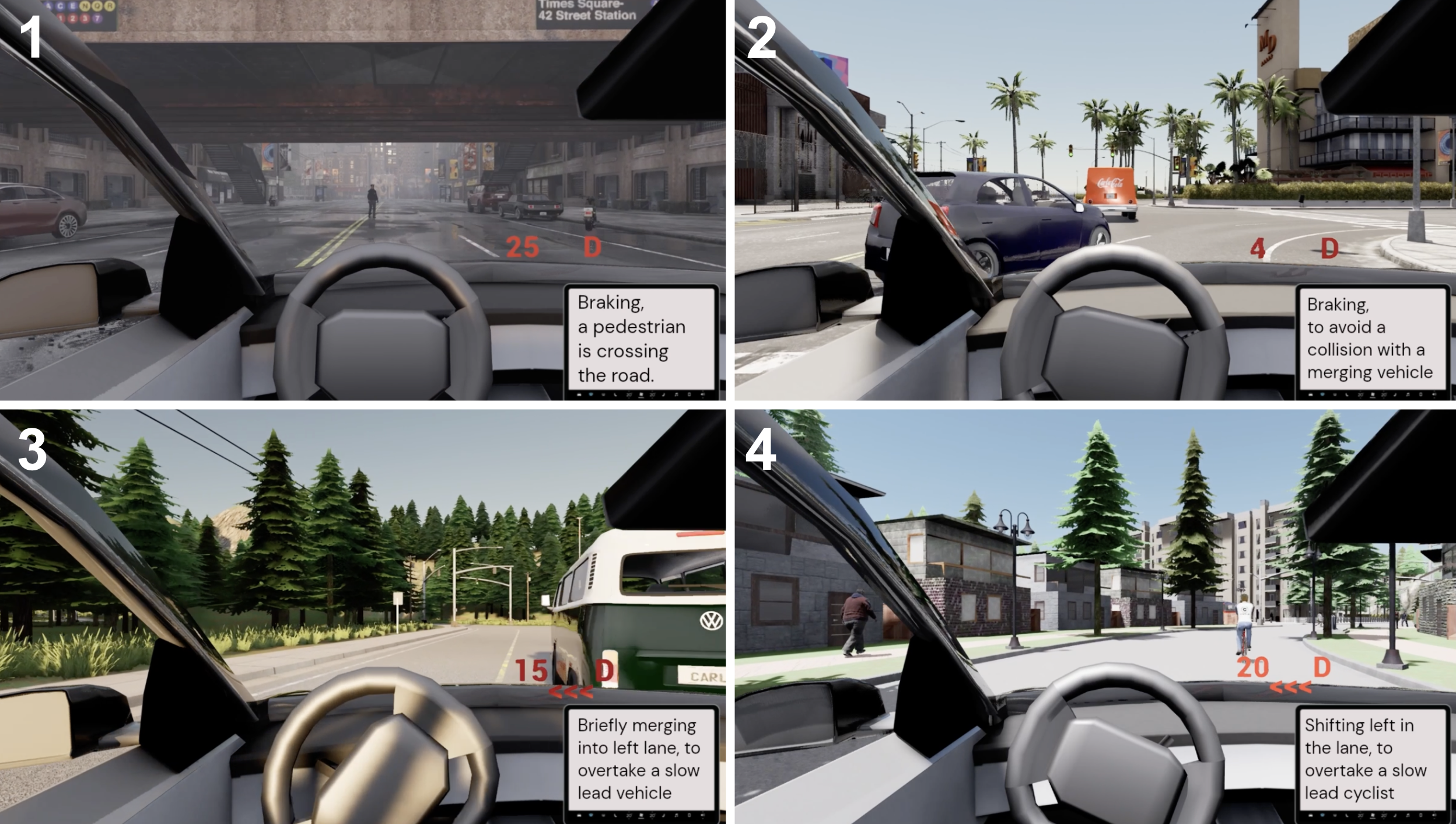}
  \caption[\textbf{Example images from four driving scenarios.}]{\textbf{Example images from four driving scenarios.} In order: (1) the AV slows for a pedestrian crossing the road on a foggy day in the city, (2) the AV is cut off by a vehicle turning right from the left lane, (3) the AV merges around a slow lead vehicle in a forested area, (4) the AV moves past a cyclist on a suburban street. In the bottom right corner of each scenario, written AV explanations appear on the vehicle's dashboard display.}
  %\Description{Example images from four driving scenarios: (1) the AV slows for a pedestrian crossing the road on a foggy day in the city, (2) the AV is cut off by a vehicle turning right from the left lane, (3) the AV merges around a slow lead vehicle in a forested area, (4) the AV moves past a cyclist on a suburban street. In the bottom right corner of each scenario, written AV explanations appear on the vehicle's dashboard display.}
  \label{scenarioexamples}
\end{figure}

\subsection{Participants}
A total of 232 participants spread throughout the United States participated in the study and completed all study procedures. Participants were recruited from the general population via existing participant email lists and via SONA,\footnote{https://www.sona-systems.com} an undergraduate study pool system where students are granted study credit for participation. The mean age of the study sample was 23.6 (SD = 11.3), with ages ranging from 18 to 85 years. The sample was 73.7\% female.

\subsection{Simulated Driving Scenarios}
All simulated driving videos were custom-made by the research team using DReye VR~\cite{silvera2022dreye}, a tool for creating realistic driving scenarios using the open-source driving simulator CARLA~\cite{dosovitskiy2017carla}. Scenario design was inspired by the National Highway Traffic Safety Administration (NHTSA) list of common driving situations that result in vehicle crashes~\cite{najm2007pre}. Examples include unexpected pedestrians in the road, collision avoidance from erratic other drivers, construction and emergency vehicle zones, parallel parking and reversing from park, navigating around stopped or slowed vehicles or cyclists, and dealing with flat tires or hydroplaning. Videos lasted between 10 and 30 seconds each (Examples in Fig. \ref{scenarioexamples}). Three scenarios were removed during analysis due to the AV’s driving determined to be imperfect (going above the speed limit, crossing over the center line during a turn, and improper yielding), making a total of 24 included in the final analysis. Data was cleaned prior to analysis to ensure that all videos were viewed in full.

\subsection{AV Explanations and Errors}
During all driving scenario videos, explanations of the AV's behavior were provided to participants at the time of AV action, \color{black} modeled by the research team after those provided by~\citet{kim2018textual}. Explanations were presented both visually in written English on the vehicle's dashboard display and audibly via spoken English produced by Amazon Polly Text-To-Speech~\cite{polly}. These modalities are reflective of current trends in AV explanation research~\cite{kaufman2024effects, koo2015did, lee2023investigating}, providing accessibility to participants and ensuring that findings can be immediately incorporated into the design of state-of-the-art AVs and XAI interfaces. \color{black} For an example of the visual presentation of explanations, see Figure~\ref{scenarioexamples}.

Explanations provide information on `what' action the AV is doing (e.g. \textit{braking}) as well as a local explanation for `why' the vehicle is doing it (e.g. \textit{to avoid a collision with a merging vehicle}). The importance of `what' and `why'-type information has been studied in past experimental work on explanations of driving behavior by~\citet{kaufman2024effects} and~\citet{koo2015did}.  Frameworks by~\citet{wang2019designing} and~\citet{lim2019these} in cognitive science and~\citet{miller2019explanation} in philosophy emphasize the importance of `what' and `why' information for transparency, trust, and understanding with AI-based systems like AVs. We examine both the impact of `why' errors alone and `why and what' errors combined.

Explanation errors were presented via three conditions \color{black} (Table~\ref{conditions}). Those with an “accurate” explanation were correctly told `what' the AV was doing and `why' it was doing it. Errors were introduced via a “low” error condition, where the AV correctly explained `what' it was doing but incorrectly explained `why', and a “high” error condition, where both `what' and `why' explanations were incorrect. \color{black} Comparing the “accurate” to “low” group shows the impact of errors related to the AV's rationale for behavior (i.e. `why'). Comparing the “low” to “high” group isolates the impact of adding errors related to the AV's description of its own action (i.e. `what'). \color{black} We did not include a condition with inaccurate `what’ but accurate `why,’ as it would be implausible for an AV to provide a correct rationale for an incorrect action; this would reduce the study's ecological validity. \color{black}

\color{black}We hypothesized that users may process AV errors not just in terms of occurrence, but also in terms of potential outcome -- mirroring real-world scenarios where the consequences of an error are as critical as the error itself. \color{black} To test this, we conducted a secondary analysis, categorizing mistakes in the “high” error condition -- where the AV is providing an incorrect description of what it is doing -- based on the potential harm that \emph{would} result should the AV have acted on the mistaken `what' description. For example, if the AV mistakenly says it is going “left” when really it is going straight, we hypothesized that the impact of this error would be greater if going left would result in an accident, as opposed to simply a wrong turn. To explore if this difference in pragmatics does impact our results, we run a secondary analysis within just the `high' condition to test for impact of \emph{potential} harm.

\newpage
\begin{table}[htbp] %change to H if i want to force positioning
  \caption[\textbf{Experimental Conditions: `What' and `Why' Errors.}]{\textbf{Experimental Conditions: `What' and `Why' Errors.} Participants saw videos across each error condition, providing ratings for each.}
  %\Description{This table shows the three experimental conditions, providing a description and example of each. Participants saw videos across each condition, providing ratings for each.}
  \begingroup % Start a new group
  \renewcommand{\arraystretch}{1}
  \label{conditions}
  \begin{tabular}{p{0.095\linewidth}|p{0.36\linewidth}|p{0.465\linewidth}}
    \toprule
    \textbf{Error Level} & \textbf{Explanation Description} & \textbf{Example}\\
    \midrule
    Accurate & Correct `what' + correct `why'& \textit{“Braking, a pedestrian is crossing the road.”} \\
    \midrule
    Low & Correct `what' + incorrect `why' & \textit{“Braking, a cyclist is crossing the road.”} \hspace{0.2in}(When the obstacle is actually a pedestrian)\\
    \midrule
    High & Incorrect `what' + incorrect `why' & \textit{“Merging right, a cyclist is crossing the road.”} (When the vehicle is slowing, not merging, and the obstacle is a pedestrian)\\
  \bottomrule
\end{tabular}
\endgroup
\end{table}

\subsection{Measures}
\color{black} The measures in this study were carefully selected to offer clear and comprehensive insight into the impact of errors, context, and personal traits, while remaining easy and intuitive for participants to understand. \color{black}
\subsubsection{Main Outcomes: Scenario Ratings} 
These measures were taken after each video to provide insight into the impact of the explanation errors. \color{black} All measures were adapted from existing work and adjusted to fit the present study. \color{black}

\begin{itemize}[itemsep=0pt, topsep=0pt]
    \item \textbf{\underline{Comfort} relying on AV} (proxy for trust). “How comfortable would you feel relying on this AV in this specific situation?” \color{black} This was adapted from the Situational Trust Scale for Automated Driving (STS-AD) by \citet{holthausen2020situational}, which itself was based on Hoff and Bashir's trust in automation framework \cite{hoff2015trust}. \color{black} (0-10 rating scale)
    \item \textbf{\underline{Reliance} on AV} (preference to take control). “If this specific situation were to happen in the real world, would you prefer to rely on an AV or take control yourself?” \color{black} This was also adapted from the STS-AD scale \cite{holthausen2020situational}. \color{black} (Binary choice: `Rely on AV' or `Take control myself'). To aid comparison, reliance data was scaled from 0-1 to 0-10 to match the other variables. This scaling does not impact interpretation.
    \item \textbf{\underline{Satisfaction} with explanation.} “How satisfied are you with the AV’s explanation?” \color{black} This was adapted from the Explanation Satisfaction Scale by \citet{hoffman2018metrics}. \color{black} (0-10 rating scale)
    \item \textbf{\underline{Confidence} in AV driving ability.} “Please rate your confidence in the AV’s driving ability.” \color{black} This was adapted from the Performance Expectancy section of the Autonomous Vehicle Acceptance Model (AVAM) \cite{hewitt2019assessing}. \color{black} (0-10 rating scale). Note: the actual driving performance was always high and never changed between error conditions.
\end{itemize}

% %THIS IS A POTENTIAL WAY TO PUT THESE MEASURES IN A TABLE. I DONT LIKE IT AS MUCH AS TEXT, BECUASE WE LOST THE ABILITY TO MAKE NOTES / COMMENTS IN A VISUALLY APPEALING WAY. 
% \begin{table}[H] %change to H if i want to force positioning
%   \caption{Measures: Scenario Ratings}
%   \label{scenario_rating_measures}
%   \renewcommand{\arraystretch}{1.1}
%   \begin{tabular}{p{0.2\linewidth}>{\raggedright}p{0.1\linewidth}p{0.5\linewidth}p{0.2\linewidth}}
%     \toprule
%     \textbf{Variable} & \textbf{Abbr.} & \textbf{Measure Description} & \textbf{Scale} \\
%     \midrule
%    Comfort Relying on AV & Comfort & “How comfortable would you feel relying on this AV in this specific situation?” (proxy for trust) & (0-10 rating scale) \\
%    Reliance on AV & Reliance & “If this specific situation were to happen in the real world, would you prefer to rely on an AV or take control yourself?” (preference to take control) & (Binary choice; scaled 0 or 10) \\
%   \bottomrule
% \end{tabular}
% \end{table}
%\vspace{-0.5em}
\subsubsection{Context Descriptors} 
These measures were taken for each scenario after the “accurate” condition video only to provide insight into the impact of driving context on main outcomes. \color{black} Both measures were based on the external variability factors impacting situational trust described by \citet{hoff2015trust}. \color{black}

\begin{itemize}[itemsep=0pt, topsep=0pt]
    \item \textbf{\underline{Harm} of Driving Situation.} “In the real world, how would you rate the risk of harm of this specific driving situation?” (0-10 rating scale)
    \item \textbf{\underline{Difficulty} of Driving.} “In the real world, how would you rate the difficulty of driving in this specific situation?” (0-10 rating scale)
\end{itemize}

\vspace{1em}
\subsubsection{Additional Outcomes}
These further contextualize our main experimental findings and were measured before and/or after the rating task.

\begin{itemize}[itemsep=0pt, topsep=0pt]
    \item \textbf{Expertise}. Expertise was measured before the rating task via three self-rated questions related to a person's knowledge and understanding of AVs, \color{black} adapted from \citet{kaufman2024effects}. \color{black} (5-point likert scale from Strongly Disagree to Strongly Agree)
    \item \textbf{Trust}. Trust was measured before and after the rating task via four questions related to adaptability, safety, overt trust, and willing to recommend a friend to ride in an AV. \color{black} Questions were summed to form a single, comprehensive composite measure. This is similar to the approach taken by \citet{hewitt2019assessing}, from whom these questions were adapted. \color{black}(5-point likert scale from Strongly Disagree to Strongly Agree)
    \item \textbf{Explicit Factors Contributing to Reliance Decisions}. After the task, participants rated the relative impact of seven aspects of the driving and explanation experience on their reliance decisions, \color{black} building an explicit understanding of which factors may be most important. \color{black}(5-point likert scale from Not At All to Very Much).
\end{itemize}

\section{Results}
\subsection{Impact of Explanation Errors: Comfort, Reliance, \newline Satisfaction, and Confidence}
\subsubsection{Summary of Main Outcomes}
\begin{figure}[H] %change to H if i want to force positioning
  \centering
  \includegraphics[width=\linewidth]{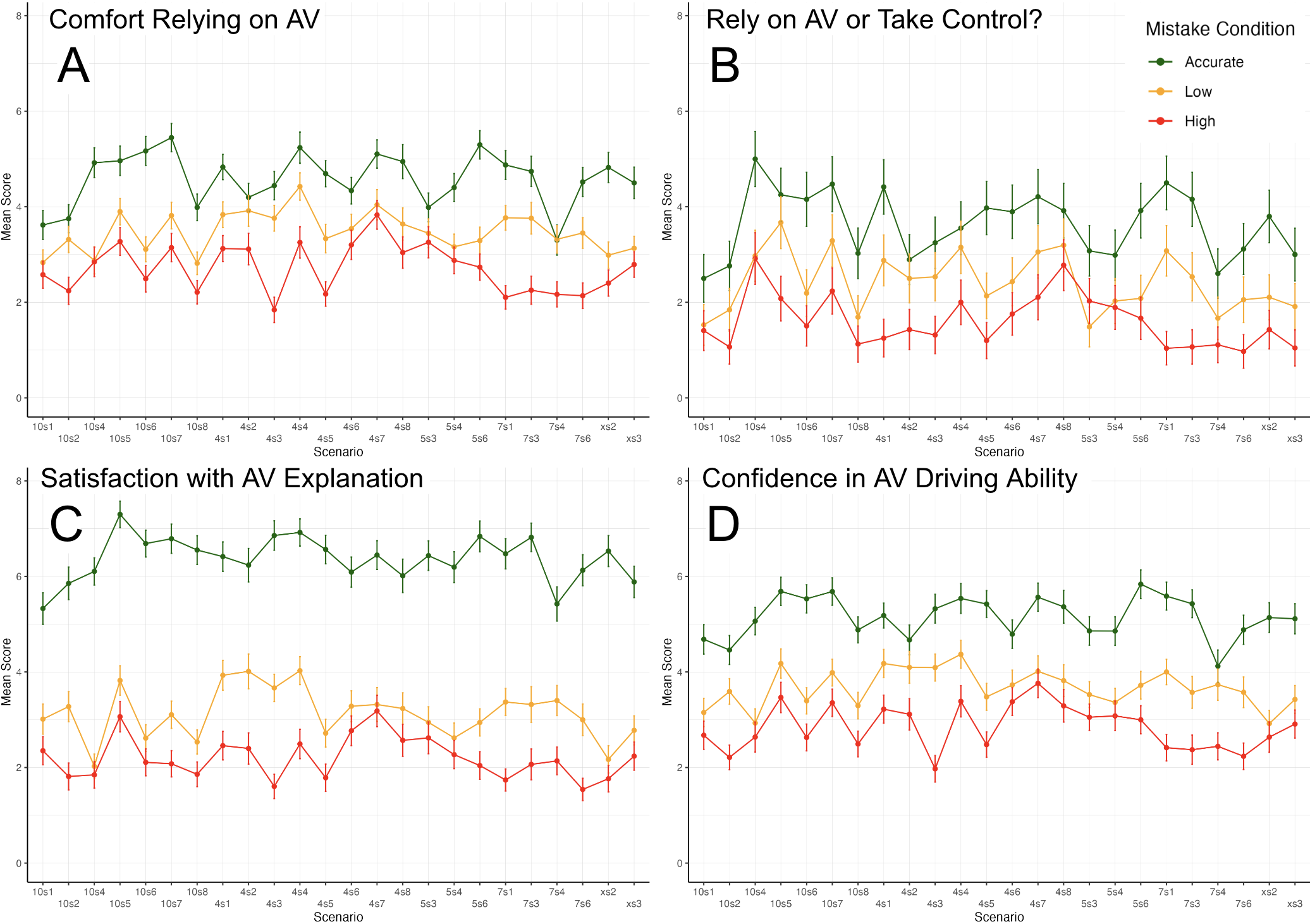} 
  \caption[\textbf{Mean Outcome by Error Condition Across All Scenarios.}]{\textbf{Mean Outcome by Error Condition Across All Scenarios.} Plot A shows mean scores for Comfort Relying on AV, Plot B shows mean scores for Reliance Preference (Binary, scaled to 1-10), Plot C shows mean scores for Satisfaction with AV Explanation, and Plot D shows mean scores for Confidence in AV Driving Ability. Error bars are standard error. Results consistently demonstrated the trend of Accurate $>$ Low $>$ High.}
 % \Description{Four visualizations, one for each major outcomes by error condition, shown for all 24 scenarios, depicting the mean outcome score by error condition across. Plot A shows mean scores for Comfort Relying on AV, Plot B shows mean scores for Reliance Preference (Binary, scaled to 1-10), Plot C shows mean scores for Satisfaction with AV Explanation, and Plot D shows mean scores for Confidence in AV Driving Ability. Results consistently demonstrated the trend of Accurate $>$ Low $>$ High.}
  \label{individualscenarios}
\end{figure}

Across our four main outcomes, segmenting the data by explanation error condition level gives us an initial impression on the impact of our experimental manipulation. We find the highest scores for comfort relying on the AV, reliance preference, satisfaction with the AV's explanation, and confidence in the AV's driving ability in the Accurate condition, followed by the Low condition and then the High condition (Table \ref{summary_of_main}). Visualizing main outcomes by scenario, we find that the effect of condition was very consistent across scenarios (Figure \ref{individualscenarios}\color{black})\color{black}.

Even when explanations were accurate, participants' overall comfort relying on the AV (comfort), and their preference to rely on the AV (reliance), were middling to low, reflecting the overall reluctance to trust AVs found in past human-AV interaction research  by~\citet{kenesei2022trust}. Participants were more positive about the AV explanations provided to them in the study, however, this satisfaction deteriorated quickly when errors were introduced. Despite the AV’s driving performance -- and therefore, demonstrated ability -- remaining consistent across all conditions, \emph{impressions} of the AV’s driving ability worsened as AV explanation errors were introduced.

\begin{table}[htbp] %change to H if i want to force positioning
\centering
  \caption[\textbf{Means and Standard Errors of Main Outcomes By Error Condition.}]{\textbf{Means and Standard Errors of Main Outcomes By Error Condition.} As errors increased, outcome ratings decreased.}
  %\Description{This table shows the Means and Standard Errors of each main outcomes, divided by Error Condition. As errors increased, outcome ratings decreased.}
  \label{summary_of_main}
  \begingroup
  \renewcommand{\arraystretch}{1}
  \begin{tabular}{p{0.32\linewidth}p{0.15\linewidth}p{0.15\linewidth}p{0.15\linewidth}}
    \toprule
     & \textbf{Accurate} & \textbf{Low} & \textbf{High}\\
    \midrule
    Comfort Relying on AV & 4.59 (0.06) & 3.48 (0.06) & 2.71 (0.06) \\
    Reliance Decision & 3.65 (0.11) & 2.43 (0.10) & 1.60 (0.09) \\
    Satisfaction w/ Expl. & 6.38 (0.06) & 3.13 (0.06) & 2.20 (0.06) \\
    Confidence in Driving & 5.16 (0.06) & 3.68 (0.06) & 2.84 (0.06) \\
    \bottomrule
    \end{tabular}
    \endgroup
\end{table}

\subsubsection{Overall Impact of Errors on Main Outcomes}
Linear mixed-effects (LME) models were used to measure the effect of errors on each major outcome \color{black} independently\color{black}. Though LME models produce similar results to mixed-model ANOVAs, they offer greater flexibility for repeated measures experiments. By incorporating random effects, they reduce the likelihood of Type 1 errors~\cite{boisgontier2016anova}. \color{black} Separate models were made for each outcome, with fixed effects for error condition (Accurate, Low, High) \color{black} and random effects for the specific scenario and individual participant. The fixed effects allow us to compare differences between study conditions, while the random effects allow us to control for differences by scenario and individual participant \cite{meteyard2020best, barr2013random}. \color{black} An example formula is: \[\textit{$ \text{outcome} \sim \text{error\_level} + (1 \mid \text{scenario}) + (1 \mid \text{participant}) $}\]

The model was fit using restricted maximum likelihood (REML) with Satterthwaite’s approximation for degrees of freedom. In our LME models, the ‘Accurate’ condition was set as the baseline, with contrasts comparing ‘Accurate’ to ‘Low’ and ‘High’ conditions. To compare the ‘Low’ and ‘High’ conditions directly, the baseline was changed to ‘Low’. This contrast structure allowed us to compare the relative impact of each error level. Using treatment coding allows for direct comparisons between each error condition, and is preferred over distributing contrasts across all levels when the primary goal is to interpret each level's effect in terms of its deviation from a meaningful baseline \cite{meteyard2020best, barr2013random}. Direct contrasts were calculated within the LME model framework, eliminating the need for additional post-hoc comparisons. Following best practices for LME model reporting \cite{meteyard2020best} and recent related work using similar models \cite{kaufman2024effects}, we provide detailed estimates (\begin{math}\beta\end{math}), degrees of freedom (df), t-values (t), and p-values (sig.) for each fixed effect to enhance reproducibility. \color{black}

Individual comparisons between each condition (Accurate-Low, Accurate-High, Low-High) show highly significant effects (p $<$ 0.001) across all four outcome measures (Figure \ref{LMEoverall}), implying error condition significantly impacted outcomes. The results for comfort relying on the AV, reliance decision, and satisfaction with the explanation follow the expected trend: we find that outcome scores decrease as error level increases. Surprisingly, we found the same effect on a person’s evaluation of the AV’s driving ability, where confidence scores also decrease as error level increases. This is unexpected, given that the driving performance shown in the videos were identical and only the explanations changed. The implication is that there is a cross-over effect between a person’s evaluation of the explanation and the person’s evaluation of the vehicle’s driving, \emph{despite} evidence suggesting that the driving performance is consistently high quality.

\begin{table}[htbp] %change to H if i want to force positioning
\centering
  \caption[\textbf{Comparative effects of each Error Condition on Main Outcome variables (LME Models).}]{\textbf{Comparative effects of each Error Condition on Main Outcome variables (LME Models).} These show highly significant outcome score differences between each error level.}
  %\Description{This table shows comparisons of each error condition for each main outcome (LME Models). These show highly significant outcome score differences between each error level.}
  \label{LMEoverall}
  \begingroup
  \renewcommand{\arraystretch}{1}
  \begin{tabular}{p{0.3\linewidth}p{0.1\linewidth}p{0.1\linewidth}p{0.1\linewidth}p{0.05\linewidth}}
    \toprule
    \multicolumn{5}{c}{\textbf{Acc. (intercept) vs. Low}} \\
    \midrule
    & \begin{math}\beta\end{math} & df & t & sig. \\
    \midrule
    Comfort Relying on AV & -1.08 & 5114 & -17.0 & *** \\
    Reliance Decision & -1.19 & 5114 & -10.2 & *** \\
    Satisfaction w/ Expl. & -3.24 & 5116 & -45.0 & *** \\
    Confidence in Driving & -1.46 & 5114 & -25.5 & *** \\ \\
    \toprule
    \multicolumn{5}{c}{\textbf{Acc. (intercept) vs. High}} \\
    \midrule
    & \begin{math}\beta\end{math} & df & t & sig. \\
    \midrule
    Comfort Relying on AV & -1.85 & 5115 & -28.9 &  *** \\
    Reliance Decision & -2.00 & 5115 & -17.1 & *** \\
    Satisfaction w/ Expl. & -4.16 & 5118 & -57.8 & *** \\
    Confidence in Driving & -2.29 & 5114 & -38.3 & *** \\ \\
    \toprule
    \multicolumn{5}{c}{\textbf{Low (intercept) vs. High}} \\
    \midrule
    & \begin{math}\beta\end{math} & df & t & sig. \\
    \midrule
    Comfort Relying on AV & -0.75 & 5110 & -11.9 &  *** \\
    Reliance Decision & -0.80 & 5108 & -6.8 & *** \\
    Satisfaction w/ Expl. & -0.93 & 5111 & -12.8 & *** \\
    Confidence in Driving & -0.83 & 5111 & -13.8 & *** \\
    \bottomrule
    \multicolumn{5}{p{.5\linewidth}}{\raggedright \textit{Sig. Codes: ‘***’ p $<$ 0.001}}
\end{tabular}
\endgroup
\end{table}

\subsubsection{Potential Harm of `What'-type Errors} 
In the high error condition group, what-type errors are incorrect descriptions of what the AV is doing. To understand the impact of the \emph{potential} harm of these errors, we conducted a secondary analysis comparing errors that would result in an accident if acted upon by the AV versus those that would not. \color{black} This analysis adds a more nuanced understanding for the impact of `what'-type errors on participant ratings of an AV. It is important to note that, though similarly named, this \textit{potential} harm categorization is unrelated to the perceived harm of the driving situation measured via participant rating. In the potential harm case, scenarios were categorized by the research team as harmful or not based on what the result would have been if the AV had driven in accordance with the mistaken `what' explanation. As the basis of our analysis, \color{black}  we use LME models with fixed effects for the categorization of potential harm (0 or 1) and random effects for driving scenario and individual differences by participant. This analysis is conducted only on data from the “high” condition in isolation.

We find significant effects for satisfaction with an explanation (\begin{math}\beta\end{math} = -0.46, t(22) = -2.6, p $<$ .05) and confidence in the AV's driving ability (\begin{math}\beta\end{math} = -0.45, t(22) = -2.7, p $<$ .05). Comfort relying on the AV showed results trending towards significance (\begin{math}\beta\end{math} = -0.38, t(22) = -1.9, p = 0.07). No significant results were found for the reliance preference measure. These results imply that the content of the error -- in this case, the potential harm that could result \emph{from} the error -- may impact the effect of the error on outcomes. Specifically, we find evidence that higher gravity errors may have a greater negative impact on some outcomes.

\subsection{Impact of Driving Context: Perceived Harm and Difficulty of Driving}
\subsubsection{Relationship Between Harm and Difficulty} 
\begin{figure}[H] %change to H if i want to force positioning
  \centering
  \includegraphics[width=\linewidth]{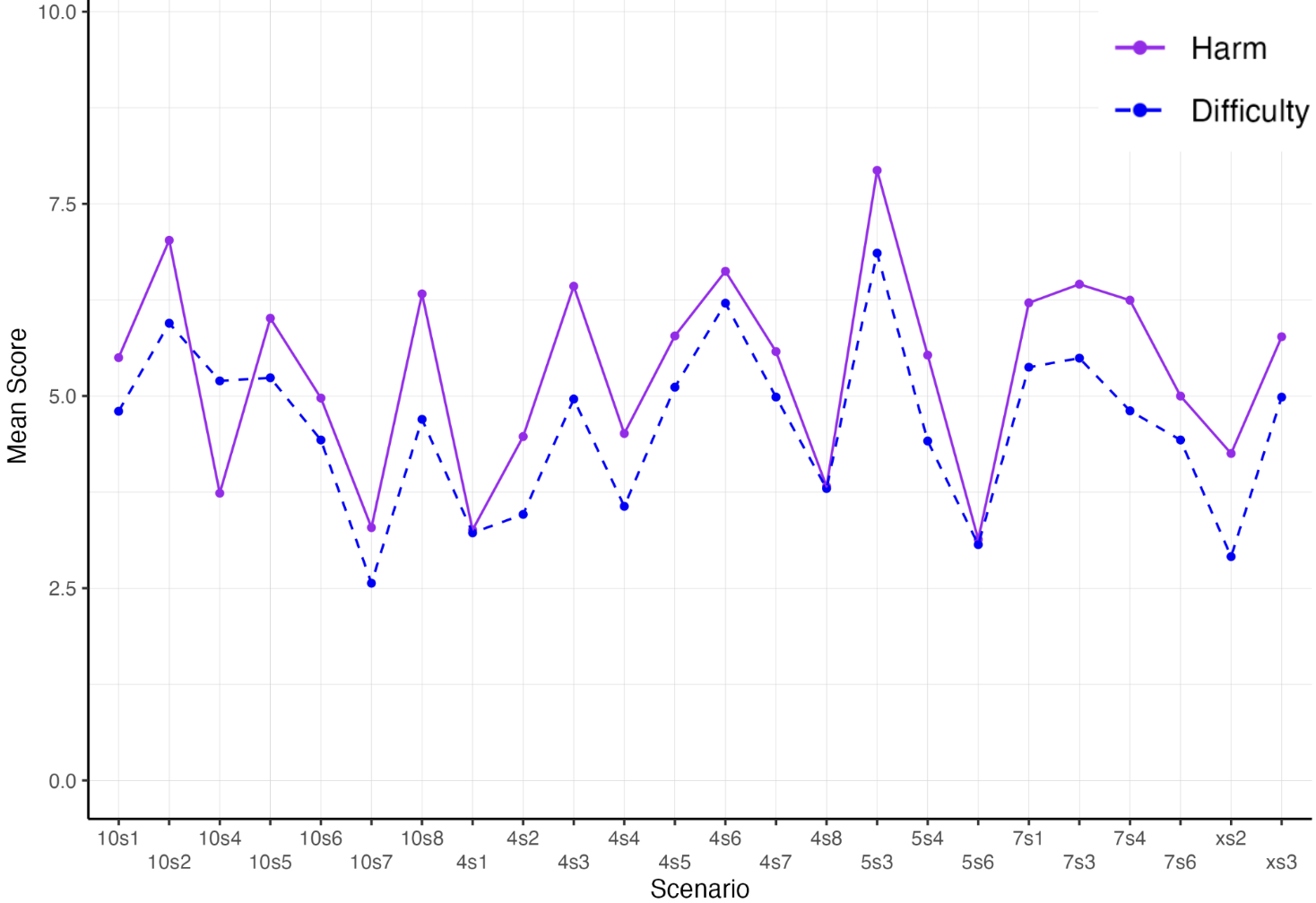}
  \caption[\textbf{Mean Harm and Difficulty Across All Scenarios.}]{\textbf{Mean Harm and Difficulty Across All Scenarios.} Results show harm and difficulty are generally correlated, but differ depending on the specific scenario. We find that these contextual characteristics directly impacted outcome ratings as well as influenced the relationship between errors and outcomes.}
  %\Description{A visualization of the mean Harm and Difficulty scores for all 24 scenarios, depicting the mean Harm and Difficulty ratings for each scenario. Results show harm and difficulty are generally correlated, but differ depending on the specific scenario. We find that these contextual characteristics directly impacted outcome ratings as well as influenced the relationship between errors and outcomes.}
  \label{individualdifficultyharm}
\end{figure}

By examining the impact of harm and difficulty on outcome ratings, we can derive an understanding of how these contextual factors influenced our results. Unsurprisingly, we found a strong, positive relationship between the perceived harm and the difficulty of driving in a particular scenario (r = 0.69, p $<$ .001; Adj R\textsuperscript{2} = .48). Figure \ref{individualdifficultyharm} shows difficulty and harm ratings by scenario.

%\vspace{2em}

\subsubsection{Impact of Harm and Difficulty On Main Outcome Ratings} 
We assess the impact of harm and difficulty on main outcome ratings using LME models with added fixed effects for the average harm and difficulty of scenarios. We maintain the fixed effect of \color{black}error \color{black} condition as well as random effects for scenario and participant \color{black} used previously, following the maximal structure described by \citet{barr2013random}. \color{black} Using this model, main effects for difficulty and harm give us an understanding of how these factors impacted outcome judgments \emph{independent} of error condition. \color{black} Once again, each outcome variable was modeled separately. \color{black} Estimates (\begin{math}\beta\end{math}) are given relative to the “accurate” condition. We find that harm significantly impacted comfort relying on the AV (\begin{math}\beta\end{math} = -0.43, t(38) = -3.7, p $<$ .001), reliance decision (\begin{math}\beta\end{math} = -0.74, t(62) = -4.4, p $<$ .001), and confidence in the AV's driving ability (\begin{math}\beta\end{math} = -0.22, t(41) = -2.2, p $<$ .05). No significant effects were found between harm and explanation satisfaction. We found that difficulty impacted only reliance decision (\begin{math}\beta\end{math} = 0.60, t(61) = 3.0, p $<$ .01). Interestingly, this was in the opposite direction as harm, where greater difficulty was associated with increased reliance. 

The implication is that contextual factors like harm and difficulty affect interaction with an AV system regardless of error condition, with harm potentially being the more influential factor. Specifically, higher harm is generally associated with lower comfort, reliance, and confidence, while higher difficulty is generally associated with higher reliance ratings. There does not appear to be an overall association between harm or difficulty and explanation satisfaction.

\subsubsection{Impact Of Harm and Difficulty On Differences Between Error Conditions (Interaction Effects)} 
Using similar LME models \color{black} as used to assess the main effect of error conditions, \color{black} we can assess if difficulty or harm impacted the relationship \emph{between} error level and outcomes by looking at interaction effects. Specifically, we examine if the \emph{differences} found between each error condition are predicted by a scenario's average difficulty or harm \color{black}(for example, for harm, we can modify our formula to look at the interaction between level and harm by looking at the effect of \textit{error\_level * mean\_harm} \cite{barr2013random})\color{black}. In this case, we use the accurate group as a common reference (model intercept) and then compare if the changes of each main outcome from accurate to low and accurate to high vary with respect to difficulty and harm. \color{black}Just as before, separate models were made for each outcome. \color{black} We find significant interaction effects for several of our outcomes, implying that contextual characteristics like harm and difficulty may be moderating how much of an impact an error may have. Table \ref{harminteraction} shows interaction effects for harm, and Table \ref{difficultyinteraction} shows interaction effects for difficulty.

\textbf{Comfort Relying on AV} -- For comfort relying on the AV, we find that scenario difficulty has more of an impact on comfort in the low error condition than in the accurate condition. Specifically, when difficulty increased, comfort decreased significantly more in the low condition compared to the accurate condition. A similar but opposite effect was found with harm: in the low error condition, comfort increases significantly more with higher harm than in the accurate condition. The implication is that comfort judgments may be more sensitive to the difficulty and harm of the scenario when there are some errors (low condition) compared to when there are no errors (accurate). We do not see difficulty or harm as more impactful on comfort judgments in the high error condition compared to the accurate condition. This implies that these judgments of comfort are likely based on the high magnitude of error (main effect) for the high group, as opposed to being influenced by the difficulty or harm of the driving context (interaction effect) in these cases. 

\textbf{Reliance Decision} -- The case is similar for reliance decision, though the effects are trending towards significance as opposed to being statistically significant. We find that scenario difficulty has more of an impact on reliance in the low error condition than it did in the accurate condition. Specifically, when difficulty increased, reliance decreased more in the low condition compared to the accurate condition. For harm in the low error condition, higher harm increases reliance significantly more than in the accurate condition. The implication is that reliance judgments may be more sensitive to the difficulty and harm of the scenario when there are some errors (low condition) compared to when there are no errors (accurate). As with comfort, we do not find difficulty or harm as more impactful on reliance in the high error condition compared to the accurate condition. This implies that judgments of reliance when errors are high are likely based on the error itself rather than difficulty or harm.

\textbf{Satisfaction w/ Expl.} -- The interaction effects change for satisfaction with an explanation. We do not see difficulty or harm as more impactful on satisfaction in the low error condition compared to the accurate condition, implying that differences in judgments of the explanation are likely based on the explanation quality itself without being influenced by the difficulty or harm of the situation in these cases. The same effect is seen when comparing differences between the accurate and high condition groups for harm. Surprisingly, we find that difficulty differentially impacted explanation satisfaction when in the high condition compared to the accurate condition. Specifically, when difficulty increased, satisfaction increased significantly more when in the high condition than when in the accurate condition. These results together imply that, in most cases, judgments of an explanation are likely based on the explanation quality itself, however, when explanation quality is exceptionally poor and driving is difficult, participants may actually be more forgiving with their judgments.

\textbf{Confidence in Driving} -- Lastly, for confidence in the AV's driving ability, we find a similar trend as comfort and reliance decision. We find that scenario difficulty has more of an impact on confidence in the low error condition than it did in the accurate condition. When difficulty increased, confidence decreased more in the low condition compared to the accurate condition (trending towards significance). For harm in the low error condition, higher harm increases confidence significantly more than in the accurate condition. The implication is that confidence judgments may be more sensitive to the difficulty and harm of the scenario when there are some errors (low condition) compared to when there are no errors (accurate). As with comfort and reliance, we do not find difficulty or harm as more impactful on confidence in the high error condition compared to the accurate condition. This implies that judgments of confidence are more likely based on the main effect of error itself, rather than difficulty or harm.

% \begin{table}[!htbp] %change to H if i want to force positioning
%   \caption{Impact of Harm on Error Condition Differences (LME Models)}
%   \label{harminteraction}
%   \begin{tabular}{p{0.2\linewidth}p{0.05\linewidth}p{0.05\linewidth}p{0.05\linewidth}p{0.02\linewidth}|p{0.05\linewidth}p{0.05\linewidth}p{0.05\linewidth}p{0.02\linewidth}|p{0.05\linewidth}p{0.05\linewidth}p{0.05\linewidth}p{0.02\linewidth}}
%     \toprule
%     & \multicolumn{4}{c|}{Acc.-Low} & \multicolumn{4}{c|}{Acc.-High} & \multicolumn{4}{c}{Low-High} \\
%     \midrule
%     & \begin{math}\beta\end{math} & df & t & sig. & \begin{math}\beta\end{math} & df & t & sig. & \begin{math}\beta\end{math} & df & t & sig. \\
%     \midrule
%     Comfort Relying on AV & 0.43 & 5108 & 4.2 & *** & 0.10 & 5108 & 0.9 &  NS & -0.34 & 5108 & -3.2 &  ** \\
%     Reliance Decision & 0.35 & 5106 & 1.8 & . & 0.19 & 5106 & 1.0 & NS & -0.16 & 5106 & -0.8 & NS \\
%     Satisfaction w/ Expl. & 0.13 & 5108 & 1.1 & NS & -0.15 & 5108 & -1.3 & NS & -0.28 & 5108 & -2.4 & *\\
%     Confidence in Driving & 0.27 & 5108 & 2.8 & ** & -0.01 & 5108 & -0.1 & NS & -0.28 & 5108 & -2.9 & ** \\
%     \hline
%     \multicolumn{13}{l}{\textit{Sig. Codes: ‘***’ p $<$ 0.001 | ‘**’ p $<$ 0.01 | ‘*’ p $<$ 0.05 | ‘.’ trending p $<$ 0.1}}
% \end{tabular}
% \end{table}

\begin{table}[htbp] %change to H if i want to force positioning
  \caption[\textbf{Impact of Harm on Error Condition Differences For Each Main Outcome Variable (LME Models).}]{\textbf{Impact of Harm on Error Condition Differences For Each Main Outcome Variable (LME Models).} These show if Harm influences the \textit{impact} of an error. We find significant effects for several of our outcomes, implying that Harm may moderate how much of an impact an error may have.}
  %\Description{This table shows the impact of Harm on Error Condition Differences (LME Models). These show if Harm influences the impact of an error. We find significant effects for several of our outcomes, implying that Harm may moderate how much of an impact an error may have.}
  \label{harminteraction}
  \begingroup
  \renewcommand{\arraystretch}{1}
  \begin{tabular}{p{0.3\linewidth}p{0.06\linewidth}p{0.05\linewidth}p{0.06\linewidth}p{0.04\linewidth}|p{0.06\linewidth}p{0.05\linewidth}p{0.06\linewidth}p{0.04\linewidth}}
    \toprule
    & \multicolumn{4}{c|}{\textbf{Acc. (intercept) vs. Low}} & \multicolumn{4}{c}{\textbf{Acc. (intercept) vs. High}} \\
    \midrule
    & \begin{math}\beta\end{math} & df & t & sig. & \begin{math}\beta\end{math} & df & t & sig. \\
    \midrule
    Comfort Relying on AV & 0.43 & 5108 & 4.2 & *** & 0.10 & 5108 & 0.9 &  NS \\
    Reliance Decision & 0.35 & 5106 & 1.8 & . & 0.19 & 5106 & 1.0 & NS \\
    Satisfaction w/ Expl. & 0.13 & 5108 & 1.1 & NS & -0.15 & 5108 & -1.3 & NS \\
    Confidence in Driving & 0.27 & 5108 & 2.8 & ** & -0.01 & 5108 & -0.1 & NS  \\
    \bottomrule
    \multicolumn{9}{p{.95\linewidth}}{\raggedright \textit{Sig. Codes: ‘***’ p $<$ 0.001 $|$ ‘**’ p $<$ 0.01 $|$ ‘*’ p $<$ 0.05 $|$ ‘.’ trending p $<$ 0.1}}
\end{tabular}
\endgroup
\end{table}

% \begin{table}[!htbp] %change to H if i want to force positioning
%   \caption{Impact of Difficulty on Error Condition Differences (LME Models)}
%   \label{difficultyinteraction}
%   \begin{tabular}{p{0.2\linewidth}p{0.05\linewidth}p{0.05\linewidth}p{0.05\linewidth}p{0.02\linewidth}|p{0.05\linewidth}p{0.05\linewidth}p{0.05\linewidth}p{0.02\linewidth}|p{0.05\linewidth}p{0.05\linewidth}p{0.05\linewidth}p{0.02\linewidth}}
%     \toprule
%     & \multicolumn{4}{c|}{Acc.-Low} & \multicolumn{4}{c|}{Acc.-High} & \multicolumn{4}{c}{Low-High} \\
%     \midrule
%     & \begin{math}\beta\end{math} & df & t & sig. & \begin{math}\beta\end{math} & df & t & sig. & \begin{math}\beta\end{math} & df & t & sig. \\
%     \midrule
%     Comfort Relying on AV & -0.26 & 5109 & -2.1 & * & 0.1 & 5109 & 0.7 &  NS & 0.35 & 5108 & 2.9 &  **\\
%     Reliance Decision & -0.42 & 5107 & -1.9 & . & -0.16 & 5107 & -0.7 & NS & 0.26 & 5106 & 1.1 & NS \\
%     Satisfaction w/ Expl. & -0.08 & 5109 & -0.6 & NS & 0.30 & 5109 & 2.1 & * & 0.38 & 5108 & 2.7 & **\\
%     Confidence in Driving & -0.21 & 5108 & -1.9 & . & 0.04 & 5109 & 0.3 & NS & 0.25 & 5108 & -2.8 & **\\
%     \hline
%     \multicolumn{13}{l}{\textit{Sig. Codes: ‘***’ p $<$ 0.001 | ‘**’ p $<$ 0.01 | ‘*’ p $<$ 0.05 | ‘.’ trending p $<$ 0.1}}
% \end{tabular}
% \end{table}

\begin{table}[htbp] %change to H if i want to force positioning
  \caption[\textbf{Impact of Difficulty on Error Condition Differences For Each Main Outcome Variable (LME Models).}]{\textbf{Impact of Difficulty on Error Condition Differences For Each Main Outcome Variable (LME Models).} These show if Difficulty influences the \textit{impact} of an error. We find significant effects for several of our outcomes, implying that Difficulty may moderate how much of an impact an error may have.}
  %\Description{This table shows the impact of Difficulty on Error Condition Differences (LME Models). These show if Difficulty influences the impact of an error. We find significant effects for several of our outcomes, implying that Difficulty may moderate how much of an impact an error may have.}
  \label{difficultyinteraction}
\begingroup
  \renewcommand{\arraystretch}{1}
  \begin{tabular}{p{0.3\linewidth}p{0.06\linewidth}p{0.05\linewidth}p{0.06\linewidth}p{0.04\linewidth}|p{0.06\linewidth}p{0.05\linewidth}p{0.06\linewidth}p{0.04\linewidth}}
    \toprule
    & \multicolumn{4}{c|}{\textbf{Acc. (intercept) vs. Low}} & \multicolumn{4}{c}{\textbf{Acc. (intercept) vs. High}} \\
    \midrule
    & \begin{math}\beta\end{math} & df & t & sig. & \begin{math}\beta\end{math} & df & t & sig. \\
    \midrule
    Comfort Relying on AV & -0.26 & 5109 & -2.1 & * & 0.1 & 5109 & 0.7 &  NS \\
    Reliance Decision & -0.42 & 5107 & -1.9 & . & -0.16 & 5107 & -0.7 & NS  \\
    Satisfaction w/ Expl. & -0.08 & 5109 & -0.6 & NS & 0.30 & 5109 & 2.1 & * \\
    Confidence in Driving & -0.21 & 5108 & -1.9 & . & 0.04 & 5109 & 0.3 & NS \\
    \bottomrule
    \multicolumn{9}{p{.95\linewidth}}{\raggedright \textit{Sig. Codes: ‘***’ p $<$ 0.001 $|$ ‘**’ p $<$ 0.01 $|$ ‘*’ p $<$ 0.05 $|$ ‘.’ trending p $<$ 0.1}}
\end{tabular}
\endgroup
\end{table}

\subsection{Trust and Expertise}
\subsubsection{Impact Of Exposure To Errors On Trust}
We measured participant trust before and after the experiment was completed to see if exposure to repeated AV explanation mistakes affected AV trust levels as a disposition. In general, using a paired-samples t-test, we did not find that exposure to errors in the experiment impacted trust. Only one result was significant pre-post experiment: participants felt that they “understand why an autonomous vehicle makes decisions” worse after the experiment as compared to before (p $<$ 0.001).

\subsubsection{Correlations of Initial Trust and Subjective Expertise with Outcomes}
Calculating correlations between initial trust and self-reported expertise level and a participants’ average outcome ratings allows us to assess if trust or expertise can predict how a person will evaluate the AV. We found a strong positive correlation between initial trust and expertise, (r = 0.58, p $<$ 0.001), implying that those who know more about AVs trust them more.

We found moderate positive correlations between expertise and comfort (r = 0.38, p $<$ 0.001), satisfaction (r = 0.35, p $<$ 0.001), and confidence (r = 0.36, p $<$ 0.001), meaning that those with more expertise reported more positively to these outcomes on average. We found high correlations between initial trust and comfort (r = 0.52, p $<$ 0.001) as well as confidence (r = 0.51, p $<$ 0.001). We found moderate correlations between initial trust and reliance (r = 0.30, p $<$ 0.001) as well as satisfaction (r = 0.49, p $<$ 0.001). These effects are consistent with linear mixed-effects models looking at the main effects of expertise and initial trust on outcomes. Together, these results imply that participants with higher initial trust or expertise tended to rate outcomes higher on average.

\subsection{Self-Reported Rationale for Reliance Decisions}
\subsubsection{Factors Contributing to Rating Decisions}
\begin{figure}[!htbp] %change to H if i want to force positioning
  \centering
  \includegraphics[width=\linewidth]{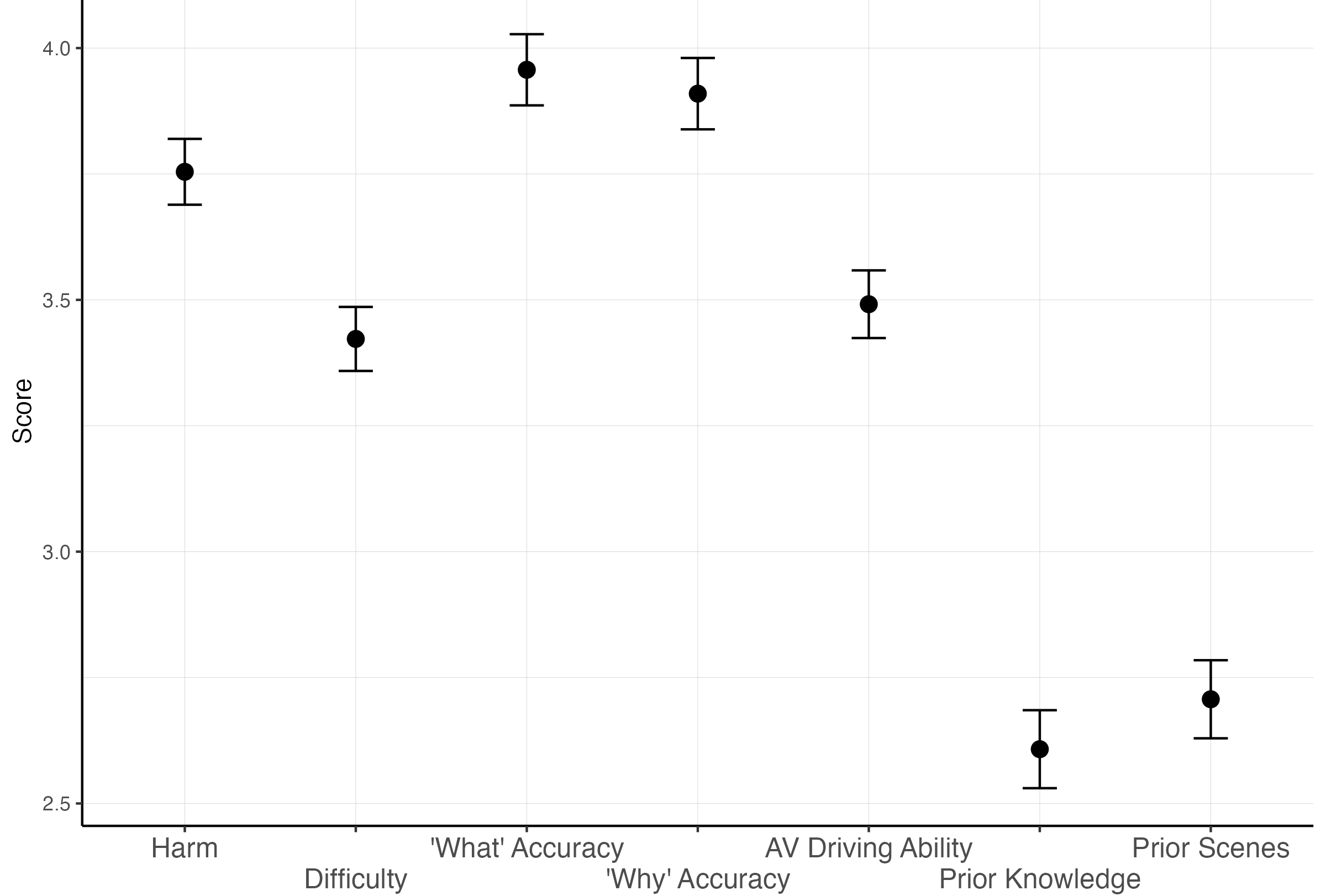}
  \caption[\textbf{Relative Importance of Factors on Reliance Decisions.}]{\textbf{Relative Importance of Factors on Reliance Decisions.} This shows the mean \color{black}self-reported \color{black} importance rating of each factor with respect to a person's reliance decision \color{black} (errors bars are standard error). \color{black} Results show \color{black} the accuracy of each type of information, perceived contextual harm and difficulty, and the driving ability of the AV were all important factors of consideration\color{black}.}
  %\Description{A visualization of relative importance (mean rating) of seven factors on reliance decisions. Results show many of the factors of interest in this study were of high importance.}
  \label{factors}
\end{figure}

We explicitly inquired upon the basis of reliance decisions by asking participants to rate the relative importance of several factors on their reliance ratings (Figure \ref{factors}). \color{black} This analysis helps clarify which elements participants explicitly prioritized, offering valuable guidance for designing explanations that align with user preferences and expectations. \color{black} Ratings were given after the main scenario experiment concluded. As a whole, we found that the accuracy of the explanation on ‘what’ the AV is doing was the most important consideration \color{black}(mean = 3.96, SE = 0.07)\color{black}, followed by the explanation accuracy of ‘why’ the AV is doing it \color{black}(mean = 3.91, SE = 0.07)\color{black}, the harm of the situation \color{black}(mean = 3.75, SE = 0.07)\color{black}, the AV’s driving ability \color{black}(mean = 3.49, SE = 0.07)\color{black}, and the difficulty of the driving situation \color{black}(mean = 3.42, SE = 0.06)\color{black}. Prior scenes viewed \color{black}(mean = 2.71, SE = 0.08) \color{black} and prior knowledge \color{black}(mean = 2.61, SE = 0.08) \color{black} did not have much reported impact, though this is unsurprising, as these implicit judgments may be difficult for participants to self-describe. Many factor differences were significant based on \color{black}Kruskal-Wallis test ($\chi^{2}$ = 285.7, df = 6, p $<$ 0.001). For pairwise comparisons, we use Dunn's test and adjust p-values using Bonferroni's method. \color{black} Notably, we find harm was significantly more important than difficulty \color{black}(p $<$ 0.05)\color{black}, while explanation accuracy (both `what' and `why' components) were significantly more important than AV driving ability \color{black}(both p $<$ 0.001). \color{black} This latter effect may be in part due to the salience of these factors based on the study's manipulation. We did not see significant differences between the importance of `what' and `why' accuracy.

\subsubsection{Feedback from Participants}
Participants were given the opportunity to express their general views on autonomous vehicles on how explanation errors impacted their decision-making specifically. Unsurprisingly, many participants reflected general concerns with AVs, including doubt that they can \textit{“adapt to any circumstance … [causing] a bigger accident.”} One participant commented that \textit{“after a lifetime of driving myself, I'm not sure if I feel comfortable giving over control to a computer.”} Desire for control was brought up multiple times. The common sentiment was that giving control to AVs is considered risky. 

Regarding exposure to explanation errors, one participant commented that they were \textit{“less confident in the reliability of [autonomous] vehicles”} after exposure to errors, while another commented that when explanations \textit{“were wrong, [it made] my confidence in the system shaky.”} These comments reiterate our findings on the negative impact of errors on reliance, as well as the holistic judgment of an AV's ability in general. Some participants broke their decision-making down in terms of individual factor priority. For instance, \textit{“if the AV described the action it was taking incorrectly I was a lot less confident and consistently chose to take control myself … if the reason was incorrect, it still impacted my confidence level but not as much.”} This reflects the general trend found in our individual factor analysis suggesting `what' information may be more important than `why' rationale, even if this trend was not found to be statically significant.

Regarding the impact of context, a participant noted, \textit{“I chose to take control myself in higher risk situations, even if the AV did a good job navigating.”} This supports this study's overall findings on the importance of contextual factors like harm on decision-making.

\color{black}
\section{Discussion}
This study aimed to assess the impact of autonomous vehicle (AV) explanation errors and driving context characteristics on participant comfort relying on an AV, preference to rely on the AV instead of taking control themselves, satisfaction with the explanation, and confidence in the AV’s driving ability. Using a mixed-methods approach with a heterogeneous sample of participants (n = 232), we found that explanation errors and contextual characteristics like driving difficulty and perceived harm have a large impact on how a person may think, feel, and behave towards AVs. These findings reinforce the overarching dissertation theme that human-AV interaction depends on dynamically balancing system reliability, user perceptions, and situational demands. By demonstrating how explanation errors can undermine trust in even error-free driving performance, this study highlights the interconnected nature of user perceptions and system design within complex human-AV systems. Driving context, particularly perceived harm and task difficulty, played a crucial role in shaping user responses, supporting the need for context-sensitive communication strategies discussed throughout this dissertation. Echoing prior work by~\citet{nourani2020role}, important consideration must also be given to personal factors like initial trust and expertise, which may further impact how a person interacts with the system. Together, these findings contribute to a deeper understanding of how AV communication strategies must evolve to meet diverse user needs across varied driving scenarios, a key focus of this dissertation. Our results provide insight into how to design effective human-AV interactions and interactions with AI-based systems more generally. We discuss key findings and their implications for future AV design in these contexts.

\textbf{Autonomous vehicle (AV) explanation errors had a detrimental effect on all outcomes \color{black}(RQ1)\color{black}.} Our results suggest that errors significantly reduced participant comfort relying on an AV, preference to rely on the AV instead of taking control themselves, and satisfaction with the explanation. These effects are unsurprising, as we would expect a person's reliance decision or satisfaction with an explanation to reflect the system's performance~\cite{hoff2015trust}. Though unsurprising, findings still reinforce the concept that an AV's traits -- including their performance -- will impact the wider behavior of the complex system within which they are embedded. This aligns with the broader argument of this dissertation that effective system design requires holistic consideration of how communication and system performance jointly influence user trust.

We were surprised, however, to find crossover effects between the \emph{explanatory} performance of the AV and a person's confidence in the AV’s \emph{driving ability}, given the AV’s actual demonstrated driving performance remained consistent across all conditions. This crossover effect alludes to the mental model of potential AV users, where evaluation of the explanatory performance of the vehicle and the driving performance of the vehicle are connected. This unexpected linkage underscores the importance of treating AV explanations as integral components of user trust, supporting the dissertation’s emphasis on designing systems where communication and functionality are seamlessly aligned. Though connecting explanatory and driving performance may be an intuitive and correct assumption, empirically observing this crossover provides evidence supporting the importance of high quality explanatory interfaces, particularly within safety-critical systems. \color{black}Further highlighting this point, the finding that explanation accuracy was explicitly rated as more important than driving ability for reliance decisions underscores the role of transparency and comprehensibility in user acceptance. This highlights the need for AV systems to prioritize clear and accurate explanations of actions (`what’) and rationales (`why’) in building trust, and suggests that users may evaluate AV systems as much by their communications as by their actual driving performance. \color{black} Taken together, in the case of AVs, this means that -- even if the AV's driving performance is perfect -- if the explanations produced by the AV are not accurate as well, people may still refuse to adopt AV technology. This insight likely generalizes to explanatory communications for other AI-based systems.

\color{black}Even with accurate explanations, Comfort, Reliance Preference, and Confidence in the AV's driving ability were nowhere near ceiling level -- hovering near the middle of the provided scale. Though the middling scores may be reflective of the overall challenging nature of the driving situations presented in the study, even the most straightforward scenarios did not have a mean reliance preference score greater than mid-range. These results reflect the trend found in a plethora of prior work supporting the premise that people do not generally trust or wish to adopt autonomous vehicles, reemphasizing this as an important area of continued study. \color{black}

\textbf{The negative impact of errors increased with error magnitude and the potential harm of the error \color{black}(RQ1 continued)\color{black}.} Expanding on the effect of errors observed in our study, we find that the impact was commensurate with the magnitude of the errors presented. In the study presented here, ‘what' and `why’ errors (high condition) in combination had worse outcomes than ‘why’-only errors (low condition). On the surface, this is unsurprising, as more errors demonstrates lower system performance, and these results could simply be the cumulative effects of seeing errors on two parts of the explanation instead of just one. 

There remains a high probability that `what' and `why' information errors play distinct roles in the evaluation of the system's performance, however. This would align with past work showing how description of action and description of justification may differentially impact the way a person interacts with autonomous systems~\cite{koo2015did, kaufman2024effects, miller2019explanation}. When asked explicitly about the factors that contributed to their reliance decisions, participants reported that explanation accuracy (both `what' and `why' components) were both important. Qualitative assessment of participant feedback, however, suggests that ‘what’ information may be more important for reliance outcomes than `why' rationale. This may be due to the potential harm that improper action (`what') can have on driving safety, in contrast to improper justification (`why'). Concretely, a car incorrectly turning right into traffic can cause physical harm, but incorrectly justifying \emph{why} it is turning right can not. \color{black} This finding highlights that explanation accuracy, particularly for ‘what’ information, is crucial not only for user comfort but also for the perception of the AV’s reliability and overall system competence. \color{black} We note that \emph{conclusively} differentiating the impact of `what' from ‘why’ errors independently will be left for future work.

Supporting the hypothesis that the \emph{implications} of an error matter in addition to the mere \emph{presence} of an error, we find that ‘what’ errors -- with more dire implications for potential harm -- had a greater negative impact on comfort, confidence, and explanation satisfaction. Specifically, in cases when `what' errors would have resulted in vehicle crashes if the AV had acted upon them, people were far less comfortable relying on the AV, had less confidence in the AV's driving, and preferred the explanation less. We posit that the only reason we did not see differences in reliance preference as well was because the presence of an error altogether already reduced reliance preference to near-minimum levels. 

Together, these results suggest that the implications of an explanation error in terms of what they may mean about the system's performance and consequences of harm, affect how a person thinks, feels, and behaves with an AV. Implications may be implicitly calculated based on the amount and type of error, as well as how these intersect with the external driving context.

\textbf{Contextual factors like the difficulty of driving in a specific situation and the perceived harm of that situation directly affect how people think, feel, and prefer to behave with AVs \color{black}(RQ2)\color{black}.} We found evidence that driving difficulty and perceived harm directly influenced outcomes regardless of error condition. Specifically, higher harm was associated with lower comfort, reliance, and confidence ratings, while higher difficulty was associated with higher reliance ratings. We did not find direct associations between harm or difficulty and explanation satisfaction.

We attribute the negative impact of harm to the common concern that AVs may malfunction, and a malfunction in a more harmful driving situation may have more severe implications. This reflects the general sentiment that many people think \emph{they} can drive better than an AV in most situations. The seemingly positive impact of difficulty on reliance may reflect that, in very high difficulty situations, people lack the confidence that they could perform better than the AV and, as such, prefer to rely on the AV in these cases. Indeed, the driving demonstrated by the AV in even the most difficult driving situations presented in the study was high quality. 

Explicitly, participants reported contextual harm as significantly more important than driving difficulty for their reliance decision. \color{black}Theoretically, this may be because difficult driving situations often entail a higher risk of harm, especially when participants are uncertain about the AV's ability to handle these challenges. We hypothesize that reliance decisions are grounded in an overall evaluation of potential harm, with driving difficulty acting as one indicator of this risk. Accordingly, \color{black} difficulty and AV driving performance matter \emph{because} harm matters, making harm itself the underlying driver of reliance decisions. 

A person's decision to rely on an AV (and other outcomes) may be a function of the demonstrated performance of the vehicle -- compounded by an implicit evaluation of the overall harm of the situation (in part as a function of difficulty) -- weighted against that person's confidence in their own ability to perform better. This is similar to the model proposed by~\citet{hoff2015trust}, with added details on the nested relationship between driving difficulty and harm.

The general implication of these findings is that there is a complex relationship between contextual factors like harm and difficulty, and outcomes like comfort, reliance, and confidence. It is clear that models of AV behavior incorporating internal and external characteristics of the driving situation can bring further insight into how people will interact with AVs~\cite{kaufman2024developing}, and future research should examine how these can be best supported by explanatory communication. 

\textbf{Contextual factors moderate the relationship between errors and outcomes in complex, sometimes counter-intuitive ways \color{black}(RQ2 continued)\color{black}.} By examining the interaction between contextual characteristics and the effects of error conditions, we found that -- in some cases -- the driving context may influence the \emph{amount} of impact an error has. The general trend we observed for comfort relying on the AV, reliance preference, and confidence in the AV's driving ability was that when difficulty increased, outcome scores decreased \emph{more} in the low error condition compared to the accurate condition, with no differences seen between the accurate and high conditions. The same but opposite effect was found for harm: higher harm increased outcomes effect in the low condition more than the accurate condition, with no difference between the accurate and high condition.

The lack of difference found in the way difficulty and harm moderate the relationship between the accurate and high conditions implies that differences in these judgments are likely based solely on the error level as opposed to difficulty or harm of the driving context. Concretely, the effect of error likely overrides the effect of context when errors are at the extremes. When errors are in the middle -- such as for the low error condition -- the impact of the error appears to be more strongly moderated by the context. This may be because the ramifications of `why'-type errors are not strong enough to drive effects fully based on the presence of the error, and instead the amount of change that the error can cause is based on how dire the implications may be in a particular context. The direction of effect, in this case, is similar to the main effects explained previously.

For explanation satisfaction, we do not see difficulty or harm as more impactful on satisfaction in the low error condition, if we compare this to the accurate condition. This implies that differences in satisfaction are likely based on the explanation quality itself in these cases. The same effect is seen between the accurate and high condition for harm, but not for difficulty. Surprisingly, we found that when difficulty increased, satisfaction increased significantly more when in the high condition than when in the accurate condition. Together, these results imply that, in most cases, judgments of an explanation are based primarily on the explanation quality itself, however, when explanation quality is especially poor and driving is difficult, participants may actually be more forgiving of the AV.

Taken together, results from this analysis on interaction effects illustrate a nuanced and often complex relationship between driving difficulty, perceived harm and error condition when predicting our main outcomes. This nuance alludes to a prioritization of how different aspects of the driving situation -- including harm, difficulty, and explanation errors -- combine to form judgments of comfort in the AV, reliance preference, satisfaction with an explanation, and confidence in an AV's driving ability.

\textbf{Trust and expertise influenced how people think, feel, and behave towards the AV \color{black}(RQ3)\color{black}.} As discussed, how a person thinks and behaves towards AVs is not just a function of their external environment. We find evidence that internal characteristics, such as a participant’s trust and expertise, may also have played a role in the study’s results. Participants with higher initial trust or expertise tended to rate study outcomes higher on average. We also found that those with higher AV expertise trusted AVs more in general. As with the function on driving reliance described previously, these findings also align with Hoff and Bashir's model of trust with AI systems, where a person's prior experience impacts their reliance with the system~\cite{hoff2015trust}. Those with higher expertise may also have other underlying characteristics, like an affinity for new technologies, which may have influenced their trust. We attribute the finding that priors were not reported as explicitly important for reliance decisions in our post-evaluation to the common difficulty articulating the impact of implicit dispositions.

Though trust may be influential, we did not find evidence that exposure to AV mistakes due to the study’s manipulation had a large effect on a person’s overall trust level. A single exception was that participants reported less understanding of why AVs make decisions once the study concluded. This makes sense given the inconsistent explanations for behavior they had received, which may have impacted a person's mental model of AV decision-making.

\subsection{Implications For Autonomous Vehicle Design and Research}
This study provides actionable insights for AV design and future research by emphasizing the importance of context-aware, personalized communication strategies, as discussed throughout this dissertation. Understanding the consequences of AV explanation errors and contextual characteristics like driving difficulty and perceived harm are necessary first steps towards designing vehicles which may be more trustworthy, reliable, and satisfactory for people interacting with the AV. \color{black} Using current, multimodal XAI methods of explanation presentation -- visual and auditory cues of both `what' and `why' information -- allows this study’s findings to directly apply to current AV explanation design. \color{black} By contributing to the foundational principles of adaptive human-AV communication systems, these findings set the stage for future investigations into optimizing explanatory interfaces, improving user experience, and fostering broader public acceptance of autonomous systems. In this section, we will discuss how our study's insights can be applied to build more trustworthy AVs.

The foremost implication of this work is to emphasize the importance of designing systems that can produce accurate explanations for AV decisions. This implication is not too surprising, as why would designers ever \emph{intentionally} produce inaccurate explanations? It becomes more meaningful, however, when testing explanatory systems before deployment. Our results indicate that even with a well-functioning driving system, if explanations are not of high quality, people still won't want to use the AV.

The crossover between a person's evaluation of \emph{explanation} performance and \emph{driving} performance may be of particular interest to AV designers. If this crossover is a concern -- as it should be in the case of deployment -- the most obvious solution would be to work on improving explanations so that they are on par with driving ability. In the meantime, specific UI features may be implemented to help users separate their evaluations of explanations from driving performance, such as by presenting distinct indicators for each system reassuring a rider that even if the explanation is messed up, the car can still drive sufficiently. This may be challenging or misleading, however, particularly if the explanatory system and driving system are indeed connected.

Though not tested in the present study, our results also have implications for \emph{conditionally} automated vehicles -- those which may require human takeover in difficult or computationally complex driving situations (SAE Level 3)~\cite{inagaki2019critique, ayoub2021investigation}. There is the possibility that explanation errors produced in conditional contexts may result in inappropriate behaviors by human passengers. If users lose trust in the AV due to explanation errors, for example, they may react unpredictably or take over control at inappropriate times, potentially leading to accidents.

For both the fully autonomous and conditionally autonomous cases, our results indicate that explanatory systems should be deployed with confidence in their accuracy or potentially not at all. Future work can estimate the exact outcome differences between no explanation and inaccurate explanations, but combining the results of our study and prior findings on the importance of AI explainability by~\citet{gunning2019darpa} and~\citet{miller2019explanation} suggests that \emph{accurate} explanations should be opted for whenever possible, and deploying untested or inconsistent systems can be disastrous.

In terms of design priority, our work suggests that if explanation designers are going to focus on improving a single aspect of the explanation, they should optimize for accurate `what' information before `why' justification. Though both may be necessary in the long run, providing a correct description of behavior may produce a higher value return on effort for teams looking to find a place to start. Just as they have for differentiating the effects of accurate `what' and `why' explanations of driving behavior, future work should delineate the potentially different impacts `why' and `what' \emph{errors} may have from each other in isolation. This can allow more detailed and conclusive conclusions on their relative importance to outcomes such as those in the present study.

It is important to note that, while explanation satisfaction scores were generally higher than other outcome scores in the accurate condition, explanations were still nowhere near ceiling level. This means that there are likely ways in which our explanations could have improved. Prior work has emphasized outcome differences based on informational content, modality (such as including visualizations or haptic feedback), timing, or interactivity~\cite{miller2019explanation, kaufman2024effects, avetisyan2022investigating, goldman2022trusting}. In this regard, future AV research and design teams should continue iterating on these aspects of explanation design and human-machine interface (HMI) development in order to fine-tune explainable systems.

Another major implication of this work is to orient AV designers to the potentially different outcomes that may result from interactions by people with different traits or while operating in driving environments. \color{black}In AVs, user familiarity with driving norms allows for immediate identification of errors, which intensifies the effect on trust and reliance. This contrasts with fields where users cannot as easily detect errors without domain expertise, highlighting the importance of context-specific trust calibration mechanisms for XAI systems. Specific to the case studied here, \color{black} context-aware or personalized explanations may be necessary in cases where the driving difficulty or perceived harm are classified as greater, or for people with little expertise or initial trust. It is likely that designers will need to focus on limitations on cognitive resources like attention and load, particularly for high difficulty scenarios. \color{black} In high-risk domains like AVs, these factors may be particularly important, as the potential consequences of explanation errors are immediate and tangible. We thus contribute to an emerging understanding of how explanation accuracy and error detectability shape user perceptions in safety-critical environments. \color{black} Other personal traits like risk preferences or personality may also be a basis for personalization~\cite{bockle2021can}. Segmenting populations or isolating particular driving conditions to test future explanation designs would be an effective method to figure out how to meet the differing needs of diverse user groups~\cite{kaufman2024developing}.

Taken in full, our results provide insight into the impact of explanation errors, and provide direction for future AV researchers and designers to improve human-AV communication. Given the serious negative repercussions that may result from errors, our results also provide a foundation for the creation of ethical or regulatory guidelines which dictate the testing and accuracy of AV explanations before they can be deployed to real-world consumers.

\subsection{Limitations and Future Study}
As with any study, this research is not without limitations. First, though study findings were generally robust and fit within logical narratives, there is the possibility that findings resulting from online data collection and based on simulations may not generalize to real-world attitudes or behavior. A large effort was put into making driving simulations as realistic as possible, however, we could not test the effect of explanation errors on real roads out of concern for participant safety. Future work should seek to test the impact of errors in a real-world environment. A similar concern may be found for study outcomes: though results on reported reliance or comfort may be clear in a research context, there is no guarantee that explicit declarations of thought or behavior will remain consistent when immersed in a real-world driving context with real consequences of bodily or financial harm. This is a concern for all studies which rely on survey measures or explicit participant statements as a proxy for real-world behavior. It is possible that seeing each scenario three times (with different explanations) may have impacted the ratings provided. This could be mitigated in the future by showing participants only one video per scenario, randomized by error condition. Finally, we examined proximal explanations for action (\textit{“braking”}) and cause of action (\textit{“… a pedestrian is crossing the road.”}) presented in written fashion and auditory fashion. Future work can examine explanations of distal causes (“pedestrian in the road … because there is an obstacle on the sidewalk”) which could provide additional context for \emph{why} a driving situation is happening in the first place. These could be explained using potentially different modalities of presentation.

\section{Conclusion}
In a simulated driving study with 232 participants, we tested how AV explanation errors, driving context characteristics (perceived harm and driving difficulty), and personal traits (prior trust and expertise) impact four driving perception and behavior-related outcomes: comfort relying on an AV, preference to rely on the AV instead of taking control themselves, satisfaction with the explanation, and confidence in the AV’s driving ability. 

Our results indicate that explanation errors, contextual characteristics, and personal traits have a large impact on how a person may think, feel, and behave towards AVs. Explanation errors negatively affected all outcomes. Surprisingly, this included reduced ratings of the AV's driving ability, despite driving performance remaining constant. The negative impact of errors increased with error magnitude and the potential harm of the error, providing evidence that \emph{implications} of an error matter in addition to the mere \emph{presence} of an error. Harm and driving difficulty directly impacted outcomes as well as moderated the relationship between errors and outcomes. In general, harm was associated with lower comfort, reliance, and confidence ratings, while driving difficulty was associated with higher reliance ratings. Overall harm was the more important contextual factor of consideration. In terms of a decision function for AV reliance, perceived harm -- influenced by driving difficulty -- may underlie the evaluation between a person's confidence in the AV's performance compared to their own ability. We found that individuals with higher expertise tended to trust AVs more, and these each correlated with more positive outcome ratings in turn. 

These findings contribute to the overarching framework of this dissertation by emphasizing the dynamic interplay between AV system performance, contextual factors, and individual user traits. They build on the insights from Chapter 2, extending the investigation of adaptive communication strategies to the realm of explanation errors, highlighting the necessity of designing systems that are sensitive to both task demands and user variability.

Overall, our results emphasize the need for accurate and contextually adaptive AV explanations to foster trust, reliance, satisfaction, and confidence. Understanding the ramifications of explanation errors can help future AV research and design teams prioritize design and better understand the impacts of their design choices. By situating these findings within the broader dissertation, this study underscores the importance of context-aware and personalized communication in fostering trust across diverse users and scenarios. It provides a foundational step toward developing explainable AI systems that are robust to the complexities of real-world human-AV interaction.

This chapter complements the broader narrative of this dissertation by illustrating how explanation errors—and their implications—serve as critical factors in shaping user trust and reliance. These results pave the way for further exploration in Chapter 4, where individual traits and trust dynamics are investigated in more depth, expanding on the interconnections between personal factors, system design, and AV adoption.

\section{Acknowledgements}
We would like to thank Saumitra Sapre and Rohan Bhide for driving simulator development, without which the videos could not have been produced, as well as Chloe Lee, Janzen Molina, and Emi Lee for simulator testing. Chapter 3, in full, has been accepted for publication using the title \textit{What Did My Car Say? Impact of Autonomous Vehicle Explanation Errors and Driving Context On Comfort, Reliance, Satisfaction, and Driving Confidence}, in the Proceedings of the 2025 CHI Conference on Human Factors in Computing Systems (CHI ’25). Kaufman, Robert; Broukhim, Aaron; Kirsh, David; Weibel, Nadir. The dissertation author was the primary investigator and author of this paper. \cite{kaufman2024didcarsayimpact}

\chapter[\underline{Personal Traits:} Predicting Trust In Autonomous Vehicles: Modeling Young Adult Psychosocial Traits, Risk-Benefit Attitudes, And Driving Factors With Machine Learning]{\underline{Personal Traits:} \\ Predicting Trust In Autonomous \\ Vehicles: Modeling Young Adult \\ Psychosocial Traits, Risk-Benefit \\ Attitudes, And Driving Factors \\With Machine Learning}
\newpage

\section{Interim Summary and Chapter 4 Overview}
In Chapter 3, I presented a study that tested the impact of explanation errors on trust and reliance perceptions of autonomous vehicles. By comparing the impact of explanations at varying levels of accuracy, we observed the disastrous consequences of errors on how people think, feel, and behave towards AVs. 

This study provided clear and obvious examples of how a person's interaction with an AV depends on multiple factors within the human-AV system, including the AV's performance and the driving context. Looking at context specifically, driving difficulty and a scenario's potential for harm had clear influences on participant trust and willingness to rely on the AV. As such, it is likely that different driving contexts will have different information requirements to elicit trust in a rider in the presently explored case involving AV errors in everyday commuting. We expect this to generalize to many driving cases and for many driving goals (including the previously discussed goal of learning from an AV)-- this is a hypothesis that could be tested in future research. 

By augmenting the situational awareness \textit{of the AV} to include an understanding of how driving context elements (like harm and difficulty) will impact the perceptions of their riders, AVs can anticipate and better support a rider's moment-to-moment needs. Driving situations could be categorized in real-time, allowing the AV to adjust its communication to match the demands of a particular situation. Communication design principles, such as those tested in Chapter 2, could be defined to optimize information delivery for each driving context. As previously discussed, the effectiveness of communication would be expected to vary based on its content and presentation, particularly in how they capture attention, meets information needs, and minimizes cognitive load. 

Of course, we would also expect variance at the level of the individual, meaning that truly optimized AV communications would adjust based on the traits and preferences of the rider as well. I take a first step towards personalization based on human traits in Chapter 4.

Chapter 3 provided initial insight into the way people incorporate external information to form an evaluation of an AV. The mere presence of explanation errors was perceived as a holistic indicator of the AV's overall quality, leading to a crossover effect between \textit{explanation} errors and \textit{driving} performance perceptions. Negative effects were intensified when errors suggested serious potential harm, even if these outcomes were never realized. Taken together, these suggest that people make meaning of AV information in the context of its practical implication on their safety, with perceived risk playing a significant role in shaping their overall judgment. Perceived risk may be one element of import, but what else determines a person's trust judgment?

Among the myriad of factors which may impact a person's trust in AVs, personal traits may be among the most important predictors to consider. Just as certain contextual factors proved relevant to trust and reliance judgments, Chapter 3 demonstrated that certain personal traits -- specifically, expertise and prior trust levels -- predicted how individuals rated the AV throughout the study. For AVs to truly meet the informational needs of different users, understanding which factors impact their trust levels is paramount. Building the situational awareness of an AV so that they are aware of the trust dispositions of different riders can enable truly personalized experiences. We can envision AVs that can connect driving context characteristics to profiles of their rider's trait profiles, enabling communications that are tailored for each person in each situation. To make this vision a reality, we start out with a foundational question: What are the most important factors that predict trust? Chapter 4 takes a first step towards addressing this question, using machine learning to identify factors predicting trust. These have strong implications for AV communication design.

\textbf{Chapter 4 Overview} -- \textit{Low trust remains a significant barrier to Autonomous Vehicle (AV) adoption. To design trustworthy AVs, we need to better understand the individual traits, attitudes, and experiences that impact people's trust judgments. We use machine learning to understand the most important factors that contribute to young adult trust based on a comprehensive set of personal factors gathered via survey (n = 1457). Factors ranged from psychosocial and cognitive attributes to driving style, experiences, and perceived AV risks and benefits. Using the explainable AI technique SHAP, we found that perceptions of AV risks and benefits, attitudes toward feasibility and usability, institutional trust, prior experience, and a person's mental model are the most important predictors. Surprisingly, psychosocial and many technology- and driving-specific factors were not strong predictors. Results highlight the importance of individual differences for designing trustworthy AVs for diverse groups and lead to key implications for future design and research.}
\newpage

\section{Introduction}
It has long been established that trust is a major determinant of adoption of AI-based technologies like AVs, where people's beliefs about a system directly inform their willingness to accept and rely on the system \cite{lee2004trust, muir1994trust}. As we established in Chapters 2 and 3, the relationship between trust and adoption may be especially strong in high-stakes contexts like driving, where users must rely on the technology for personal safety \cite{hoff2015trust}. Highly influential models by \citet{choi2015investigating} and \citet{hoff2015trust} demonstrate how trust is a necessary pre-condition for intention to use autonomous vehicles: if a person does not trust an AV, they will not be willing to ride in it. \citet{liao2022designing} discuss how trust impacts users’ perception of risk and shapes their interactions with AI-based systems, meaning that even a perfectly safe AV may face adoption challenges if users do not trust it. \emph{In essence, this means that no matter how safe or high performing an AV is, without trust, adoption efforts will fail.}

Trust reflects the subjective, psychological confidence that users place in the technology, and does not automatically follow when a system is high-performing or safe \cite{muir1994trust}. In this way, trust serves as a bridge between vehicle performance and user acceptance -- without this bridge, AVs may not reach widespread use. Understanding user trust perceptions in AV research (as opposed to solely improving AV safety or performance) enables designers and researchers to address specific user perceptions and concerns that may hinder adoption, including those that move beyond driving performance \cite{kraus2021s}. By identifying which personal factors most influence trust, this study offers insights on the foundation of AV trust perceptions, informing the design of AVs that align with user needs beyond technical safety alone. \color{black}

There is substantial theoretical and experimental evidence supporting that a person's trust in AVs is influenced by their individual traits and experiences~\cite{hoff2015trust, kaufman2024developing, lee2004trust}. To develop more trustworthy AVs, it is therefore necessary to understand \emph{which} specific factors influence trust judgments, and by \emph{how much}. This can empower AV designers to address the unique needs of different individuals.

Much prior research has examined personal traits that may predict trust and adoption attitudes toward AVs~\cite{mosaferchi2023personality, choi2015investigating, liu2019public, ferronato2020examination, he2022modelling, li2020personality, bockle2021can}. However, many of these studies lack comprehensive coverage of relevant characteristics, limiting their utility. By focusing on only a narrow set of traits, potentially critical factors are left unexamined. For instance, several personal factors discussed by theoretical frameworks and trust research in other AI domains have not yet been explored in the context of AVs, including certain driving behaviors, cognitive traits, and institutional attitudes among others ~\cite{hoff2015trust, kaufman2024developing, lee2004trust}. Further, studying traits in isolation or in small sets prevents a clear understanding of the \emph{relative importance} of different factors in trust development. While various traits -- ranging from demographic characteristics to personality, risk assessment, and experience with AVs -- have been associated with trust, without a comprehensive approach that measures and models diverse factors together, our understanding of their relative significance remains incomplete. This makes it difficult for AV designers and researchers to prioritize factors of importance and develop personalized approaches to the human-AV trust problem. These two research gaps -- (1) unexplored factors and (2) lack of information on relative importance -- underscore the need for studies that account for a broader range of traits while using methods that can capture the interplay between them. In this work, we aim to address both gaps.

A concrete result of securing a more comprehensive understanding of how traits impact trust (in particular, which traits are more important) is designing AVs that can meet the specific needs of diverse individuals, as opposed to taking a one-size-fits-all approach to human-AV interaction. For example, explainable AI (XAI) explanations meant to increase trust via transparency for “black box” AI (and AV) decisions can provide \emph{tailored} communications to meet the needs of specific groups~\cite{gunning2019darpa, liao2021human, kaufman2024developing}. Tailoring communications to individual users has been shown to more effectively address informational needs, helping to calibrate system reliance to appropriate and optimal levels~\cite{schneider2019personalized, ma2023analysing, wang2019designing, kaufman2022cognitive}. Knowing \emph{who} has \emph{what} needs to address can guide the creation of more accessible and equitable AVs~\cite{hassija2024interpreting}, including tailored educational campaigns~\cite{pataranutaporn2023influencing} and inclusive policy guidelines\cite{jobin2019global}. Without meeting the needs of specific individuals or subpopulations, a future with ubiquitous AV adoption cannot be realized. 

In the present study, we use survey methods (n = 1457) to measure a broader range of variables than any previous research on this topic, bridging a diverse spectrum of driving and non-driving related research. We include variables ranging from cognitive and psychological traits to cultural values, driving style, and technology attitudes. We also include a wide range of specific risks and benefits in AVs, as these may be important contributors to trust perceptions~\cite{ayoub2021modeling}. Including such a wide range of variables allows us to build a more nuanced and comprehensive understanding of the factors that influence trust than has been previously possible.

\color{black} We are able to capture the complex interactions between variables and assess the \emph{relative} importance of the factors in our analysis by using machine learning (ML) to predict trust in AVs. The rationale for using ML as the basis of our inquiry stems from its ability to capture complex, nonlinear interactions among our chosen personal traits, attitudes, and trust that traditional regression methods may miss \cite{wang2020survey}. \citet{yarkoni2017choosing} argue that, unlike linear models which assume independent, additive effects, ML can reveal nuanced, multidimensional relationships and should be opted for in research where predictors have the potential to be interrelated, such as in the work presented here. ML models also handle high-dimensional data very effectively, addressing multicollinearity and retaining diverse features better than traditional regression, without overfitting \cite{breiman2001random}. By using ML, we can retain a comprehensive set of trust-related features without compromising model validity, offering a more holistic view of a wide range of factors potentially impacting trust. We apply SHapley Additive eXplanations (SHAP)~\cite{lundberg2020local} to “peek beneath the hood” of our models, adding interpretability to the predictive power of our ML models by showing precisely how important different factors were to predictions. This transparency makes our ML insights actionable, guiding AV design and policy toward strategies that directly address users’ trust-building needs. By utilizing these advanced analysis methods, we can build deeper insights into how a wide range of traits relate to trust judgments. 
\color{black}

Our research also highlights the importance of isolating subpopulations when predicting trust in AVs. Most studies in this area analyze data across general populations~\cite{ferronato2020examination, he2022modelling, li2020personality, bockle2021can}, which can obscure unique traits and concerns relevant to specific groups. For example, the common finding that age predicts a person's trust in AI may be more of a superficial correlation rather than a true driving factor~\cite{araujo2020ai, choung2023trust, abraham2017autonomous}. Though age may be helpful for predicting trust, age often correlates strongly with underlying variables, such as technical expertise, familiarity with AI systems, and openness to innovation~\cite{chen2011review, steinke2012trust}. These underlying variables are far \emph{more} informative of \emph{why} people of different ages may trust AI systems differently, giving insight into \emph{how} to design better systems for them. Subpopulation-focused approaches have proven effective in other areas of computing, such as online interventions for social media, where deeper cognitive traits outperformed broader indicators like age or political affiliation in predicting behavior~\cite{kaufman2022s, kaufman2024warning}. We apply similar techniques to this work by focusing on a specific subpopulation -- in this case, young adults. By holding broader factors constant, we can identify the more nuanced, informative factors that shape trust in this subpopulation.

We sought to prioritize early adopters and the next generation of transportation users, as these groups will likely play a key role in the adoption and shaping of AV design. Young adults, in particular, fit this profile due to their high comfort with new technologies~\cite{choung2023trust} and their potential to influence future transportation trends~\cite{berliner2019uncovering}. By studying young adults, we can better understand their unique perspectives and concerns -- a critical step towards aligning AV systems with the needs of these early adopters. Our findings not only inform AV design and communication strategies for this group but also offer a methodological template and approach for investigating other subpopulations, ultimately supporting the development of AV systems that support appropriate trust and acceptance across diverse user bases.

The results of this study show that young adult trust in autonomous vehicles (AVs) can be effectively predicted from personal factors, using machine learning. Our models achieved high accuracy (85.8\%) in classifying trust levels. By applying the explainable AI technique SHAP, we identified which factors most strongly predicted trust and how the value of each factor contributes to the model's predictions. Notably, perceptions of AV risks and benefits, attitudes toward AV feasibility, institutional trust, and prior experience were the most significant trust predictors, while personality, culture, cognitive preferences, driving style, and driving cognition were less impactful than expected, despite previous research suggesting their importance.

\vspace{1em}
\noindent
The work presented here builds new understanding on how personal factors can be used to predict trust in AVs, a crucial step towards designing future, more trustworthy human-AV interactions. In Summary, we contribute: 
\begin{itemize}[topsep=0pt, nolistsep]
    \item A comprehensive study on personal factors predicting young adult AV trust, including relative factor importance.
    \item A methodology and set of important factors that can be replicated for study in different subpopulations.
    \item A set of design implications based on this work, including high-priority areas for future research and design.
\end{itemize}

This chapter aligns with the overarching themes of this dissertation by emphasizing the necessity of a personalized and contextually adaptive approach to human-AV interaction. This study reinforces the argument that trust is a multifaceted construct deeply influenced by individual traits and contextual factors. The findings here extend the discussion of Chapters 2 and 3 by showing how trust can be systematically analyzed and predicted through machine learning, providing actionable insights for tailoring AV designs to specific user groups. This approach complements the broader systems perspective advocated throughout this dissertation, demonstrating that understanding the dynamic interplay of personal, contextual, and technological factors is essential for creating AV systems that can foster trust, meet diverse user needs, and ultimately achieve widespread adoption.

\section{Related Work}
\subsection{Personal Factors Impacting Trust in AVs}
\label{prior_personal}
\textbf{Individual Differences ---} Hoff and Bashir's~\cite{hoff2015trust} influential analysis of modern human-AI trust research posits that dispositional factors including culture and age, as well as situational and learned factors such as expertise and prior experience, impact how much a person will trust and rely on automated systems like AVs. According to Hoff and Bashir, these factors influence the system in conjunction with the context of use (e.g. driving from location A to location B) and the system's demonstrated performance in the moment. Similarly, \citet{kaufman2024developing} propose a framework for human-AV interaction based on situational awareness and joint action, emphasizing the role personal traits and contextual characteristics play in guiding interactions and achieving goals. Lee and See's~\cite{lee2004trust} review of trust foundations posits that trust success is shaped by a trustor's predispositions and environment. By all three accounts, personal traits clearly influence trust in the system. 

\textbf{Demographics ---} Prior work examining demographic factors like age have shown that younger individuals tend to adopt autonomous systems earlier than older individuals~\cite{hulse2018perceptions, abraham2017autonomous}, however, differences in age may be caused by underlying variables, including familiarity with AI, concern over risks, and self-efficacy~\cite{chen2011review, steinke2012trust, mosaferchi2023personality}. Gender differences commonly show that men are more likely to adopt AVs than women~\cite{hulse2018perceptions, mosaferchi2023personality}, however, these effects may also be moderated by underlying variables like risk preferences~\cite{mosaferchi2023personality} or vehicle anxiety~\cite{hohenberger2016and}. Education may also play a role: people with more education may be more willing to adopt autonomous vehicles~\cite{hudson2019people}. Recent work has even shown that people who lean politically liberal in the United States may be more willing to adopt AVs -- and see greater benefit to adoption -- than those who lean politically conservative, though this finding too may be motivated by underlying driving variables like cultural values and belief systems~\cite{mack2021politics, mosaferchi2023personality}. Socioeconomic status (SES) is understudied in the realm of AV trust prediction; SES may be important given its negative association with interpersonal trust in other domains~\cite{stamos2019investigating}. In the present study, we measure all of these demographic variables: age, gender, education, political orientation, and socioeconomic status to see how important they are for our trust models. Our assessment of trust in young individuals primarily associated with an undergraduate institution means we can hold age and education relatively constant, allowing us to examine deeper underlying factors which may drive trust in AVs.

\textbf{Personality and Culture ---} Personality and culture may impact a person's trust in automated vehicles, however, prior findings are conflicting. Using the Big Five Inventory (BFI)~\cite{rammstedt2007measuring}, \citet{ferronato2020examination} found that extroversion negatively correlated with trust, while openness, conscientiousness, and agreeableness showed positive associations. In contrast, \citet{bockle2021can} reported positive correlations for both extroversion and agreeableness with trust in automation, but none for openness. \citet{chien2016relation} found that agreeable and conscientious individuals were more trusting. They also measured Hofstede's Five Dimensions of Cultural Values via the Cultural Values Scale (CVSCALE)~\cite{yoo2011measuring}, which did not seem to have an impact on trust. \citet{chien2018effect}, however, found that Hofstede's Values such as power distance, uncertainty avoidance, and collectivism did impact automation trust. These often conflicting findings suggest that we should reevaluate the assumption that personality and culture are key drivers of trust, highlighting the need for clarification on their importance in relation to other factors. Cultural identity and race may also impact AI trust, one study attributing trust differences between black and white participants to prior experiences with institutional discrimination~\cite{lee2021included}. This highlights the important impact prior experiences and institutional trust may have on AI trust.

\textbf{Knowledge and Prior Experience with AVs ---} There is substantial evidence suggesting that a person's domain knowledge and past experiences with a system will impact their level of trust~\cite{hoff2015trust}. \citet{mosaferchi2023personality} find that individuals with higher AV knowledge tend to have higher trust in autonomous vehicles. Higher trust is not always a good thing; \emph{calibrating} trust to cases where trust is warranted is critical for sustained adoption. Preventing over-trust is an often neglected secondary benefit of knowledge, since it can inform when to rely on an AV and when to not~\cite{khastgir2018calibrating}. Prior experience interacting with AVs and AV technology may also be an important contributor to trust judgments: individuals with positive past experiences with AV technology -- even if they have never ridden in a fully autonomous AV itself -- may be more willing to trust and adopt AVs in the future~\cite{hoff2015trust}. By contrast, exposure to performance errors can deteriorate trust~\cite{luo2020trust, kaufman2024didcarsayimpact}. In the present study, we examine how AV expertise and prior experience with AV and other autonomous vehicle technologies may impact trust judgments. We also examine prior experiences with driving collisions in general, as these may impact a person's judgments of driving safety and risk~\cite{butler1999post}. 

\textbf{Technology and Institutional Attitudes ---} Technology attitudes may also impact trust: those with a stronger technology affinity tend to have higher trust~\cite{mosaferchi2023personality, bennett2019attitudes, huangexploring}. We assess this using questions from the commonly used Affinity for Technology Interaction (ATI) scale~\cite{franke2019personal}. As previously noted, institutional trust may impact trust in AI-based systems~\cite{lee2021included}. We examine a person's trust in AV-specific institutions, including tech companies, automakers, and government regulators, as these may impact a person's trust in AVs~\cite{kaufman2022s}.

\textbf{Driving Behavior, Attitudes, Cognitions, and Abilities ---} The way a person drives on their own and how they think about themselves as a driver may impact how much they trust an AV. A person's own driving style -- such as whether they are an aggressive, passive, or distracted driver -- has demonstrated importance on rider satisfaction and willingness to rely on an AV's driving~\cite{zhang2024effects}. Specifically, defensive drivers showed better calibrated AV reliance than aggressive drivers. Similar factors have not been thoroughly examined for AV trust. The widely-used Multidimensional Driving Style Inventory (MDSI)~\cite{taubman2004multidimensional} is used in the present study to assess a person's driving style. \citet{kraus2020scared} study self-esteem, driving self-efficacy, and driving anxiety, showing a negative relationship between anxiety and trust in AVs. This effect was moderated by a person's confidence in their own ability. Similarly, \citet{hohenberger2016and} show driving anxiety as a major driver of trust. Anxiety specific to driving and driving concerns has been understudied in AV trust research. We leverage the Driving Cognitions Questionnaire (DCQ)~\cite{ehlers2007driving} to assess several dimensions of social and accident-related anxieties. The effects of self-esteem and self-efficacy have been contested in past AV trust research: some studies claim they are important~\cite{kraus2020scared, hegner2019automatic} while other studies claim they are not~\cite{lohaus2024automated}. We seek to clarify conflicting prior work by assessing driving anxiety, specific concerns and cognitions, self-esteem, self-efficacy, and style.

\textbf{Cognitive and Psychological Traits ---} Common AV adoption concerns involve perceptions of increased risk and worry over giving up driving control~\cite{howard2014public}. Traits such as a person's willingness to take risks~\cite{ajenaghughrure2020risk} and their desire for control~\cite{hegner2019automatic} have been shown to positively correlate with trust in AVs~\cite{choi2015investigating, liu2019public}. These are examined in the present study. Given that most interactions with fully autonomous vehicles require giving up agency to the vehicle, it is relevant to understand a person's tolerance towards decision ambiguity and rationalization. Preference for predictability and ambiguity can be measured via the Need for Closure~\cite{roets2011item} scale; these factors have been highlighted as potentially important for the design of autonomous vehicles~\cite{hamburger2022personality} and have clear implications for explainable AI~\cite{wang2019designing}. At present, they are understudied in AV trust research and are included in our analysis. Similarly, a person's preference for rational or intuitive decision-making may also impact their trust in AVs, as AVs operate using complex, logic-based algorithms for perception, recognition, and classification~\cite{levinson2011towards}. We study these cognitive and psychological traits and preferences to further understand their impact and importance for AV trust.

\textbf{Risks and Benefits of AVs ---} Perhaps the most obvious determinant of AV trust and adoption decisions are a driver's personal estimates of the risks and benefits of AV's~\cite{choi2015investigating, hoff2015trust}. In their model of trust in autonomous vehicles, \citet{ayoub2021modeling} found that perceptions of risks and benefits were the most important predictors of trust. Unfortunately, they failed to measure specifically which risks and benefits are important. As a whole, there are a plethora of potential benefits and risks that may impact a person's willingness to adopt an AV~\cite{fagnant2015preparing}. \citet{acheampong2019capturing} specify perceived benefits related to safety, self-image, and the ability to do other activities while the AV is driving, among others. The Autonomous Vehicle Acceptance Model (AVAM)~\cite{hewitt2019assessing} includes several risks and benefits in their questionnaire, including measures of performance expectancy, ease of use, realism of deployment, safety, and social influence. Likewise, the Checklist for Trust between People and Automation (CTPA)~\cite{jian2000foundations} includes automation concerns surrounding the potential harm, dependability, and reliability of the system, suggesting higher concerns will deteriorate trust. Additional concerns over cost, usefulness, and ease of use have also been highlighted~\cite{zhang2023human}. Several risks and benefits were included in the present study to delineate specifically which are the best predictors of trust.

\vspace{1em}
\noindent
\textbf{Summary ---} Despite the range of studies on trust, knowledge gaps remain on how certain untested variables impact \emph{AV} trust specifically, and how different variables impact AV trust \emph{relative} to each other. Most prior studies on AV trust utilize a small selection of variables and limited analysis techniques, impeding their ability to bring the community a comprehensive understanding of relative variable importance. Further, very little prior work is done on isolated populations, making it difficult to apply insights to specific populations of interest, such as young adults. The present study seeks to address these knowledge gaps.

\subsection{Predicting Trust Using Machine Learning}
The application of machine learning (ML) techniques to predict trust in autonomous vehicles is a relatively new but promising research area. Using large datasets, advanced ML-based predictive models may provide more accurate predictions than traditional analysis and modeling approaches like ANOVA or linear regression when used for challenges like predicting trust~\cite{wang2020survey}. One of the main obstacles with using advanced ML techniques is a lack of explainability -- it is difficult to know how features of a dataset individually contribute to a model's conclusion. Without this information, we cannot know how important different variables are to predicting an outcome. Recent techniques developed for introducing transparency for AI-based systems -- such as SHapley Additive exPlanations (SHAP)~\cite{lundberg2020local} -- can be used to provide insight into a model's behavior, providing justification for decisions via coefficients for relative feature importance. In the case of predicting trust from personal traits, the use of advanced ML models combined with model explainability via SHAP allows us to understand \emph{if} specific personal factor variables are important for trust judgments as well as the relative importance of \emph{each} variable to the model's prediction. 

Traditional analyses like linear regression and ANOVA are by far the most common methods to understanding the personal factors which may predict trust. Some recent works, however, have used advanced machine learning techniques like Random Forest, XGBoost, Naive Bayes, decision trees, or SVM for such goals~\cite{wang2020survey}. For example, \citet{zolfaghar2012syntactical} used decision trees to successfully predict a person's trust in social web applications based on social trust factors like knowledge and personality. \citet{liao2022driver} use personality traits and driver style to categorize drivers into profiles. \citet{kraus2021s} build a hierarchical model to show the relationships between elemental traits (e.g. personality and self-esteem), situational traits (affinity for technology and interpersonal trust), and surface traits (e.g. trust and technology acceptance). \citet{liu2014generic} use decision trees to classify trust in large-scale systems like recommendation agents. Other studies have proposed frameworks for classifying trust using Support Vector Machines (SVMs)~\cite{lopez2015towards}.

For AVs specifically, a few recent studies have used ML techniques to predict trust from personal traits. \citet{ma2024understanding} predict reliance decisions in a simulated driving task using a random forest model. Unsurprisingly, they find that a person's prior situational trust and experience with AVs predicted their current reliance decision. \citet{bennett2019attitudes} model autonomous vehicle adoption attitudes for people with disabilities using demographics, anxiety, locus of control, and other attitudes towards AVs and technology as a whole. \citet{huangexploring} use Additive Bayesian Networks (ABNs) to explore factors related to trust in an Advanced Driver Assistance System (ADAS), finding prior system and technological experience were the two highest predictors. 

Outside of prediction from personal traits, some recent work has sought to assess trust from physiological signals using machine learning. \citet{yi2023measurement} use a multi-modal feature fusion network to predict trust from physiological signals like galvanic skin response and heart rate variability. Other studies have used XGBoost~\cite{ayoub2023real} and discriminant classification~\cite{akash2018classification} to accomplish similar ends. 

Using similar methods as the present study, \citet{ayoub2021modeling} used demographics, knowledge, feelings, and behavioral intentions to predict how a person responded to the question “how much would you trust an autonomous vehicle?” They included general ratings of the overall “risk” and “benefit” of AVs, however, they did not specify which risks or benefits, nor did they gather precise concerns. Nonetheless, they found that the risk and benefit scores were the highest contributors to their XGBoost model. We seek to build upon this effort by incorporating a much larger range of features, a more comprehensive measure of “trust” based on modern trust theory, and a more specific analysis focusing on a particular user group of interest (young adults). In addition, we include a wide range of \emph{specific} risks and benefits based on prior work, allowing us to understand \emph{which} risks and benefits may be most important.

In sum, the integration of machine learning and explainable AI techniques like SHAP for predicting trust in AVs provides a significant opportunity to advance our understanding of trust development in complex sociotechnical systems like autonomous driving~\cite{fraedrich2015transition}. These models not only offer high predictive accuracy but also provide valuable insights for datasets with large numbers of variables not available using other methods. As AV technology continues to evolve, the ability to predict and enhance user trust will be crucial for successful adoption and widespread acceptance of AVs. In the present work, we seek to build on prior research to advance the study of individual differences in trust, addressing knowledge gaps that will allow researchers and designers to build AVs that can meet the needs of diverse groups.

\section{Method}
\begin{figure}[H]
    \centering    \includegraphics[width=\linewidth]{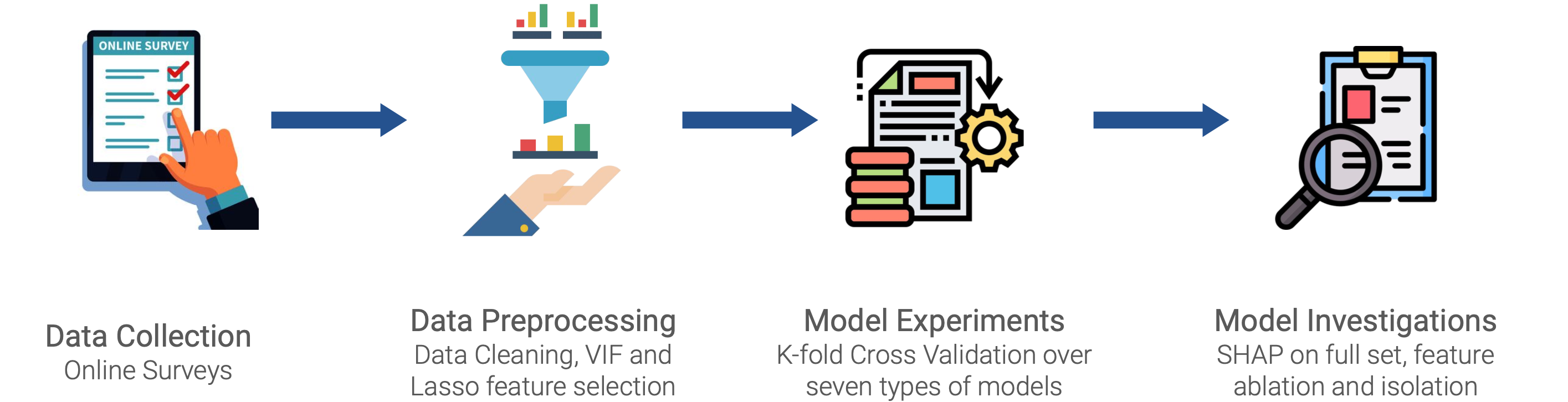}
    \caption[\textbf{Overall process architecture.}]{\textbf{Overall process architecture.} Our pipeline moves from data collection and processing to model development and explanations using SHAP.}
    %\Description{Overall process architecture diagram for our study, moving from data collection and preprocessing to model development (experiments) and investigations (explanations using SHAP).}
    \label{fig:pipeline}
\end{figure}

In the present study, we use survey methods to assess a wide range of personal factors from a pool of young adults. Then, we build models predicting a person's trust in autonomous vehicles from those factors, using a variety of machine learning techniques. Following, we use SHAP~\cite{lundberg2020local} to make the top performing models explainable, giving us measures of feature importance. We add additional nuance to our study with ablation experiments knocking out specific feature groups from the models in order to derive greater insight into how factors contribute to AV trust. Figure~\ref{fig:pipeline} describes the overall modeling process. To look at trust differences between the “high” and “low” groups at the level of each factor, we also perform a more traditional descriptive analysis using Kruskal-Wallis tests as a means to provide comparison to our ML results.

\vspace{-1em}

\subsection{Participants and Data Collection}
A total of 1457 participants completed the study and passed all quality control requirements. The survey was designed and deployed using Qualtrics~\cite{Qualtrics2024}, taking a median of 17.9 minutes to complete. Recruitment occurred via email lists and via SONA,\footnote{https://www.sona-systems.com} an undergraduate study pool where students are granted study credit for participation. As such, the vast majority of participants were enrolled at or affiliated with a large undergraduate university in the United States. The mean age of the study sample was 20.7 (SD = 2.4), with ages capped between 18 and 35 years. The sample was 73.9\% female. Table~\ref{demographic_table} shows distributions of age and education for our sample. Only participants who passed three attention checks were included. As an additional means of quality control, we screened out participants with low variability in survey responses (e.g. used the same pattern of answers) and those who had unrealistically low completion time. As a result of this quality control process, we are confident that the data used in our analysis is of high quality.

\begin{table}[htbp] % !htbp change to H if i want to force positioning
  \caption{\textbf{Distributions of participant age and highest completed education.}}
  %\Description{Table showing the distributions of participant age and highest completed education.}
  \label{demographic_table}
  \renewcommand{\arraystretch}{1}
  \vspace{-.5em}
  \begin{tabular}{p{0.15\linewidth}p{0.15\linewidth}p{0.15\linewidth}p{0.2\linewidth}p{0.22\linewidth}}
    \toprule
    \textbf{Age} & \textbf{18-19} & \textbf{20-21} & \textbf{22-23} & \textbf{24+}\\
    & 29.3\% & 48.2\% & 15.2\% & 7.3\% \\
    \midrule
    \textbf{Education} & \textbf{{\footnotesize $<$2 years \newline of college}} & \textbf{{\footnotesize 2 years \newline of college}} & {\footnotesize \textbf{4-year degree \newline (e.g. bachelor's)}} & {\footnotesize \textbf{Grad. Degree \newline (e.g. master's, PhD)}}\\
    & 27.3\% & 59.3\% & 12.6\% & 0.8\% \\
    \bottomrule
\end{tabular}
\end{table}

\subsection{Survey Design \color{black} and Administration \color{black}}
The survey was designed to be as comprehensive as possible while considering limits on survey duration and participant attention. \color{black} Our questions were heavily based on validated questionnaires and survey measures to enhance the reliability and interpretability of responses; we relied on previously tested measures to ensure that our questions were well-aligned with established constructs and our own expectations. Given the use of validated questions, we deemed that a formal cognitive pretest was not necessary. However, we conducted a pretest with approximately 10 participants and qualitative feedback on the survey was solicited to ensure that questions were straightforward, clear, free of errors, and not too burdensome. Results from these pilot participants were not included in the final analysis. The pretest allowed us to refine any questions that appeared ambiguous or confusing to participants. %The survey itself can be found in Supplementary Material.

Participants were provided with a brief description of autonomous vehicles at the beginning of the survey, specifying AVs as \textit{“vehicles that use technology, such as artificial intelligence and special cameras, to drive without the need for human intervention. They sense their environment, adjusting speed and direction to bring a passenger from one location to another safely.”} This was accompanied by a simple and clear cartoon visualization of an autonomous sedan in an everyday driving scenario (commuting amongst other vehicles). This introduction aimed to establish the context of interest as AVs focused on typical, day-to-day applications, as opposed to specialized uses like autonomous trucking, racing, or industrial AVs. We chose this commuting-oriented context to ensure that participants were evaluating AVs intended for personal, everyday driving tasks. By avoiding references to specific brands or technical specifications, we allowed participants to respond based on their own knowledge and attitudes toward AVs used in common, relatable scenarios. This approach was designed to capture perceptions of AVs in the specific context of everyday driving, reflecting a more realistic and relevant use case for general public adoption.
\color{black} 

\subsubsection{Survey Scales and Questionnaires}
Table \ref{driving_factors} shows factors related to driving-specific attitudes and behaviors, Table \ref{demo_psych_factors} shows demographic, psychosocial, and general attitude measures, and Table \ref{risk_benefit_factors} details risk and benefit measures as well as the trust measures used to form our composite trust outcome. Given the large number of variables, variable descriptions that are not relevant to the reported results are omitted from these tables; all variable names were designed to be descriptive and intuitive.

\subsubsection{Trust as an Outcome}
We predict a person's AV trust attitudes by creating a composite score of 10 individual trust questions designed to encompass several dimensions of trust, based on  current models of trust formation~\cite{jian2000foundations, lee2004trust, hoff2015trust}. These can be found in the “trust” section of Table \ref{risk_benefit_factors}; \texttt{Danger to Self} and \texttt{Danger to Others} are reverse-scored. Despite a large corpus of cognitive science and philosophy work demonstrating the multidimensional nature of trust formation~\cite{hoff2015trust, lee2004trust, kaufman2024developing, jian2000foundations}, most present studies use single-item outcomes (e.g. “do you trust AVs?”)~\cite{ayoub2021modeling}. People are notoriously bad at articulating complex implicit judgments like trust in systems~\cite{miller2019explanation}, and thus multidimensional measures provide a more valid approach to measurement. Scores above or below the middle value of scores were categorized as “high” and “low”, respectively. We note that these groups were of slightly different sizes: 44.5\% of participants were in the “high” condition. This was accounted for during model construction.

\begin{table}[htbp] %change to H if i want to force positioning
  \caption{\textbf{Measures: Driving and Technology Behaviors, Attitudes, and \newline Perceptions.}}
  %\Description{Explanation of measures used in the study: Driving and Technology Behaviors, Attitudes, and Perceptions}
  \label{driving_factors}
  \begingroup
  \renewcommand{\arraystretch}{1.1}
  \small
  \begin{tabular}{p{0.11\linewidth}>{\raggedright}p{0.12\linewidth}p{0.45\linewidth}p{0.21\linewidth}}
    \toprule
    \textbf{Category} & \textbf{Factor} & \textbf{Description} & \textbf{Scale and Source} \\
    \midrule
    \textbf{Driving Behavior} & Driving Style & 43-question Multidimensional Driving Style Inventory (MDSI) assessing driving behaviors and thoughts. & Strg. Disagree- Strg. Agree (1-5).~\cite{taubman2004multidimensional} \\ \hdashline
    & Driving Freq. & How often do you drive a car? & Daily - Never (1-7) * \\ \hdashline
    & Driving Collisions & How many driving collisions have you been a part of in the past 3 years?  & * \\
    \midrule
    \textbf{Driving Priors} & AV Knowledge and Expertise & 10 questions related to knowledge of autonomous vehicles and AI-based systems. & Strg. Disagree- Strg. Agree (1-5). Adapted from~\cite{kaufman2024effects} \\ \hdashline
    & AV Experience & Have you ever ridden in an autonomous vehicle? & Fully auto., Partially auto, No. * \\ \hdashline
    & Assistive Tech. & Have you ever used any of the following autonomous driving functionalities? List of 8 (e.g. Adaptive Cruise Control,  Lane Assist). & Never - Often (1-3), n/a.~\cite{abraham2017case} \\
    \midrule
    \textbf{Driving Attitudes} & Tech. Self-Efficacy & 5 questions similar to “I am good with technology.” & Strg. Disagree- Strg. Agree (1-5). Adapted from~\cite{teo2012understanding} \\ \hdashline
    & Affinity for Tech. & 4 questions from the Affinity for Technology Interaction (ATI) scale measuring how much a person enjoys new technologies, e.g. “I am quick to incorporate new technologies into my life.” & Strg. Disagree- Strg. Agree (1-5).~\cite{franke2019personal} \\ \hdashline
    & AV Feasibility & The AV and the infrastructure necessary to use the AV are practically feasible. & Strg. Disagree- Strg. Agree (1-5).~\cite{hewitt2019assessing} \\
  \bottomrule
  \multicolumn{4}{p{.95\linewidth}}{\raggedright \textit{* For simple questions, the research team came up with question wording and scaling.}}
\end{tabular}
\endgroup
\end{table}

\begin{table}[htbp] %change to H if i want to force positioning
  \caption{\textbf{Measures: Demographic and Psychosocial.}}
  %\Description{Explanation of measures used in the study: Demographic and Psychosocial}
  \label{demo_psych_factors}
  \begingroup
  \renewcommand{\arraystretch}{1.1}
  \small
  \begin{tabular}{p{0.11\linewidth}>{\raggedright}p{0.12\linewidth}p{0.45\linewidth}p{0.21\linewidth}}
    \toprule
    \textbf{Category} & \textbf{Factor} & \textbf{Description} & \textbf{Scale and Source} \\
    \midrule
   \textbf{General} & Age & What is your age in years? & * \\ \hdashline
    & Education & Highest level of education completed. &  * \\ \hdashline
    & SES & What is your yearly household income (pre-tax)? &  * \\ \hdashline
    & Gender & What is your gender? & Adapted from~\cite{lindqvist2021gender} \\ \hdashline
    & Politics & Politically speaking, how would you best describe yourself? (US politics, N/A optional). & Strong Dem. - Strong Rep. \newline (1-5).~\cite{kaufman2022s} \\ \hdashline
    & Trust in Institutions & Trust in technology companies, traditional automakers, and the government. & Strg. Disagree- Strg. Agree (1-5). Adapted from~\cite{funk2019trust} \\
    \midrule
    \textbf{Psycho-social} & Personality & Big Five Inventory (BFI): extroversion, agreeableness, conscientiousness, neuroticism, and openness. & Strg. Disagree- Strg. Agree (1-5).~\cite{rammstedt2007measuring}  \\ \hdashline
    & Culture & Hofstede's Dimensions of Cultural Values via the Cultural Values Scale (CVSCALE). 17 questions, 3 sub-scales: power distance, uncertainty avoidance, and collectivism. & Strg. Disagree- Strg. Agree (1-5).~\cite{yoo2011measuring} \\ \hdashline
    & Risk Willingness & How willing are you to take risks, in general? & Not at all - \newline Very (1-10).~\cite{dohmen2005individual} \\ \hdashline
    & Need For Control & 5 questions from the Desirability Of Control Scale, e.g. “I prefer to be in control of situations involving my safety”. & Strg. Disagree- Strg. Agree (1-5).~\cite{burger1979desirability} \\ \hdashline
    & Decision Style & 3 questions on intuitive (gut feeling) vs rational (logic-based) decision-making style preferences. & Strg. Disagree- Strg. Agree (1-5).~\cite{scott1995decision} \\ \hdashline
    & Self-esteem & 5-question scale with questions similar to “On the whole, I am satisfied with myself.” & Strg. Disagree- Strg. Agree (1-4).~\cite{monteiro2022efficient} \\ \hdashline
    & Need For Closure & NFC is a 15-question scale assessing a person's preferences for ambiguity and unpredictability. Results are split into these two sub-categories. & Strg. Disagree- Strg. Agree (1-5).~\cite{roets2011item} \\
    \midrule
    \textbf{Driving-specific} & Driving Cognitions & We assess dimensions of social and accident-related thoughts that occur during driving via 13 questions of the Driving Cognitions Questionnaire (DCQ). & Strg. Disagree- Strg. Agree (1-5).~\cite{ehlers2007driving} \\ \hdashline
    & Driving Risk & How risky do you consider driving? & Not at all - \newline Extremely (1-5) * \\ \hdashline
    & Passenger Anxiety & How anxious do you feel as a passenger when someone else is driving? & Not at all - \newline Extremely (1-5) * \\
  \bottomrule
  \multicolumn{4}{p{.95\linewidth}}{\raggedright \textit{* For simple questions, the research team came up with question wording and scaling.}}
\end{tabular}
\endgroup
\end{table}

\begin{table}[htbp] % !htbp change to H if i want to force positioning
  \caption{\textbf{Measures: Risks, Benefits, and Trust.}}
  %\Description{Explanation of measures used in the study: Risks, Benefits, and Trust}
  \label{risk_benefit_factors}
  \begingroup
  \renewcommand{\arraystretch}{1.05}
  \small
  \begin{tabular}{p{0.12\linewidth}>{\raggedright}p{0.2\linewidth}p{0.6\linewidth}}
    \toprule
    \textbf{Category} & \textbf{Factor} & \textbf{Description} \\
    \midrule
    \textbf{Perceptions} & AV Feasibility & The AV and infrastructure necessary to use the AV are feasible.  \\ \hdashline
    & Overall Risk-Benefit & How do you perceive the risk-benefit trade-off of using an AV? \textit{[Scale: All Risk, No Benefit- No Risk, All Benefit (1-7)]}  \\ \hdashline
    & Mental Model & AVs make decisions like humans do. \\
    \midrule
    \textbf{Benefits} & Reduce Accidents & AVs can reduce traffic crashes. \\ \hdashline
    & Reduce Traffic & AVs can reduce traffic congestion. \\ \hdashline
    & Reduce Emissions & AVs can reduce vehicle emissions and pollution. \\ \hdashline
    & Improve Fuel Econ. & AVs can improve fuel economy. \\ \hdashline
    & Reduce Trans. Cost & AVs can reduce transport costs. \\ \hdashline
    & Improve Mobility & AVs can increase mobility for those who are unable to drive. \\ \hdashline
    & Free Time & AVs will give me more free time (e.g. to text, play games). \\ \hdashline
    & Increase Fun & An AV would make driving more interesting or fun. \\ \hdashline
    & Improve Efficiency & An AV would enable me to reach my destination quickly. \\
    \midrule
    \textbf{Risks} & System Failure & I am concerned about equipment and system failures in AVs. \\ \hdashline
    & Legal Liability & I am concerned about the legal liability of drivers or owners of AVs. \\ \hdashline
    & Hacking & I am concerned that the computer systems of AVs may be hacked. \\ \hdashline
    & Data Privacy & I am concerned about sharing driving data with a company or government agency. \\ \hdashline
    & Performance (varied) & I am concerned that AVs cannot handle varied weather and terrain. \\ \hdashline
    & Cost & I am concerned that AVs will be too expensive. \\ \hdashline
    & Losing Control & I am concerned about giving up driving control to an AV. \\ \hdashline
    & Lacking Understanding & I am afraid that I would not understand an AV's decisions. \\ \hdashline
    & Ease of Use & I would find an AV easy to use. \\
    \midrule
    \textbf{Trust} & Dependability & AVs are more dependable than human drivers. \\ \hdashline
   \textbf{(Outcome)} & Adaptability & AVs will adapt well to new environments. \\ \hdashline
    & Goal Alignment & The goals for AVs are aligned with my own. \\ \hdashline
    & Danger to Self & I believe that using an AV would be dangerous to me. \\ \hdashline
    & Danger to Others & I believe that AVs would be dangerous to other drivers/pedestrians. \\ \hdashline
    & Safety & I would feel safe while using an AV. \\ \hdashline
    & General Trust & I would trust an AV. \\ \hdashline
    & Good Idea & Using an AV would be a good idea. \\ \hdashline
    & Positivity & Overall, I feel positive about AVs. \\ \hdashline
    & Recommendation & I will recommend family members / friends to ride in AVs. \\
  \bottomrule
  \multicolumn{3}{p{.95\linewidth}}{\raggedright \textit{All questions across categories were adapted from~\cite{hewitt2019assessing, fagnant2015preparing, acheampong2019capturing, jian2000foundations, zhang2023human}. Unless otherwise specified, scale: Strg. Disagree- Strg. Agree (1-5).}}
\end{tabular}
\endgroup
\end{table}

% \begin{itemize}
%     \item Demographics (age, education SES, Gender, politics)
%     \item Personality (BFI) 
%     \item Culture (CVSCALE) Power distance, uncertainty avoidance, and collectivism
%     \item Driving Frequency and number of collisions.
%     \item Perceived Driving Risk, Risk willingness, and Anxiety as passenger
%     \item Self-assessment of knowledge and expertise (AV priors)
%     \item Past experience with autonomous or fully autonomous vehicles, or assistive technologies like lane control and parking assist.
%     \item Self-assessment of knowledge and expertise (AV priors 10 questions)
%     \item Need and desirability for control
%     \item rational or intuitive decision-making 
%     \item self-esteem and self-efficacy with technology (5 questions)
%     \item Need for closure (ambiguity and predictability)
%     \item Trust in institutions (3)
%     \item Affinity for technology (4)
%     \item Behavioral attitudes towards AVs including: 

%av feasibility, 

%overall risk-benefit assessment
%     \item Benefits of AVs including: 
%     \item Risks including:
%     \item MDSI Driving Style (43 questions)
%     \item Driving Cognitions (13 questions)
% \end{itemize}

\subsection{Model Development}
\label{sec:model_development}
After deriving an exploratory understanding of the dataset and its features using descriptive statistics and Kruskal-Wallis tests (discussed in Section \ref{sec:descriptive}), we investigated the potential for machine learning systems to predict users trust in AVs, \color{black} as these advanced modeling methods provide a powerful and nuanced way to derive deep insights from complex data \cite{wang2020survey, yarkoni2017choosing}. \color{black} We discuss our approach to designing such a model below.

\textbf{Feature Selection ---}
Our survey data resulted in a set of features that could be used as predictors for our models predicting user trust. For survey questions that contributed to validated composite scores (like Big 5 personality or CVSCALE culture values), only the composites were kept in the dataset. For measures that provide meaning on their own (for example, each risk/benefit or specific driving behaviors), these were included individually as their own feature. 

Since this data comes from surveys, we took multiple steps to achieve a concise feature set and mitigate the possibility that some features could be correlated with each other. Following recent work around modeling personality traits~\cite{saha2021person}, we first filtered features using their Variance Inflation Factor (VIF) with respect to the raw trust scores, using a threshold of 10. This helped us detect and reduce multicollinearity. To further reduce multicollinearity, next we ran feature selection through Lasso regression~\cite{tibshirani1996regression} on the raw trust scores. \color{black} Lasso was opted for because it not only regularizes the model but also performs feature selection by driving less important coefficients to zero, making it more effective than ridge regression for identifying relevant predictors in high-dimensional datasets \cite{tibshirani1996regression}. \color{black} This resulted in a final set of 130 features, which was used for all our machine learning experiments and is henceforth referred to as the \textit{full feature set}.

\textbf{Machine Learning Pipeline ---}
Our machine learning pipeline consisted of two components: (1) feature transformation and (2) classification modeling. For effective modeling, we reduce dimensions through Linear Discriminant Analysis (LDA)~\cite{tharwat2017linear} on the feature set to attain 10 transformed feature vectors. Since the number of “high” and “low” people in the dataset was not perfectly balanced, we over-sample the minority in the training data to match the number of majority samples. We used the Simple Minority Oversampling TEchnique (SMOTE)~\cite{chawla2002smote} to achieve this synthetic balanced dataset, a common method used in similar ML analyses.

Next, we applied several classical machine learning models on the derived feature set in order to find the best performing method. Given our task of binary classification (“low” vs. “high” trust), we used models similar to \color{black} \citet{bedmutha2024conversense} and \citet {ayoub2021modeling} \color{black} to achieve a balance of interpretability and performance. Specifically, we applied Logistic Regression (LR), Support Vector Machines (SVM) with linear and radial kernels, Decision Tree Classifiers (DTC), Random Forest (RF), Naive Bayes, and Gradient-Boosted Decision Trees. We used XGBoost implementation~\cite{chen2016xgboost} for the gradient-boosted trees.

\textbf{Training Experiments ---}
We conducted cross-validation experiments to quantify the performance of different models. The pipeline was evaluated across a five-fold split on the overall dataset. For each instance, the data was first normalized using Standard Scaling based on the training data and then fed into the machine learning pipeline. Each type of model was evaluated using balanced accuracy. We also tracked the precision, recall and macro-F1 scores for each iteration given their robustness to imbalance~\cite{plotz2021applying}. For each model, we experimented with the various parameters possible for that model type. Final model selection was conducted on the mean balanced accuracy performance across all cross-validation folds.

\subsection{Model Investigations}
\label{sec:model_investigations}
After training different possible models, our goal was to understand how features contribute to their overall performance. This explainability brings insight into the most important personal factors that predict a person's trust judgment. To achieve this, we used SHapley Additive exPlanations (SHAP)~\cite{lundberg2020local} to investigate dimensions indicative of trust. SHAP scores are game-theoretic measures of importance for how each feature contributes to the overall prediction, providing local interpretability of the pipeline. SHAP not only allowed us to understand \emph{which} features are important, but also \emph{how} important via a relative importance coefficient. In our training process, we tracked SHAP scores for each feature for every fold of all models. We created explainer plots for each of these sub-models for every fold to identify key descriptors influencing trust. While some features ranked relatively higher or lower across folds, to get a sense of the overall contribution, we report on the mean absolute contribution of each feature across the folds similar to the method used by \citet{nepal2024moodcapture}.

We present two types of SHAP visualizations. First, \emph{bar graphs} show the mean absolute SHAP scores for the top 20 features contributing to model predictions of high and low trust together (high/low trust predictions are mirrors of each other); these give us a breakdown of overall feature importance for predicting trust. Second, we present SHAP \emph{beeswarm plots} that show how individual feature values contribute to the model's prediction of trust. These are also organized based on feature importance, but provide additional information allowing us to see how high and low scores for a particular feature predict the high or low trust classification (e.g. Fig. \ref{fig:fullset_shap}). For example, these allow us to make claims like higher \texttt{Overall Risk-Benefit} assessment is associated with higher trust.

\subsection{Deeper Feature Experiments (Systematic Ablation \newline Studies)}
\label{sec:feature_investigations}
We conducted a series of experiments to derive a deeper understanding of how key features impact our trust models. Given how the risk and benefit dimensions used in this study showed to be the top predictors of trust in our full model (Section \ref{full_model_SHAP_sec}), we conducted two follow-up ablation studies to test (1) the accuracy of predicting trust from risk and benefit features only, and (2) the impact of removing risk and benefit features from our full model (leaving only other types of features like psychosocial and driving). For each ablation (systematic elimination) study, we followed a pipeline and training process similar to Section~\ref{sec:model_development} and report the metrics averaged across folds. We compare them to the scores found in the larger full feature model; we consider the net decrease in performance to be an indicator of importance.

\section{Results}
We present results from our survey of 1457 young adults. First, we show descriptive statistics and Kruskal-Wallis tests comparing high and low trust individuals, generating initial dataset understanding. Next, we introduce the results of our ML analysis, with model explanations provided via SHAP. Finally, we deepen the ML analysis by performing two ablation (systematic elimination) studies, which compare results with and without risk-benefit contributions. 

\subsection{Descriptive Analysis: Comparison of Factors Using Kruskal–Wallis}
\label{sec:descriptive}
Descriptive statistics and initial comparisons using Kruskal–Wallis tests for the high and low trust groups can be found in Table \ref{kw_large} (factors with large effect size only) and Appendix \ref{appendixA.trust} (all factors). Kruskal–Wallis results show if individual predictor variable scores differ between people of high trust and people of low trust. Predictably, given the large sample size, nearly all group differences were significant. To give a more functional comparison, we report effect size (Cohen's D) and provide an interpretation based on accepted standards~\cite{fritz2012effect}.

We find 8 variables to have large effect sizes: \texttt{Overall Risk-Benefit}, \texttt{Ease of Use}, \texttt{Reduce Accidents}, \texttt{Trust in Tech Companies}, \texttt{Increase Fun}, \texttt{AV Feasibility}, \texttt{Improve Efficiency}, and \texttt{Reduce Traffic}. Each of these may be important for predicting how much a person trusts AV, and indeed, each of these factors were in the top 20 most important factors for our full feature ML model. However, we note that in the Kruskal-Wallis case, each factor was tested individually, limiting our ability to understand complex patterns and non-linear relationships from the larger, combined set of predictors. Using the more sophisticated and predictive machine learning approaches described next, we get a more nuanced understanding of how each variable contributes to a person's trust assessment.

\begin{table}[htbp] %change to H if i want to force positioning
\centering
  \caption[\textbf{Descriptive Statistics and Kruskal–Wallis Tests For Low and High Trust Groups.}]{\textbf{Descriptive Statistics and Kruskal–Wallis Tests For Low and High Trust Groups.} This table shows the factors that had the largest individual differences between the high and low trust groups (top 15 by effect size). Each predictor was tested independently.}
  %\Description{Descriptive Statistics and Kruskal–Wallis Tests For Low and High Trust Groups. This table shows the factors that had the largest individual differences between the high and low trust groups (top 15 by effect size). Each predictor was tested independently.}
  \label{kw_large}
  \begingroup
  \renewcommand{\arraystretch}{1}
    \begin{tabular}{p{0.23\linewidth} | p{0.06\linewidth} | p{0.045\linewidth} | p{0.06\linewidth} | p{0.045\linewidth} | p{0.07\linewidth} | p{0.08\linewidth} | p{0.085\linewidth} | p{0.06\linewidth}}
  \toprule
 & \multicolumn{2}{c|}{\textbf{Low Trust}} & \multicolumn{2}{c|}{\textbf{High Trust}} & \multicolumn{4}{c}{\textbf{Comparison}}\\
 \midrule
\textbf{Predictor} & \textbf{{\footnotesize Mean}} & \textbf{{\footnotesize SD}} & \textbf{{\footnotesize Mean}} & \textbf{{\footnotesize SD}} & \textbf{{\footnotesize H-stat}} & \textbf{{\footnotesize P-value}} & \textbf{{\footnotesize Cohen's D}} & \textbf{{\footnotesize Effect}} \\ 
  \midrule
Overall Risk-Benefit & 3.23 & 0.87 & 4.63 & 0.81 & 595.25 & $<$ 0.01 & 1.66 & large \\ 
  Ease of Use & 2.55 & 1.13 & 3.72 & 0.88 & 345.75 & $<$ 0.01 & 1.14 & large \\ 
  Reduce Accidents & 2.79 & 1.07 & 3.89 & 0.86 & 338.97 & $<$ 0.01 & 1.12 & large \\ 
  Trust Tech Comp. & 2.19 & 0.99 & 3.09 & 0.99 & 241.61 & $<$ 0.01 & 0.91 & large \\ 
 Increase Fun & 2.38 & 1.22 & 3.46 & 1.16 & 237.14 & $<$ 0.01 & 0.91 & large \\ 
  AV Feasibility & 2.58 & 1.02 & 3.43 & 0.87 & 235.40 & $<$ 0.01 & 0.88 & large \\ 
  Improve Efficiency & 2.46 & 1.06 & 3.36 & 1.03 & 214.80 & $<$ 0.01 & 0.86 & large \\ 
 Reduce Traffic & 2.82 & 1.17 & 3.71 & 1.04 & 194.96 & $<$ 0.01 & 0.80 & large \\
   \bottomrule
    \end{tabular}
    \endgroup
\end{table}

% Descriptive statistics and variable comparisons between the high and low \textit{predictor} groups can be found in Appendix \ref{appendixB.predictor}. This table allows us to compare trust scores between people high and low in each predictor. For example, this allows us to see if trust ratings differ between people high and low in their assessment of AV feasibility. We report the same statistical comparisons, including effect size.

\subsection{Machine Learning Analysis Using The Full Feature Set}
\label{full_set_results}
\subsubsection{Model Comparisons and Best Performance}
We conducted a series of experiments to identify the best performing overall models using the full feature set, optimizing for different sets of parameters. We report the best achieved scores for each model averaged across all five cross-validation folds in Table~\ref{tab:model_main}. 

We find that all models performed well in general, strengthening support for the hypothesis that personal factors can predict trust dispositions for AVs. Overall, Random Forest produced our most predictive model for trust using the full feature set, resulting in 85.8\% balanced accuracy. We also note that precision and recall are higher than chance and have similar magnitudes. This implies that the model is intrinsically robust and consistent across both high and low trust groups. The area under ROC curve observed in the best performing model (0.934) shows that our model pipelines and features can create a strong class separation and are indifferent to label imbalance.

\begin{table}[htbp] %!htbp
\caption[\textbf{Model performances using the full feature set, reporting the mean (standard deviation) validated across five folds.}]{\textbf{Model performances using the full feature set, reporting the mean (standard deviation) validated across five folds.} All measures are calculated as macro weighted scores between the two classes of trust, ensuring balance between the high and low trust groups. All models demonstrated generally good predictability -- Random Forest models showed the highest performance and were thus selected to be used for the remainder of the analysis.}
%\Description{This table shows model performances using the full feature set, reporting the mean (standard deviation) validated across five folds. All measures are calculated as macro weighted scores between the two classes of trust, ensuring balance between the high and low trust groups. All models demonstrated generally good predictability -- Random Forest models showed the highest performance and were thus selected to be used for the remainder of the analysis.}
\begingroup
\renewcommand{\arraystretch}{1}
\small
\begin{tabular}{p{0.2\linewidth} | p{0.125\linewidth} | p{0.125\linewidth} | p{0.125\linewidth} | p{0.125\linewidth} | p{0.125\linewidth}}
\toprule
\textbf{Model} & \textbf{Balanced\newline Accuracy} & \textbf{Precision} & \textbf{Recall} & \textbf{F1-score} & \textbf{AUROC} \\
\midrule
% Majority Voting (Baseline)  &   &   &   &  & \\
Naive Bayes & 0.797 (0.03) & 0.799 (0.03) & 0.797 (0.03) & 0.797 (0.03) & 0.875 (0.03)\\
Logistic Regression & 0.840 (0.02) & 0.859 (0.02) & 0.857 (0.02) & 0.840 (0.02) & 0.922 (0.02) \\
SVM (Linear) & 0.840 (0.02) & 0.842 (0.02) & 0.840 (0.02) & 0.840 (0.02) & 0.922 (0.03) \\
SVM (Radial) & 0.838 (0.02) & 0.840 (0.02) & 0.838 (0.02) & 0.838 (0.02) & 0.888 (0.03) \\
Decision Tree Class. & 0.800 (0.03) & 0.801 (0.03) & 0.800 (0.03) & 0.800 (0.03) & 0.800 (0.03) \\
\underline{\textit{Random Forest}} & \underline{\textit{0.858 (0.02)}} & \underline{\textit{0.859 (0.02)}} & \underline{\textit{0.858 (0.02)}} & \underline{\textit{0.858 (0.02)}} & \underline{\textit{0.934 (0.01)}} \\
XGBoost & 0.825 (0.03) & 0.825 (0.03) & 0.825 (0.03) & 0.824 (0.03) & 0.906 (0.03) \\
\bottomrule
\end{tabular}
\label{tab:model_main}
\endgroup
\end{table}

\subsubsection{Explaining the Full Feature Model using SHAP}
\label{full_model_SHAP_sec}
\begin{figure}[H]
    \centering
        \centering
        \includegraphics[width=.9\linewidth]{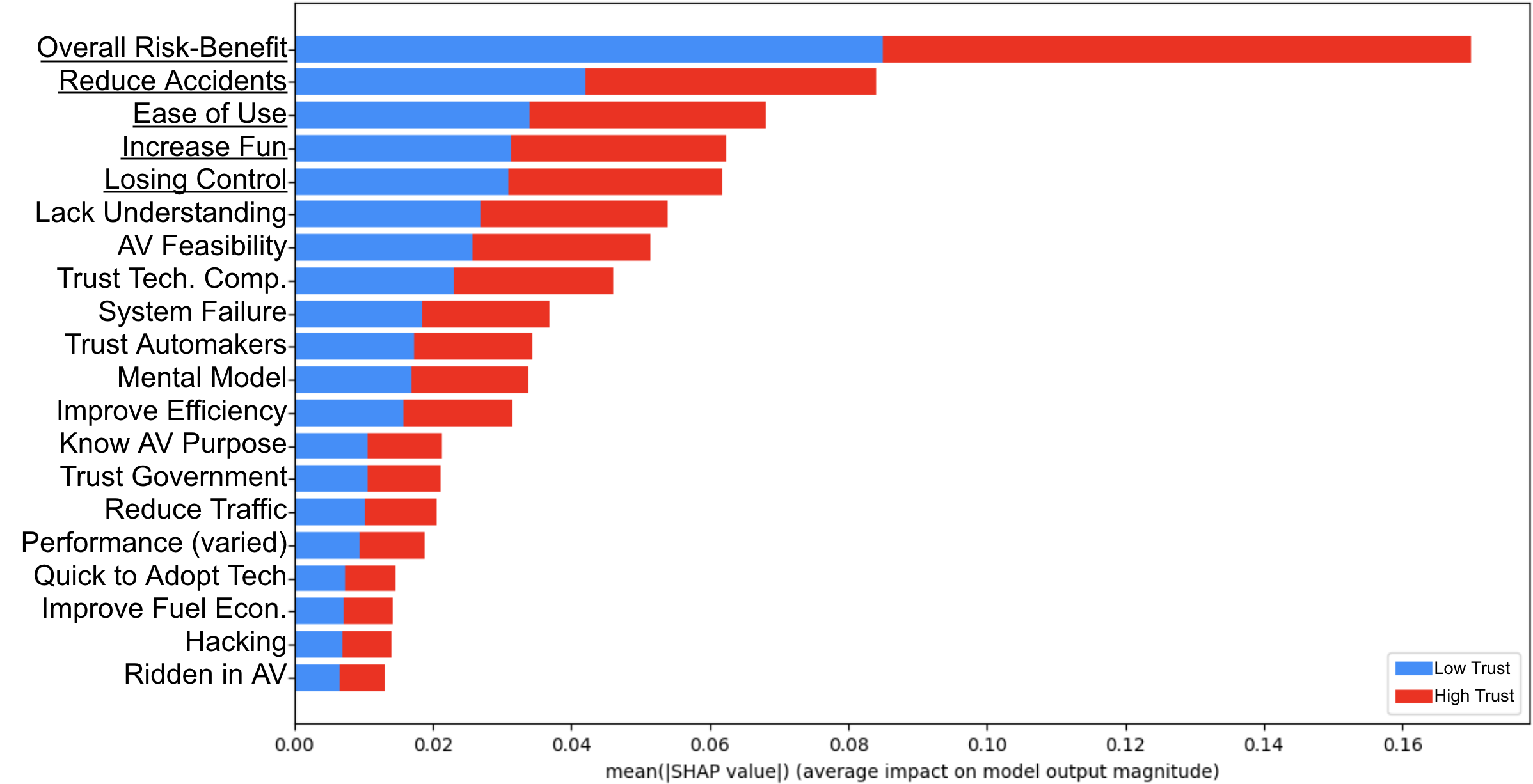}
        \caption[\textbf{Full Feature Model: mean absolute SHAP scores, averaged across 5 folds.}]{\textbf{Full Feature Model: mean absolute SHAP scores, averaged across 5 folds.} This shows the most important features for predicting both high and low trust (shown in combination, as they are mirrors of each other). Risk and benefit factors were the most important.}
        %\Description{This figure shows the results of the Full Feature Model: mean absolute SHAP scores, averaged across 5 folds. This shows the most important features for predicting both high and low trust (shown in combination, as they are mirrors of each other). Risk and benefit factors were the most important.}
        \label{fig:full_shap_mean}
\end{figure}

\begin{figure}[H]
    \centering
        \includegraphics[width=.5\linewidth]{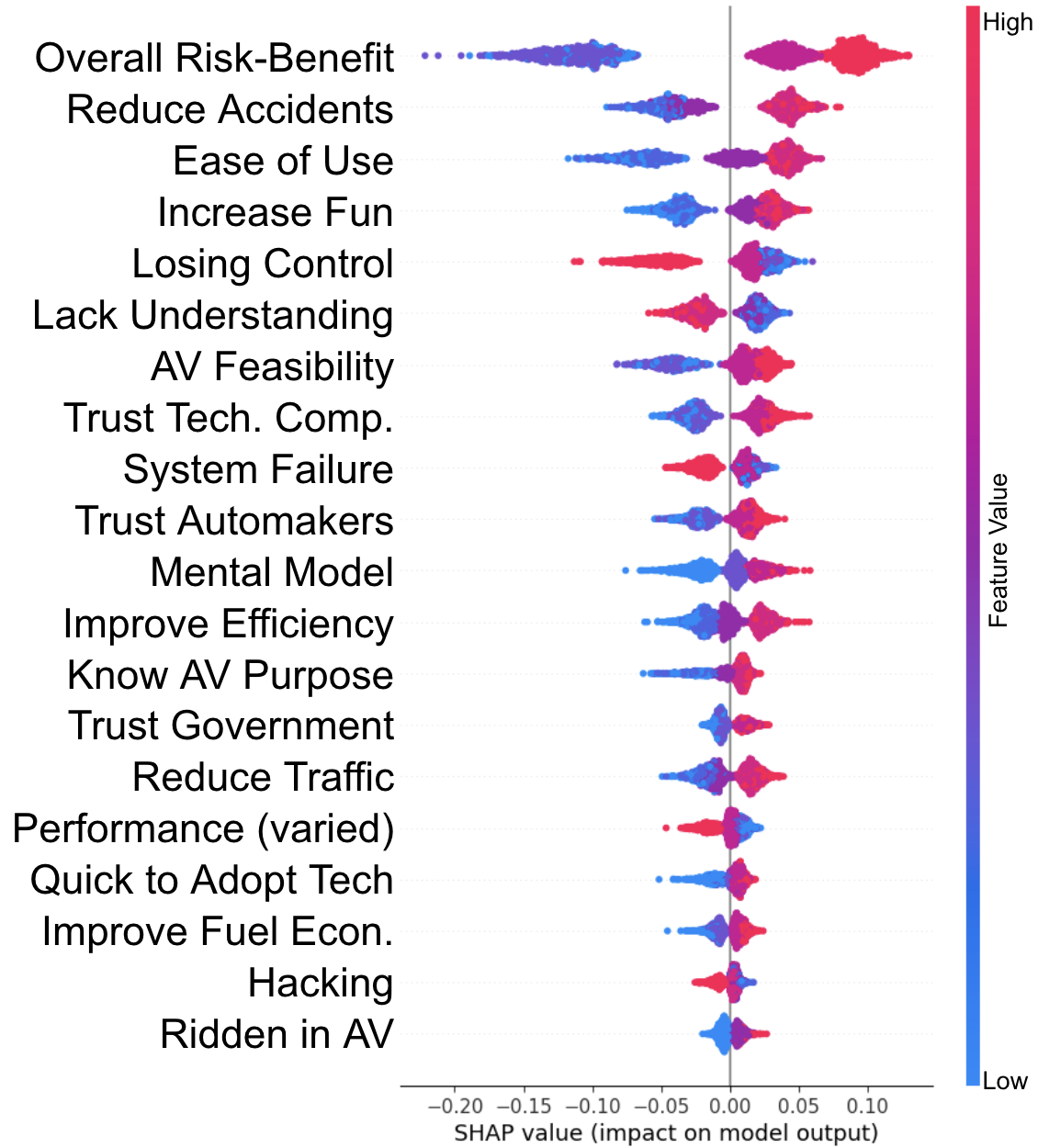}
        \caption[\textbf{Full Feature Model SHAP summary plot showing feature value impact on trust.}]{\textbf{Full Feature Model SHAP summary plot showing feature value impact on trust.} This plot shows how the \emph{value} of a feature impacts the model's output. The more extreme the SHAP value, the more indicative that value was of being high trust (positive SHAP values) or low trust (negative SHAP values). Features are organized by overall importance.}
%        \Description{This figure shows the Full Feature Model SHAP summary plot, highlighting feature value impact on trust. This plot shows how the \emph{value} of a feature impacts the model's output. The more extreme the SHAP value, the more indicative that value was of being high trust (positive SHAP values) or low trust (negative SHAP values). Features are organized by overall importance.}
        \label{fig:fullset_shap}
\end{figure}

Once the most predictive models were made, we identified the most important features via SHAP. We find that \texttt{Overall Risk-Benefit}, \texttt{Reduce Accidents}, \texttt{Ease of Use}, \texttt{Increase Fun} and \texttt{Losing Control} were the most important factors that influence trust in AVs (Fig.~\ref{fig:full_shap_mean}). Notably, 12 of the top 20 features were related to risk and benefit judgments. Fig.~\ref{fig:fullset_shap} shows how individual feature values contribute to the model's output. We observe that higher scores for \texttt{Overall Risk-Benefit}, \texttt{Reduce Accidents}, \texttt{Ease of Use}, and \texttt{Increase Fun} are positively associated with trust. \texttt{Losing Control} was negatively associated with trust.

% \begin{figure}[H]
%     \centering
%     \begin{minipage}[t]{0.59\textwidth}
%         \centering
%         \includegraphics[width=1\linewidth]{figures/top_20_shap_features_BLOCK1.1.ini.png}
%         \caption{Full Feature Model Mean Absolute SHAP scores, averaged across the five folds.}
%         \label{fig:full_shap_mean}
%     \end{minipage}
%     \hfill
%     \begin{minipage}[t]{0.39\textwidth}
%         \centering
%         \includegraphics[width=1\linewidth]{figures/Kfold-1_Class_1_RFC_SHAP_BLOCK1.1.ini.png}
%         \caption{Full Feature Model SHAP summary plot (representative example fold) showing how individual feature values impact model predictions.}
%         \label{fig:fullset_shap}
%     \end{minipage}
% \end{figure}

\vspace{-1em}

\subsection{Ablation Studies: Systematic Isolation and Elimination of Risk-Benefit Factors}
Using our full feature model, risk and benefit-related factors were consistently among the most important for predicting a person's trust level. As such, we conducted two ablation (systematic elimination) studies to derive a deeper understanding of how risk and benefit factors specifically impact trust predictions. We followed a similar approach as we did while building and evaluating the full feature model (Section~\ref{full_set_results}). The first ablation study looks at only the risk and benefit factors from Table \ref{risk_benefit_factors} in isolation (henceforth referred to as the \emph{feature subset}). The second ablation study is this feature subset's complement, which instead \emph{eliminates} all risk-benefit factors from the full feature model. Table~\ref{tab:ablation} shows a summary of performance between these two ablation models compared to our full model.

We find that the full feature model outperforms both ablation models. The difference between the accuracy of the full feature model (85.8\%) and the full feature set without the risk and benefit factors (73.8\%) is quite large, dropping 12\%. We find much smaller differences between the full feature set and the isolated feature subset (84.6\%), losing only 1.2\% accuracy. These results imply that risk and benefit factors are very important for the prediction of trust in AVs.

\begin{table}[htbp]
\caption[\textbf{Summary of the three experiments on our dataset.}]{\textbf{Summary of the three experiments on our dataset.} The full feature set showed the best performance, followed closely by the feature subset (only risk and benefit features). The full feature set \emph{without} risk and benefit features performed substantially worse.}
%\Description{This table shows a summary of the three experiments on our dataset. The full feature set showed the best performance, followed closely by the feature subset (only risk and benefit features). The full feature set without risk and benefit features performed substantially worse.}
\begingroup
\renewcommand{\arraystretch}{1}
\begin{tabular}{p{0.4\linewidth}  p{0.15\linewidth}  p{0.15\linewidth}  p{0.19\linewidth}}
\toprule
\textbf{Models} & \textbf{Balanced \newline Accuracy} & \textbf{F1 score} & \textbf{Net Accuracy \newline Difference} \\
\midrule
Full Feature Set & 0.858 (0.02) & 0.858 (0.02) & -- \\
Risk + Benefits (\textit{feature subset})  & 0.846 (0.02) & 0.843 (0.02) & (-) 0.012 \\
Full Feature Set \textit{w/o} Risk + Benefits & 0.738 (0.02) & 0.738 (0.02) & (-) 0.120 \\
\bottomrule
\end{tabular}
\endgroup
\label{tab:ablation}
\end{table}

\subsubsection{Feature Subset: Risk-Benefit Features Only}
The first ablation study, which we call the \emph{feature subset}, consisted only of risk and benefit measures. As discussed, the performance of this model was quite high. Investigating the top performing features in this subgroup using SHAP allows us to derive a better understanding of how different risks and benefits impacted our trust prediction.

\begin{figure}[H]
        \centering
        \includegraphics[width=.9\linewidth]{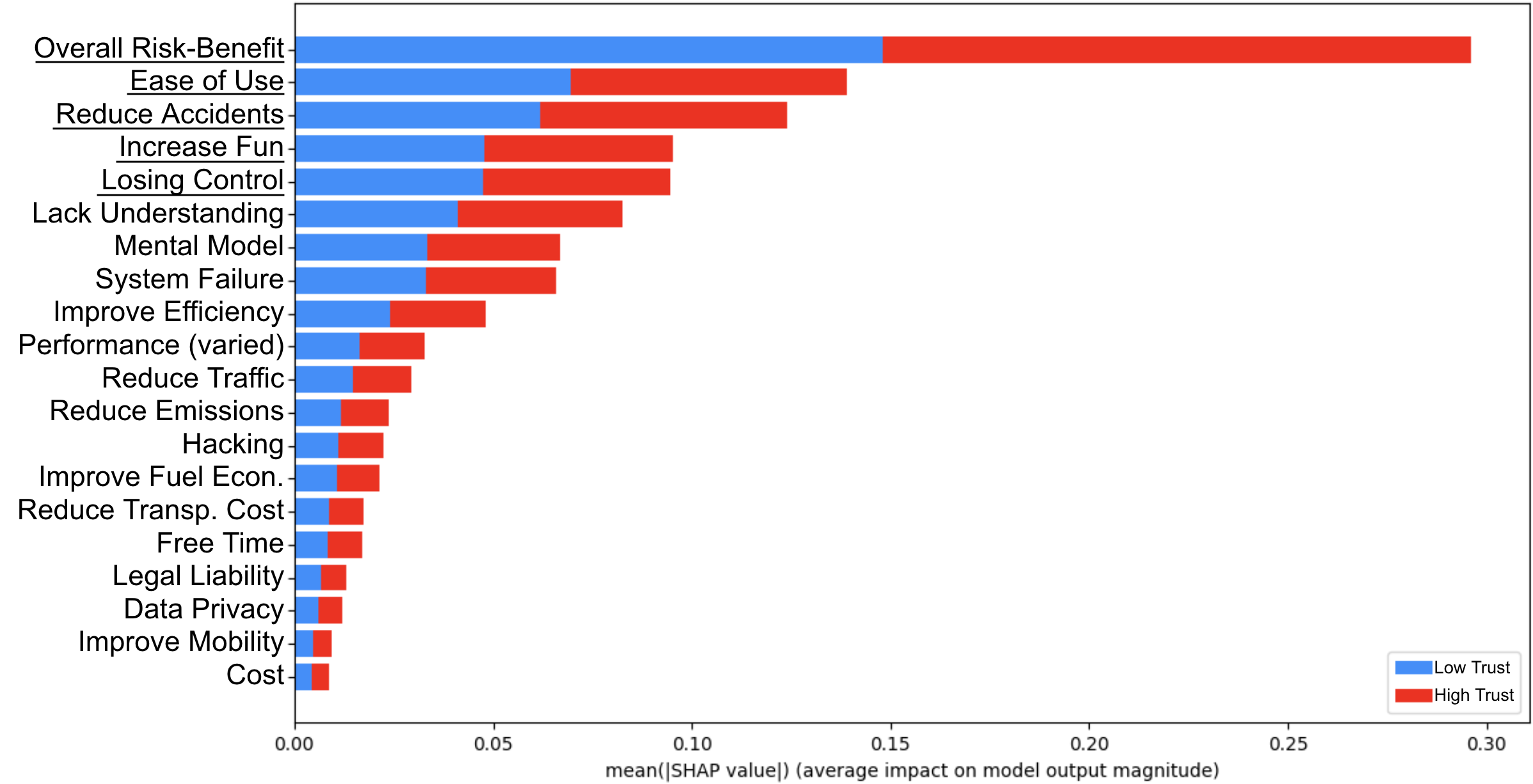}
        \caption[\textbf{Feature Subset (risk and benefit only): mean absolute SHAP scores, averaged across 5 folds.}]{\textbf{Feature Subset (risk and benefit only): mean absolute SHAP scores, averaged across 5 folds.} This shows the most important features for predicting both high and low trust (shown in combination, as they are mirrors of each other). Results are similar to the full feature set model, supporting risk and benefit features as the most important predictors of trust.}
%        \Description{This figure shows the results of the Feature Subset (risk and benefit only): mean absolute SHAP scores, averaged across 5 folds. This shows the most important features for predicting both high and low trust (shown in combination, as they are mirrors of each other). Results are similar to the full feature set model, supporting risk and benefit features as the most important predictors of trust.}
        \label{fig:ablation2_shap_mean}
        \vspace{-.5em}
\end{figure}

\begin{figure}[H]
        \centering
        \includegraphics[width=.5\linewidth]{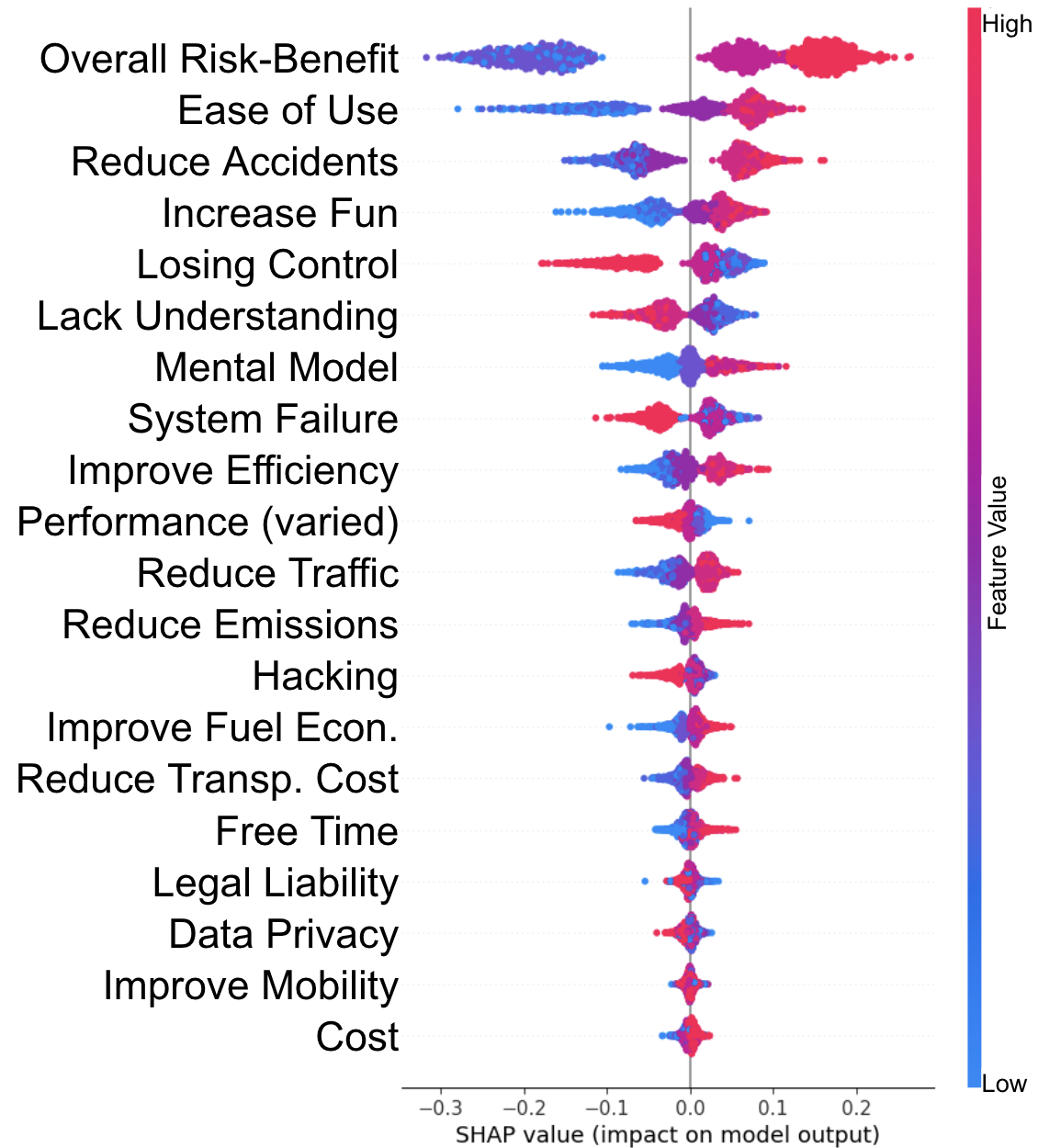}
        \caption[\textbf{Feature Subset (risk and benefit only) SHAP summary plot showing feature value impact on trust.}]{\textbf{Feature Subset (risk and benefit only) SHAP summary plot showing feature value impact on trust.} This plot shows how the \emph{value} of a feature impacts the model's output. The more extreme the SHAP value, the more indicative that value was of being high trust (positive SHAP values) or low trust (negative SHAP values). Features are organized by overall importance.}
%        \Description{This figure shows the Feature Subset (risk and benefit only) SHAP summary plot, highlighting feature value impacts on trust. This plot shows how the value of a feature impacts the model's output. The more extreme the SHAP value, the more indicative that value was of being high trust (positive SHAP values) or low trust (negative SHAP values). Features are organized by overall importance.}
        \label{fig:ablation2_shap}
\end{figure}

Fig.~\ref{fig:ablation2_shap_mean} shows the top contributing factors in terms of feature importance via mean absolute contribution, averaged across five folds. Fig.~\ref{fig:ablation2_shap} illustrates how different values of a feature predict that a person will be either high or low trust. We find \texttt{Overall Risk-Benefit}, \texttt{Ease of Use}, \texttt{Reduce Accidents}, \texttt{Increase Fun} and \texttt{Losing Control} are the most impactful features, with high scores in all showing a positive association with trust except \texttt{Losing Control}. This is near identical to the full feature model, with only slight differences in relative importance. The implication of this ablation study is that -- even when removing all demographic, psychosocial, driving, and tech-related factors -- factors specific to the risk-benefit evaluation of AVs can still predict trust quite well on their own.

% \begin{figure}[H]
%     \centering
%     \begin{minipage}[t]{0.59\textwidth}
%         \centering
%         \includegraphics[width=1.0\linewidth]{figures/top_20_shap_features_ablation 2.1.ini.png}
%         \caption{Feature Subset (risk and benefit only) Mean Absolute SHAP scores, averaged across the five folds.}
%         \label{fig:ablation2_shap_mean}
%     \end{minipage}
%     \hfill
%     \begin{minipage}[t]{0.39\textwidth}
%         \centering
%         \includegraphics[width=1\linewidth]{figures/Kfold-1_Class_1_RFC_SHAP_ablation 2.1.ini.png}
%         \caption{Feature Subset (risk and benefit only) SHAP summary plot (representative example fold) used to show how individual feature values impact the categorization of high and low trust.}
%         \label{fig:ablation2_shap}
%     \end{minipage}
% \end{figure}

\subsubsection{Full Model Without Risk-Benefit Features}
\label{ablation2section}
We also investigated the effect of eliminating the \emph{feature subset} from the full feature model. This second ablation experiment gives us insight into two additional areas of inquiry. First, if we do not have access to a person's risk-benefit assessment, can we still predict trust from other personal factors? Given the plethora of prior research on dispositional traits and attitudes discussed in Section \ref{prior_personal}, we hypothesized that this answer should be \emph{yes}. Second, \emph{if} we can predict trust from other personal factors, what factors are most important? 

To answer these questions, we ran experiments similar to the development of other models in this study, to find the best possible performance with this group of features. Unlike the full feature model and the feature subset (ablation experiment 1), we found that balanced model accuracy of this second ablation study was quite a bit lower, dropping to 73.8\%, a full 12\% dip compared to the full feature model. This further emphasizes the importance of risk-benefit features.

\begin{figure}[H]
    \centering
        \centering    
        \includegraphics[width=.9\linewidth]{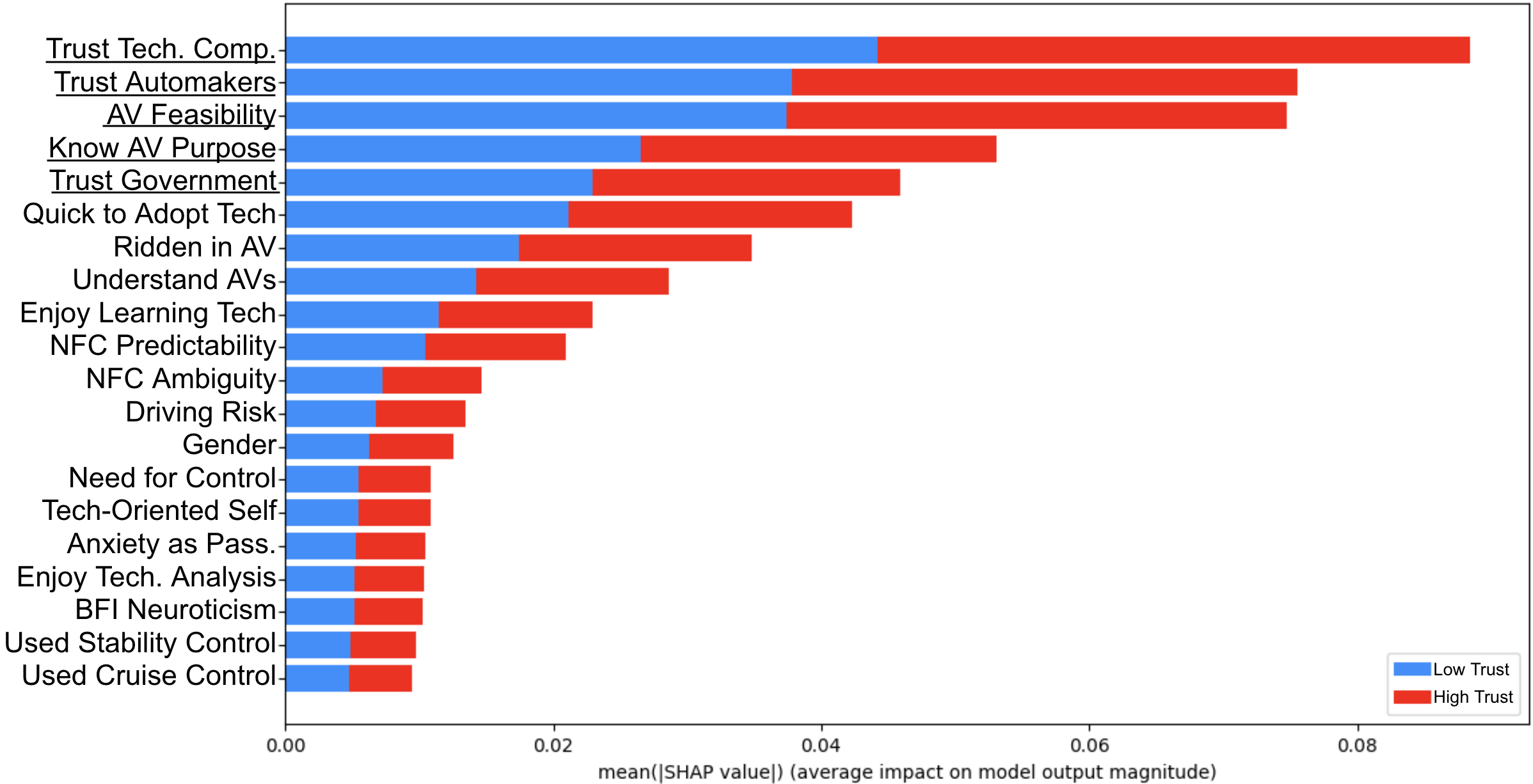}
        \caption[\textbf{Full Feature Model \emph{without} risk and benefit factors: mean absolute SHAP scores, averaged across 5 folds.}]{\textbf{Full Feature Model \emph{without} risk and benefit factors: mean absolute SHAP scores, averaged across 5 folds.} This shows the most important features for predicting both high and low trust (shown in combination, as they are mirrors of each other). Without risks and benefits, factors such as trust in institutions, AV feasibility, knowledge, and experience were the most important predictors of trust.}
%        \Description{This figure shows the results of the Full Feature Model without risk and benefit factors: mean absolute SHAP scores, averaged across 5 folds. This shows the most important features for predicting both high and low trust (shown in combination, as they are mirrors of each other). Without risks and benefits, factors such as trust in institutions, AV feasibility, knowledge, and experience were the most important predictors of trust.}
        \label{fig:ablation1_shap_mean}
\end{figure}

\begin{figure}[H]
        \centering
        \includegraphics[width=.5\linewidth]{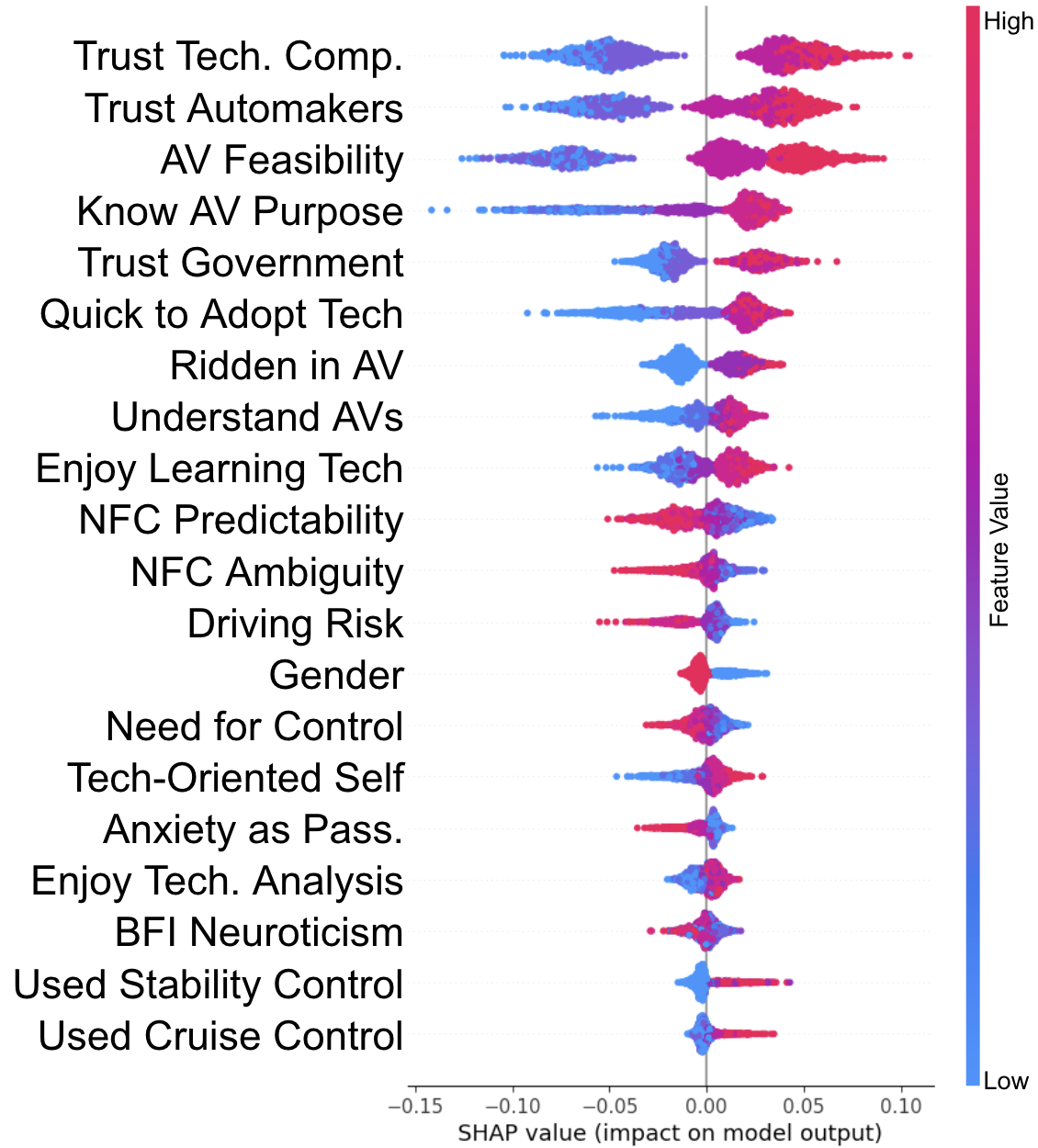}
        %\vspace{-.5em}
        \caption[\textbf{Full Feature Model \emph{without} risk/benefit factors: SHAP summary plot showing feature value impact on trust.}]{\textbf{Full Feature Model \emph{without} risk/benefit factors: SHAP summary plot showing feature value impact on trust.} This plot shows how the \emph{value} of a feature impacts the model's output. The more extreme the SHAP value, the more indicative that value was of being high trust (positive SHAP values) or low trust (negative SHAP values). Features are organized by overall importance.}
%        \Description{This figure shows the Full Feature Model without risk/benefit factors: SHAP summary plot, highlighting feature value impact on trust. This plot shows how the value of a feature impacts the model's output. The more extreme the SHAP value, the more indicative that value was of being high trust (positive SHAP values) or low trust (negative SHAP values). Features are organized by overall importance.}
        \label{fig:ablation1_shap}
\end{figure}

Though 73.8\% is not excellent predictive accuracy, from a theoretical lens it is still \emph{predictive enough} to draw initial insights on how non-risk-benefit features impact a person's judgment of high and low trust. We found that \texttt{Trust in Tech Companies}, \texttt{Trust in Automakers}, \texttt{AV Feasibility}, \texttt{Know AV Purpose}, and \texttt{Trust in Government Authorities} were the strongest predictors (Fig. \ref{fig:ablation1_shap_mean}); higher values of each were associated with higher trust (see Fig. \ref{fig:ablation1_shap}).
 
% \begin{figure}[H]
%     \centering
%     \begin{minipage}[t]{0.59\textwidth}
%         \centering
%         \includegraphics[width=1.0\linewidth]{figures/top_20_shap_features_ablation 1.1.ini.png}
%         \caption{Full Feature Model \textit{without} risk and benefit factors: Mean Absolute SHAP scores, averaged across the five folds.}
%         \label{fig:ablation1_shap_mean}
%     \end{minipage}
%     \hfill
%     \begin{minipage}[t]{0.39\textwidth}
%         \centering
%         \includegraphics[width=1\linewidth]{figures/Kfold-1_Class_1_RFC_SHAP_ablation 1.1.ini.png}
%         \caption{Full Feature Model \textit{without} risk and benefit factors: SHAP summary plot (representative example fold) used to show how individual feature values impact the categorization of high and low trust.}
%         \label{fig:ablation1_shap}
%     \end{minipage}
% \end{figure}

%\vspace{3em}
\section{Discussion}
We provide evidence that young adult trust in autonomous vehicles (AVs) can be predicted via machine learning. We used a comprehensive set of personal factors with 130 distinct input features, derived from survey. We compared the predictive power of several machine learning models, all of which performed quite well. Random forest models performed best -- 85.8\% (Table \ref{tab:model_main}) -- at categorizing people into high and low trust groups. By applying the explainable AI technique SHAP to our best performing models, our analysis not only shows \emph{which} factors were most predictive of trust, but also \emph{how} each feature uniquely contributed to the model's conclusion. Our findings underscore a central argument of this dissertation: trust in autonomous systems is not merely a byproduct of technical performance but is deeply shaped by user-specific perceptions, traits, and contextual factors. These results complement the broader themes of the dissertation by advancing our understanding of how personalized and adaptive communication strategies can bridge the gap between AV capabilities and user acceptance. Ultimately, this study strengthens the systems-based framework that positions trust as an essential, dynamic element of successful human-AV collaboration.

In the following sections, we contextualize our findings within the scope of human-AV interaction research and provide insights for how our analysis can be used to build trustworthy vehicles for diverse groups. Many of these insights may generalize to other AI-driven domains or for groups beyond young adults.

A general principle that resulted from our study is that \emph{perceptions of risks and benefits} are the most important factors for predicting trust. \color{black}The importance of risks and benefit evaluation for trust and acceptance has been highlighted in the past, especially for high-stakes contexts like AVs~\cite{ayoub2021modeling, hewitt2019assessing, zhang2023human}. Unlike prior studies, however, our analysis allowed us to isolate \emph{specific} types of AV-related risks and benefits, such as individual concerns over system failures and usability. As such, the present study is the first to articulate the \emph{relative importance} of these risk and benefit perceptions compared to other types of personal factors for predicting trust. \color{black} In our full feature model, 12 of the 20 most important features (based on mean SHAP value) were related to a person's assessment of the risks and benefits of using AVs. Using ablation studies, the importance of risk-benefit perceptions was confirmed: risk and benefit factors alone had similar predictive ability as our full model, whereas a knockout model \emph{without} risk and benefit factors performed far worse.  
We find a person's overall assessment of the risks and benefits of an AV was the most highly predictive feature for predicting trust level (importance rank 1) in both the full feature model and feature subset. This is unsurprising, as we would assume that the ratio of benefits to risks -- as discussed at length in cost-benefit analyses and (ir)rational decision-making~\cite{wang2019designing} -- would impact a person's decision to trust an AV. Indeed, \citet{ayoub2021modeling} also found that assessment of benefits and risks were the most important factors predicting trust in their model, though they do not specify which risks and benefits these may be. In the present study, we are able to articulate precisely \emph{which} risk-benefit factors are most important. From an AV design perspective, the importance of overall risk-benefit assessment on trust judgments implies that addressing high priority risks and benefits using strategic design elements can drastically impact a person's trust judgments. In the following sections, we will discuss risks and benefits, respectively.

\vspace{.5em}
\noindent
\textbf{Risks associated with system performance and usability failures were the most consequential ---}
Six of the top 20 most important features to our full model were related to risks. These included concerns with: Ease of Use (rank 3), Losing Control (5), Lacking Understanding of AV decision-making (6), System Failures (9), Performance of AVs across varied terrain (16), and Hacking (19). High ratings of these concerns were associated with lower trust. 

These concerns highlight areas of priority for human-AV interaction design and research. For instance, concerns over ease of use, giving up control, and lack of decision-understanding are fundamental to the development of explainable AI systems for AVs~\cite{koo2015did, miller2019explanation}, which aim to increase transparency and usability with complex AI-based systems. Other specific concerns with vehicle performance and security -- such as concern with equipment failures, performance, and hacking -- may also be addressed through a mix of UI enhancements and education campaigns (assuming that the systems are, in fact, performing well and secure). These may include visualizations reassuring riders that the vehicle is safe, private, and operating within the bounds of its ability. Importantly, elements should be designed to empower a rider by providing decision agency, not to mislead them by artificially minimizing risks.

One factor -- performance of AVs across varied terrain (16) -- is of particular note to this dissertation as this concern highlights people's intrinsic worry that AVs will not be able to adjust their behavior to successfully drive (or to put it in the terms of a success function, meet a rider's needs) in different driving environments. This concern was empirically demonstrated in Chapter 3. The implication is that even lay riders inherently recognize that different contexts require different AV actions, strengthening the claim that contextually-adaptive design is necessary.

\vspace{.5em}
\noindent
\textbf{Benefits need to add value beyond what can be achieved by human-controlled driving ---}
Five of the top 20 features were related to benefits. Perceiving that AVs can Reduce Accidents (rank 2), Increase the Fun of driving (4), Improve Efficiency (12), Reduce Traffic (15), and Improve Fuel Economy (18) were all associated with higher trust. In all cases, these are benefits that move beyond what is possible with current human-controlled driving and vehicle infrastructure. Our findings support prior work that trust and adoption of AVs are likely tied closely together \cite{hewitt2019assessing}, meaning that AVs will only be opted-for when their benefits far outpace what is possible by the current status quo of non-autonomously driven vehicles (as well as the risks of AVs discussed previously).

For future AV design and research, these findings suggest that communicating the unique benefits of AVs, particularly those that exceed the capabilities of human-controlled driving, is crucial for building trust. This emphasis should be coupled with strategies to mitigate perceived risks, ensuring that the value proposition of AVs is compelling enough to encourage widespread adoption. Design elements highlighting specific elements -- such as showing fuel savings or estimations of emission reduction -- can help build confidence in potential riders.

\vspace{.5em}
\noindent
\textbf{Attitudes towards AV feasibility, affinity for technology, and institutional trust helped to predict trust, but to a smaller degree than risk and benefit perceptions ---} Though as a whole less important than risk and benefit factors, we found that certain perceptions towards AVs, technology, and institutions as a whole helped predict trust. Unsurprisingly, we found that people who considered AVs more practically Feasible (rank 7), understood the Purpose of AVs (13), and who are generally Quick to Adopt New Technology (17) were more likely to trust AVs. These make sense intuitively; to trust AVs, a person would need to understand why AVs exist and believe that the technology is realistic. Given that widespread adoption has yet to be accomplished, it also makes sense that people who are quick to adopt new technologies would be more likely to trust AVs, as these people may be interested in new technologies -- like AVs -- and may have a higher tolerance for uncertainty regarding use of less established systems. This latter finding aligns with prior work on affinity for technology predicting AV trust~\cite{kraus2021s}.

Interestingly, all three Trust in Institutions measures were important predictors of trust. Trust in Tech Companies ranked 8 in importance, Trust in Automakers ranked 10, and Trust in Government Authorities ranked 14. We provide two potential explanations for these findings. First, many people likely recognize that AVs will function well \emph{if and only if} the groups that produce and regulate them function well. This strengthens the importance of brand reputation, regulation transparency, and thorough testing criteria. Effects may also be explained by differences between people who are generally more or less trusting as a dispositional characteristic: for some people, trust may simply be easier than for others~\cite{hancock2023and}. Special care should be taken by designers to reach individuals who do not trust easily. In these cases, we would expect heightened sensitivity to errors during use~\cite{luo2020trust} and a higher benefit to risk ratio necessary for adoption. We will discuss specific design implications -- including what these mean for future AV policy -- in Section \ref{design_implications}.

\vspace{.5em}
\noindent
\textbf{Psychosocial and many driving and technology-related factors were far less important than expected ---}
In Section \ref{prior_personal}, we discussed a plethora of prior work suggesting that psychosocial factors -- including personality, self-esteem, risk perception, and need for control -- will impact a person's trust judgment for AVs. \color{black} Our results generally contrast with these prior studies, as many of these factors did not appear to be important for our model's prediction of trust. \color{black} Further, exempting the few factors noted above, technology and driving factors -- including driving style, driving cognitions, and technology self-efficacy -- also showed little impact. 

There are a few explanations for this discrepancy. First, it is possible that the risk-benefit factors and others previously noted were fully sufficient to predict trust on their own, and thus less predictive (but still potentially relevant) factors were prioritized less by our ML models. If this were true, we would likely still see effects in our traditional high-low trust group comparison using Kruskal-Wallis tests. According to the descriptive analysis in \ref{sec:descriptive} and continued in Appendix \ref{appendixA.trust}, however, not a single one of these factors showed large or even medium effect size differences comparing the high and low trust groups. Some -- including certain prior experiences, perceptions of driving risk, Need For Closure (Predictability), and some technology self-efficacy measures -- showed small effects. The rest showed negligible effects. Even in our second ablation study (knocking out risk-benefit factors, Section \ref{ablation2section}), it is not until the tenth most important factor that we start to see psychosocial or demographic variables: Need for Closure Predictability and Ambiguity (importance ranks 10 and 11, respectively), driving risk assessment (12), gender (13), need for control (14), etc. Not a single driving style or driving cognitions factor appeared in the top 20, even with ablations.

As such, our results generally conflict with past work suggesting that these psychosocial, driving, and tech factors may be \emph{important} predictors of AV trust. This is not to say that they should be omitted from future AV trust research, nor that the relationships found in prior work are false. Instead, our results suggest that other, more important factors should be prioritized in future AV trust research and design for our specific subpopulation of interest. With less data available to a research team (such as without risks/benefits) or for more general populations of study, these factors may become more important. Further, psychosocial and driving-related factors may still inform other areas of AV design and research, such as cases where personalization based on personality and other traits have improved interactions~\cite{alves2020incorporating}.

\vspace{.5em}
\noindent
\textbf{Similarity between AV and human decision-making was positively associated with trust ---}
Interestingly, we found that people who viewed AV decision-making as more similar to human decision-making (Mental Model, ranked 10th in importance) were more likely to trust AVs. This suggests that a person's mental model of AV decision-making plays a key role in their trust. When AV decision-making aligns more closely with how they perceive human decision-making, trust tends to increase. From a design perspective, this implies that making AV interactions more human-like, with communication that is easy for people to understand, could help bolster trust. Our results support current efforts to make AI explanations more similar to human explanations, particularly those given by experts~\cite{miller2019explanation, pazzani2022expert, kaufman2022cognitive, koo2015did}.

\vspace{.5em}
\noindent
\textbf{Isolating only the subpopulation of young adults allowed us to uncover deeper predictors of trust, and can be used as a methodological template for the study of future subpopulations ---}
By accounting for specific characteristics of the subpopulation, such as age or education, we focused this study on young adults -- the next generation of AV adopters. While age and education have been used as surface-level predictors of trust in previous research, isolating this group allowed us to explore deeper, more meaningful factors that are particularly relevant to this subpopulation. Specifically, instead of replicating prior findings on the importance of age~\cite{hulse2018perceptions, abraham2017autonomous} or education~\cite{hudson2019people}, our models based their predictions on potentially more important characteristics, like their risk-benefit assessment and attitudes towards AVs. The one exception in our findings was gender: similar to prior work by \citet{hulse2018perceptions} and \citet{mosaferchi2023personality}, our findings also suggest that those who identify as women are less likely to trust AVs than those who do not identify as women. This was not an important factor for our full model, but did show moderate importance for the second ablation study. As such, our findings support the premise that focusing on subpopulations may be an effective (though more time intensive) strategy for deriving deeper insights about a particular group of interest. In the present case, our work provides a deeper insight into designing for young adults, and may serve as a methodological template for future research on other subpopulations.

\vspace{.5em}
\noindent
\textbf{Prior AV experience matters, but less than what prior work suggests ---} At the very last rank of our top 20, we find that people who have prior experience riding in an AV were more likely to trust AVs. This aligns with prior work by \citet{choi2015investigating} and \citet{mosaferchi2023personality}, who also found a positive association between prior experience and trust. Prior experience was ranked number 20 in terms of importance, however, meaning that there are many features that were better predictors of trust. This has important implications, as AV companies like Waymo~\cite{WaymoOne} often claim that people will trust AVs as soon as they gain experience with them. Though this may be true to a degree, our findings show that other factors should be prioritized for AV design and research focused on building trust and increasing adoption.

% \textbf{Comparing traditional and machine learning analysis results show why leveraging ML techniques should be opted for by future research teams.}
% Perhaps also-- kruskal wallis was decent, but ML was different and should be oped for in the future.
%\vspace{-2em}
\subsection{Implications for Trustworthy Human-AV Interaction Design and Research}
\label{design_implications}
Our results have important implications for the design and study of autonomous vehicles. In this study, we successfully predicted AV trust from the personal factors of young adults. As expected, some features in our model were more important than others. We can use these results as a basis for personalized and adaptable designs based on specific factors of importance, as well as specific insights for trustworthy Explainable AI (XAI) design, communication and education campaign guidelines, and policy suggestions.

\vspace{.5em}
\noindent
\textbf{Explainable AI and Transparency ---}
The high importance of concerns around ease of use, giving up control, and lack of understanding around AV decision-making highlights the critical need for continued research and design efforts on AV explainability and explainable AI (XAI) more generally. \color{black} Our results echo prior work on XAI, emphasizing the necessity for explainability contexts where transparency can mitigate user concerns over safety and control (see Section \ref{prior_personal}). For example, these results support the design and study of clear, real-time explanations for AV decisions. \color{black} A special focus should be put on XAI explanations that are understandable by riders from diverse communities with a wide range of backgrounds and expertise levels. Failure to satisfy concerns over ease of use, vehicle performance, and safety and privacy may result in refusal to ride with an AV.

\vspace{.5em}
\noindent
\textbf{Prioritize Safety and Security Features ---}
One area of focus for XAI in AVs should be on effectively communicating driving performance metrics, particularly when they can reassure a rider that the AV is safe and protected from both physical harm and privacy risks. Explanations for \emph{what} action an AV is taking and {why}~\cite{koo2015did} may give a sense of control and agency back to a driver. These may be effectively paired with visualizations emphasizing that the AV is performing properly and trained well for the driving environment, mitigating concerns over performance adaptability and calibrating trust to an appropriate level. Visualizations and communications should be realistic and not misleading, as under-communicating performance errors could lead to over-reliance.

\vspace{.5em}
\noindent
\textbf{Communicate Unique Benefits Effectively ---}
Given the large number of potential benefits that AVs offer~\cite{fagnant2015preparing} and the importance of \emph{believing} in these benefits to trust judgments, our research suggests that educational campaigns and mid-drive interactions should strive to highlight benefits to increase trust and adoption. In particular, emphasizing the \emph{added value} AVs can bring \emph{beyond} human-controlled driving may be most effective, as the alternative to AV adoption are non-AVs (status quo) which still function quite well in many contexts. Specific benefits which should be prioritized include reducing crashes and traffic, increasing driving efficiency, and increasing fun.

\vspace{.5em}
\noindent\textbf{Anthropomorphizing AV communications may be beneficial, but should be done with caution ---}
We found that people who conceptualized AV decision-making as closer to human decision-making had higher AV trust. This alludes to the power that humanizing AV communications may have on increasing trust, suggesting similar approaches to AV explanations could be potentially useful. For instance, expert-informed explainable AI has been studied in the past with promising results~\cite{bayer2022role, pazzani2022expert, soltani2022user, kaufman2023explainable}. This approach should be taken with caution, however, as certain theoretical frameworks such as the Computers Are Social Actors (CASA) paradigm~\cite{nass1994computers} show that people often treat computerized systems the way they would treat other humans, meaning systems which are too humanized may be assumed to have human-like biases and flaws in judgment and decision-making.

\vspace{.5em}
\noindent
\textbf{Personalization and tailoring of AV communications and behavior may be a promising future direction ---}
Key traits found in our current study and in prior work emphasizes that \emph{people differ} in their underlying traits, attitudes, and experiences, and that these differences impact trust. In the present study, we highlight some of the most important traits that should be looked at for predicting a young adult's trust in AVs. This work can be used as a foundational step towards tailoring AV communications and behaviors to the specific individual who may be riding in the AV. Once the most important traits have been identified for a specific subpopulation of interest, design elements can be tailored and tested specifically for building trust for that group. By pairing a certain “trait profile” with certain design and explanation choices (e.g. different amounts, types, or modalities of information), an AV may better meet the needs of different types of individuals. For example, groups with certain concerns or backgrounds can be shown certain types of information in a display, while groups with \emph{other} concerns or backgrounds can be shown \emph{other} UI or explanation elements. If done correctly, tailoring can be done automatically based on who is interacting with the AV.

\vspace{.5em}
\noindent
\textbf{Contextually-adaptive AV behavior may mitigate concerns over AV adaptability ---}
We found that trust may be influenced by a general concern that AVs will not be able to perform well across different driving terrains. This finding highlights that riders inherently recognize that their needs (driving, communication, or otherwise) may differ in different situations, and want to see that AVs can adapt to meet them. This single concern provides a strong case for contextually-adaptive designs, further supporting evidence from previous chapters.

\vspace{.5em}
\noindent
\textbf{\color{black}Implications for Young Adults and Generalizability to Other Populations ---}
\color{black}Though this study focused on young adults, it is likely that many insights may generalize to broader populations of interest. Certain features -- such as ease of use, transparency, and institutional trust -- were likely predictive for young adults due to their high exposure to and reliance on technology in daily life. As a generation that has grown up with digital and AI-based tools, prior work has demonstrated that young adults are more accustomed to the presence of autonomous systems and may have greater baseline expectations around usability, control, and information transparency \cite{anderson2018future}. Moreover, young adults may rely on a combination of brand reputation, ethical standards, and transparency practices when evaluating AV trustworthiness \cite{twenge2014declines}, potentially explaining why institutional trust emerged as a strong predictor for this group. Focusing on young adults in isolation may also explain the lack of importance of psychosocial, driving, and technology-related factors compared to prior studies on more general populations. Future work segmenting and comparing population clusters could help clarify whether this reduced importance of psychosocial factors is specific to young adults, or if it reflects limitations in the generalizability of prior work over diverse user groups. Though these are potential explanations for why certain findings in this study were found \textit{specifically} for a young adult population, preference for usability, transparency, as well as many risk and benefit factors are likely important to a wide number of populations. Though future research is necessary to support this claim, we have no reason to suspect that concerns over safety or losing control, for example, would only be relevant to young adults. However, it is probable that different predictors (or at least, different relative importance of predictors) would emerge for other age groups or cultural contexts \cite{hancock2011meta}. \color{black}

\vspace{.5em}
\noindent
\textbf{Policy, Regulation, and \color{black}Ethical \color{black} Implications ---}
Given the importance of institutional trust (in technology companies, automakers, and government agencies), our results suggest that clear and consistent regulatory standards are crucial for building public trust in AVs. Policymakers should focus on establishing transparent safety protocols, rigorous testing standards, certification processes, and data sharing policies that are easy for the public to understand. Clear regulations can reassure the public that AVs meet high safety standards and are regularly audited for compliance. These are similar to the recommendations made by \citet{fagnant2015preparing}, though current realization of these recommendations leaves much to be desired in this regard. Our work highlights \emph{transparency} in regulation as a way to increase trust, particularly for cases when a person has low trust in institutions in general. Broader regulation focused on the most important areas of risk discussed previously -- vehicle performance, usability, and security -- should be prioritized.

\color{black}
The development of industry guidelines and regulatory oversight can be a practical way to promote the ethical application and use of trust insights for AV technology. Understanding the individual needs of potential AV users can allow equitable adoption of and access to AV technology. While our research aims to improve user-centered design by identifying factors that predict trust, insights on specific needs could be used for targeted marketing without parallel investments in AV safety and reliability. Such practices could result in users adopting AVs based on trust-building marketing tactics rather than well-informed, safety-centered choices. To prevent misuse, we recommend responsible application of trust-predictive insights, emphasizing transparency, user autonomy, and informed choice. Beyond institutional regulation, companies should ensure that trust-building efforts go hand-in-hand with clear and accessible communication about AV capabilities and limitations. For example, AV manufacturers could provide detailed information on independent safety evaluations, test outcomes, and known limitations of their systems, such as performance in inclement weather or in complex urban settings. This transparency helps users form realistic mental models of AVs, enabling them to trust based on accurate expectations rather than solely on positive marketing.\color{black} 

%\vspace{-.75em}
\subsection{Limitations and Future Work}
This study was not without limitations. First, our study leveraged survey methods, which -- though standard practice -- may be subject to bias as it is based on self-reported data. To mitigate this concern, we employed a large sample size, used validated questionnaires, and replicated variable selection from prior work whenever possible. Self-report concern may be particularly strong with measurements of trust, as (though quite common) self-reported trust may not be as effective as measuring trust via behavioral reliance~\cite{hoff2015trust}. To mitigate this risk, we used a composite trust measure to capture many aspects of a trust judgment. \color{black}Variability in participants' familiarity with AV technology may introduce latent confounds, as those with more exposure might respond differently to questions about AV trust, risks, or benefits. However, we believe this limitation is minimized by focusing on a clear, every day commuting context, which provided a shared baseline for interpretation. This range in familiarity also reflects the diversity of real-world users, enhancing the generalizability of our findings. Future work may segment participants by familiarity levels to help control for baseline knowledge differences. Likewise, we recognize that institutional trust measures like “trust in tech companies” may be interpreted as broad constructs, encompassing varying levels of trust that could be influenced by specific companies or historical practices. While these widely used variables capture overarching attitudes toward technology and institutions, which are valid for the predictive task in this study as they reflect general perceptions relevant to user trust in AVs, future work could benefit from a more nuanced approach to provide clearer insights. For example, exploring sub-features within each institutional construct could help differentiate the effects of brand or institutional identity, transparency practices, or innovation history. \color{black} Future work may seek to use behavioral trust measures. We focused on young adults, a majority of whom were enrolled at an undergraduate institution in the United States. As such, results presented in this study may not widely generalize. Future work should replicate on other populations if results in other specific settings are sought. Finally, testing the design implications through actual implementation of the guidelines outlined in Section \ref{design_implications} will be necessary to assure their validity.

%\vspace{-.75em}
\section{Conclusion}
In this study, we predicted young adult trust in autonomous vehicles from a comprehensive set of personal factors. Using machine learning and explainablity techniques, we uncovered \emph{which} factors were most predictive of trust, and \emph{how} each feature uniquely contributed to the model's conclusion. We found that perceptions of AV risks and benefits, attitudes toward feasibility and usability, institutional trust, and prior AV experience were the most important trust predictors. A person's mental model also played a role, where those who believed AV decision-making as closer to human decision-making were more trusting. Contrary to previous findings, psychosocial factors -- including personality, self-esteem, risk perception, and need for control -- as well as most technology and driving-specific factors, including driving style, driving cognitions, and technology self-efficacy -- were not strong predictors of trust. Our results build a new understanding on how personal factors can be used to predict trust in AVs. We conclude our discussion with a set of design, research, and policy implications that can be used to improve future trust and adoption of AVs that can meet the needs of diverse user bases.

This study further supports the central themes of this dissertation by illustrating how a systems-level approach can reveal the nuanced and dynamic factors shaping human-AV interaction. By identifying the most critical predictors of trust and demonstrating their relative importance, this work highlights the essential interplay between individual traits, system attributes, and contextual factors. These findings reinforce the broader argument that designing trustworthy AVs requires a shift away from one-size-fits-all approaches toward systems that are personalized, adaptable, and context-sensitive. By augmenting an AV's awareness to encompass the needs of their riders, we can facilitate better human-AV interactions. Moreover, this study complements earlier chapters by demonstrating the practical value of integrating advanced methodologies, such as machine learning, to better understand and address user-specific needs. Together, these insights provide a framework for building AV systems that empower diverse populations and ensure successful adoption across varied real-world contexts.

%%
%% If your work has an appendix, this is the place to put it.
%\appendix
\newpage
\section{Appendix}
\subsection{Comparison of High and Low Trust Groups For Each Predictor}

\setcounter{table}{0}
\renewcommand{\thetable}{A\arabic{table}}

\label{appendixA.trust}
Table~\ref{tab:appendix} contains the list of all variable names used as predictor inputs. Participants are categorized as Low Trust and High Trust based on if they are on the bottom half of possible values or the top half of possible values for the trust composite score scale. \textbf{This allows us to compare predictor variable scores between people high and low trust.} Comparisons use the non-parametric Kruskal-Wallis test. Variable descriptions are included in Tables \ref{driving_factors}, \ref{demo_psych_factors}, and \ref{risk_benefit_factors}. The top large effect sizes are included in Table \ref{kw_large}.
\newline

% \begin{tabular}{p{0.2\linewidth} | p{0.05\linewidth} | p{0.05\linewidth} | p{0.05\linewidth} | p{0.05\linewidth} | p{0.05\linewidth} | p{0.05\linewidth} | p{0.07\linewidth} | p{0.1\linewidth}}
%   \hline
% Predictor & Low Trust (mean) & Low Trust (SD) & High Trust (mean) & High Trust (SD) & H-stat & P-value & Cohen's D & Effect \\ 
%   \hline

%    \hline
% \multicolumn{9}{l}{\textit{All 0.00 reported p-values are significant at p $<$ 0.01 or less}}
% \end{tabular}

\begingroup
\renewcommand{\arraystretch}{1}
\footnotesize
\begin{longtable}{p{0.35\linewidth} | p{0.045\linewidth} | p{0.04\linewidth} | p{0.04\linewidth} | p{0.04\linewidth} | p{0.05\linewidth} | p{0.06\linewidth} | p{0.05\linewidth} | p{0.05\linewidth}}
\caption{\textbf{List of all variable names used as predictor inputs to compare high and low trust groups.}}
\label{tab:appendix}
  \\ \hline
 & \multicolumn{2}{c|}{\textbf{Low Trust}} & \multicolumn{2}{c|}{\textbf{High Trust}} & \multicolumn{4}{c}{\textbf{Comparison}}\\
 \hline
\textbf{Predictor} & \textbf{M} & \textbf{SD} & \textbf{M} & \textbf{SD} & \textbf{H} & \textbf{P} & \textbf{D} & \textbf{Effect} \\ 
  \hline
\endfirsthead
\multicolumn{9}{r}{\raggedright \textit{Table \thetable\ continued from previous page\newline}}
\\ \hline
& \multicolumn{2}{c|}{\textbf{Low Trust}} & \multicolumn{2}{c|}{\textbf{High Trust}} & \multicolumn{4}{c}{\textbf{Comparison}}\\
 \hline
\textbf{Predictor} & \textbf{M} & \textbf{SD} & \textbf{M} & \textbf{SD} & \textbf{H} & \textbf{P} & \textbf{D} & \textbf{Effect} \\ 
  \hline
\endhead
Overall Risk-Benefit & 3.23 & 0.87 & 4.63 & 0.81 & 595.25 & $<$ 0.01 & 1.66 & large \\ 
Ease of Use & 2.55 & 1.13 & 3.72 & 0.88 & 345.75 & $<$ 0.01 & 1.14 & large \\ 
Reduce Accidents & 2.79 & 1.07 & 3.89 & 0.86 & 338.97 & $<$ 0.01 & 1.12 & large \\ 
Trust Tech Companies & 2.19 & 0.99 & 3.09 & 0.99 & 241.61 & $<$ 0.01 & 0.91 & large \\ 
Increase Fun & 2.38 & 1.22 & 3.46 & 1.16 & 237.14 & $<$ 0.01 & 0.91 & large \\ 
AV Feasibility & 2.58 & 1.02 & 3.43 & 0.87 & 235.40 & $<$ 0.01 & 0.88 & large \\ 
Improve Efficiency & 2.46 & 1.06 & 3.36 & 1.03 & 214.80 & $<$ 0.01 & 0.86 & large \\ 
Reduce Traffic & 2.82 & 1.17 & 3.71 & 1.04 & 194.96 & $<$ 0.01 & 0.80 & large \\ 
\midrule
Trust Automakers & 2.47 & 1.07 & 3.28 & 0.96 & 191.45 & $<$ 0.01 & 0.79 & med. \\ 
Mental Model & 1.70 & 0.88 & 2.42 & 1.04 & 191.01 & $<$ 0.01 & 0.76 & med. \\ 
Losing Control & 4.18 & 1.06 & 3.37 & 1.13 & 208.25 & $<$ 0.01 & -0.74 & med. \\ 
Reduce Emissions & 3.02 & 1.15 & 3.76 & 0.99 & 142.02 & $<$ 0.01 & 0.69 & med. \\ 
Know AV Purpose & 3.16 & 1.22 & 3.91 & 0.89 & 137.61 & $<$ 0.01 & 0.69 & med. \\ 
Improve Fuel Econ. & 3.26 & 1.07 & 3.94 & 0.89 & 145.74 & $<$ 0.01 & 0.68 & med. \\ 
Reduce Trans. Cost & 2.80 & 1.11 & 3.52 & 1.07 & 133.91 & $<$ 0.01 & 0.66 & med. \\
Quick to Adopt Tech & 3.04 & 1.11 & 3.71 & 0.93 & 131.84 & $<$ 0.01 & 0.65 & med. \\ 
Lack Understanding & 3.55 & 1.20 & 2.86 & 1.10 & 127.35 & $<$ 0.01 & -0.59 & med. \\ 
Trust Government & 2.02 & 0.99 & 2.63 & 1.07 & 115.61 & $<$ 0.01 & 0.59 & med. \\ 
System Failure & 4.40 & 0.92 & 3.87 & 0.99 & 144.07 & $<$ 0.01 & -0.56 & med. \\ 
Enjoy Learning Tech & 2.74 & 1.21 & 3.39 & 1.16 & 96.65 & $<$ 0.01 & 0.55 & med. \\ 
Free Time & 2.63 & 1.30 & 3.29 & 1.30 & 83.72 & $<$ 0.01 & 0.51 & med. \\ 
Understand How AVs Work & 2.32 & 1.19 & 2.92 & 1.21 & 83.46 & $<$ 0.01 & 0.50 & med. \\ 
Performance (varied) & 3.93 & 1.06 & 3.40 & 1.09 & 92.62 & $<$ 0.01 & -0.50 & med. \\ 
\midrule
Ridden in AV & 1.39 & 0.57 & 1.69 & 0.68 & 79.10 & $<$ 0.01 & 0.49 & small \\ 
Tech-Oriented Self & 2.99 & 1.24 & 3.53 & 1.13 & 66.30 & $<$ 0.01 & 0.46 & small \\ 
Enjoy Tech Analysis & 2.72 & 1.28 & 3.13 & 1.23 & 37.00 & $<$ 0.01 & 0.33 & small \\ 
Hacking & 3.87 & 1.19 & 3.51 & 1.20 & 40.61 & $<$ 0.01 & -0.31 & small \\ 
Improve Mobility & 4.08 & 0.98 & 4.35 & 0.81 & 29.67 & $<$ 0.01 & 0.30 & small \\ 
Use Automatic Stability Control & 1.25 & 0.65 & 1.48 & 0.90 & 29.31 & $<$ 0.01 & 0.30 & small \\ 
Legal Liability & 4.04 & 1.03 & 3.74 & 1.04 & 36.57 & $<$ 0.01 & -0.29 & small \\ 
Understand Computer Vision & 1.81 & 1.14 & 2.15 & 1.29 & 24.92 & $<$ 0.01 & 0.28 & small \\ 
Understand AV Algorithms & 1.60 & 1.03 & 1.90 & 1.17 & 30.56 & $<$ 0.01 & 0.27 & small \\ 
Data Privacy & 3.48 & 1.30 & 3.14 & 1.29 & 25.06 & $<$ 0.01 & -0.26 & small \\ 
Driving Risk & 2.94 & 1.05 & 2.68 & 0.97 & 21.13 & $<$ 0.01 & -0.26 & small \\ 
NFC Predictability & 11.87 & 2.57 & 11.21 & 2.57 & 26.55 & $<$ 0.01 & -0.26 & small \\ 
Have Built AI Systems & 1.19 & 0.66 & 1.39 & 0.95 & 22.70 & $<$ 0.01 & 0.25 & small \\ 
Used Adaptive Cruise Control & 1.71 & 0.91 & 1.96 & 1.08 & 14.47 & $<$ 0.01 & 0.25 & small \\ 
Gender (Female vs. Not) & 0.79 & 0.41 & 0.69 & 0.46 & 19.99 & $<$ 0.01 & -0.24 & small \\ 
Good With Technology & 3.54 & 1.10 & 3.79 & 1.02 & 18.34 & $<$ 0.01 & 0.23 & small \\ 
Use Collision Avoidance System & 1.30 & 0.75 & 1.50 & 0.95 & 16.16 & $<$ 0.01 & 0.23 & small \\ 
Use ML For Work & 1.40 & 0.93 & 1.63 & 1.11 & 21.45 & $<$ 0.01 & 0.23 & small \\ 
Lots of Tech Experience & 2.84 & 1.26 & 3.12 & 1.22 & 17.03 & $<$ 0.01 & 0.22 & small \\ 
Experience Training/Evaluating ML & 1.33 & 0.87 & 1.54 & 1.09 & 16.96 & $<$ 0.01 & 0.22 & small \\ 
Use Parallel Parking Assist & 1.18 & 0.63 & 1.34 & 0.81 & 20.72 & $<$ 0.01 & 0.22 & small \\ 
Use Vehicle Guidance System & 1.20 & 0.65 & 1.36 & 0.85 & 17.03 & $<$ 0.01 & 0.21 & small \\ 
Use Blindspot Detection & 1.95 & 1.28 & 2.22 & 1.36 & 14.45 & $<$ 0.01 & 0.21 & small \\ 
\midrule
Confidence Learning Tech & 2.99 & 1.24 & 3.23 & 1.21 & 13.38 & $<$ 0.01 & 0.20 & neg. \\ 
CVSCALE Power Distance & 7.51 & 3.06 & 8.13 & 3.42 & 13.06 & $<$ 0.01 & 0.19 & neg. \\ 
Injury Concerns (DC) & 2.02 & 1.13 & 1.82 & 0.95 & 6.43 & 0.01 & -0.19 & neg. \\ 
I help solve tech probs. & 2.90 & 1.31 & 3.14 & 1.28 & 11.97 & $<$ 0.01 & 0.19 & neg. \\ 
I am knowledgeable about tech. & 3.37 & 1.13 & 3.57 & 1.09 & 11.29 & $<$ 0.01 & 0.18 & neg. \\ 
Familiar with AVs & 3.08 & 1.53 & 3.34 & 1.43 & 8.98 & $<$ 0.01 & 0.18 & neg. \\ 
I am AV Expert & 1.20 & 0.56 & 1.31 & 0.72 & 8.62 & $<$ 0.01 & 0.18 & neg. \\ 
I work with AVs & 1.05 & 0.28 & 1.11 & 0.49 & 9.30 & $<$ 0.01 & 0.17 & neg. \\ 
Anxiety as Passenger & 2.36 & 1.07 & 2.18 & 0.97 & 8.58 & $<$ 0.01 & -0.17 & neg. \\ 
NFC Ambiguity & 11.96 & 2.40 & 11.56 & 2.26 & 13.89 & $<$ 0.01 & -0.17 & neg. \\ 
BFI Neuroticism & 0.90 & 2.00 & 0.56 & 2.01 & 10.71 & $<$ 0.01 & -0.17 & neg. \\ 
Thrill from breaking law (MDSI) & 1.40 & 0.79 & 1.55 & 0.98 & 5.36 & 0.02 & 0.17 & neg. \\ 
Risk Willingness & 5.61 & 1.79 & 5.90 & 1.86 & 8.89 & $<$ 0.01 & 0.16 & neg. \\ 
Die in accident (DC) & 2.05 & 1.15 & 1.88 & 1.02 & 5.20 & 0.02 & -0.16 & neg. \\ 
Plan routes badly (MDSI) & 1.65 & 0.90 & 1.79 & 0.98 & 6.60 & 0.01 & 0.15 & neg. \\ 
Enjoy dangerous driving (MDSI) & 1.59 & 1.00 & 1.75 & 1.13 & 7.07 & 0.01 & 0.15 & neg. \\ 
BFI Conscientiousness & 1.09 & 1.69 & 0.84 & 1.58 & 8.60 & $<$ 0.01 & -0.15 & neg. \\ 
BFI Extroversion & -0.26 & 1.99 & 0.04 & 1.96 & 7.10 & 0.01 & 0.15 & neg. \\ 
Self Esteem & 4.29 & 3.20 & 4.75 & 2.95 & 6.20 & 0.01 & 0.15 & neg. \\ 
Relax activities mid-drive (MDSI) & 2.57 & 1.23 & 2.75 & 1.23 & 6.55 & 0.01 & 0.15 & neg. \\ 
Use Lane Control & 1.68 & 1.05 & 1.84 & 1.15 & 5.82 & 0.02 & 0.15 & neg. \\ 
Use Auto E-Brake & 1.61 & 0.94 & 1.76 & 1.06 & 4.99 & 0.03 & 0.14 & neg. \\ 
SES & 4.55 & 5.43 & 5.32 & 5.69 & 7.23 & 0.01 & 0.14 & neg. \\ 
Honk horn at others (MDSI) & 2.21 & 1.26 & 2.05 & 1.16 & 3.44 & 0.06 & -0.13 & neg. \\ 
Need for Control & 20.31 & 3.34 & 19.87 & 3.35 & 6.24 & 0.01 & -0.13 & neg. \\ 
Feel frustrated while driving (MDSI) & 2.61 & 1.28 & 2.45 & 1.24 & 4.56 & 0.03 & -0.13 & neg. \\ 
Meditate while driving (MDSI) & 1.63 & 0.98 & 1.76 & 1.00 & 7.28 & 0.01 & 0.13 & neg. \\ 
Distracted driver (MDSI) & 2.58 & 1.24 & 2.73 & 1.26 & 4.19 & 0.04 & 0.12 & neg. \\ 
Forget high bean lights on (MDSI) & 1.53 & 0.91 & 1.64 & 0.96 & 6.10 & 0.01 & 0.12 & neg. \\ 
Thrill of flirting with death (MDSI) & 1.30 & 0.78 & 1.40 & 0.87 & 4.67 & 0.03 & 0.11 & neg. \\ 
CVSCALE Collectivism & 18.18 & 4.56 & 18.71 & 4.92 & 5.54 & 0.02 & 0.11 & neg. \\ 
Drive through red lights (MDSI) & 1.63 & 0.98 & 1.74 & 1.05 & 4.50 & 0.03 & 0.11 & neg. \\ 
Nervous while driving (MDSI) & 2.69 & 1.39 & 2.54 & 1.35 & 3.37 & 0.07 & -0.11 & neg. \\ 
Passengers will be hurt (DC) & 1.96 & 1.17 & 1.84 & 1.04 & 1.47 & 0.23 & -0.11 & neg. \\ 
I have control over driving (MDSI) & 4.06 & 0.94 & 4.16 & 0.91 & 3.77 & 0.05 & 0.11 & neg. \\ 
Age & 20.58 & 2.22 & 20.83 & 2.54 & 2.12 & 0.15 & 0.11 & neg. \\ 
Worry of bad weather (MDSI) & 3.78 & 1.16 & 3.66 & 1.17 & 4.13 & 0.04 & -0.10 & neg. \\ 
Swear at other drivers (MDSI) & 2.57 & 1.40 & 2.43 & 1.39 & 3.30 & 0.07 & -0.10 & neg. \\ 
Attempt to drive from park (MDSI) & 1.62 & 0.98 & 1.72 & 1.07 & 2.95 & 0.09 & 0.10 & neg. \\ 
Comfortable while driving (MDSI) & 3.86 & 1.12 & 3.97 & 1.01 & 1.58 & 0.21 & 0.10 & neg. \\ 
Mix wiper and light switch (MDSI) & 1.71 & 1.07 & 1.82 & 1.12 & 2.68 & 0.10 & 0.10 & neg. \\ 
High beam when annoyed (MDSI) & 1.52 & 0.99 & 1.61 & 1.06 & 3.63 & 0.06 & 0.09 & neg. \\ 
Enjoy taking driving risks (MDSI) & 1.64 & 0.91 & 1.73 & 1.02 & 1.41 & 0.24 & 0.09 & neg. \\ 
Daydream while driving (MDSI) & 2.59 & 1.32 & 2.71 & 1.27 & 3.07 & 0.08 & 0.09 & neg. \\ 
Lack control over accidents (DC) & 2.93 & 1.26 & 2.82 & 1.26 & 2.37 & 0.12 & -0.09 & neg. \\ 
Muscle relaxation mid-drive (MDSI) & 1.93 & 1.15 & 2.03 & 1.18 & 2.50 & 0.11 & 0.09 & neg. \\ 
Fix appearance mid-drive (MDSI) & 2.39 & 1.26 & 2.50 & 1.31 & 2.06 & 0.15 & 0.09 & neg. \\ 
Plan long trips in advance (MDSI) & 4.01 & 1.10 & 3.91 & 1.15 & 2.26 & 0.13 & -0.09 & neg. \\ 
Others will notice I'm anxious (DC) & 2.08 & 1.26 & 1.98 & 1.18 & 1.78 & 0.18 & -0.09 & neg. \\ 
BFI Agreeableness & 5.23 & 2.26 & 5.42 & 2.15 & 2.49 & 0.11 & 0.09 & neg. \\ 
People will think bad driver (DC) & 2.48 & 1.26 & 2.37 & 1.21 & 1.82 & 0.18 & -0.09 & neg. \\ 
I will not react quickly enough (DC) & 2.44 & 1.18 & 2.35 & 1.11 & 1.40 & 0.24 & -0.09 & neg. \\ 
BFI Openness & 1.03 & 1.82 & 0.90 & 1.74 & 2.46 & 0.12 & -0.08 & neg. \\ 
People will laugh at me (DC) & 1.84 & 1.13 & 1.76 & 1.02 & 0.38 & 0.54 & -0.07 & neg. \\ 
I will injure someone (DC) & 1.83 & 1.09 & 1.76 & 1.02 & 1.31 & 0.25 & -0.07 & neg. \\ 
Get lost in thoughts (MDSI) & 1.92 & 1.08 & 2.00 & 1.07 & 2.53 & 0.11 & 0.07 & neg. \\ 
Drive cautiously (MDSI) & 4.15 & 0.79 & 4.09 & 0.92 & 0.12 & 0.72 & -0.07 & neg. \\ 
I will cause accident (DC) & 1.92 & 1.07 & 1.85 & 0.98 & 0.40 & 0.53 & -0.06 & neg. \\ 
Drive below speed limit (MDSI) & 2.09 & 1.19 & 2.16 & 1.25 & 0.69 & 0.41 & 0.06 & neg. \\ 
Impatient in rush hour (MDSI) & 3.38 & 1.24 & 3.32 & 1.28 & 0.62 & 0.43 & -0.06 & neg. \\ 
Misjudge space mid-drive (MDSI) & 2.40 & 1.41 & 2.47 & 1.40 & 1.13 & 0.29 & 0.05 & neg. \\ 
“Better safe than sorry” (MDSI) & 4.03 & 0.96 & 3.98 & 0.99 & 0.71 & 0.40 & -0.05 & neg. \\ 
Strategize get through traffic (MDSI) & 3.20 & 1.32 & 3.27 & 1.29 & 0.73 & 0.39 & 0.05 & neg. \\ 
Distressed while driving (MDSI) & 2.44 & 1.25 & 2.38 & 1.24 & 0.65 & 0.42 & -0.05 & neg. \\ 
Enjoy driving on limit (MDSI) & 2.66 & 1.13 & 2.71 & 1.23 & 0.56 & 0.46 & 0.04 & neg. \\ 
Decision Style & 2.04 & 2.17 & 1.95 & 2.12 & 1.01 & 0.32 & -0.04 & neg. \\ 
Patiently wait at green lights (MDSI) & 3.62 & 1.17 & 3.67 & 1.14 & 0.39 & 0.53 & 0.04 & neg. \\ 
Always ready to react (MDSI) & 3.85 & 1.00 & 3.81 & 0.97 & 1.05 & 0.30 & -0.04 & neg. \\ 
Impatient at green lights (MDSI) & 2.38 & 1.22 & 2.42 & 1.23 & 0.47 & 0.49 & 0.04 & neg. \\ 
Switch b/n lanes in traffic (MDSI) & 2.71 & 1.24 & 2.76 & 1.23 & 0.49 & 0.48 & 0.04 & neg. \\ 
Driving Frequency & 43.5 & 39.0 & 45.0 & 39.8 & 0.31 & 0.58 & 0.04 & neg. \\ 
Misjudge speed w/ passing (MDSI) & 2.03 & 1.01 & 2.07 & 1.03 & 0.32 & 0.57 & 0.04 & neg. \\ 
Politics & 3.72 & 0.86 & 3.69 & 0.85 & 0.16 & 0.69 & -0.03 & neg. \\ 
Fear of losing self-control (DC) & 1.57 & 0.95 & 1.60 & 0.96 & 0.21 & 0.65 & 0.03 & neg. \\ 
Try to relax when driving (MDSI) & 3.54 & 1.07 & 3.57 & 1.07 & 0.20 & 0.66 & 0.02 & neg. \\ 
Number of Collisions & 0.35 & 0.63 & 0.34 & 0.60 & 0.04 & 0.83 & -0.02 & neg. \\ 
Patience yielding (MDSI) & 4.27 & 0.93 & 4.25 & 0.89 & 0.59 & 0.44 & -0.02 & neg. \\ 
Education & 1.75 & 1.29 & 1.76 & 1.42 & 0.04 & 0.84 & 0.01 & neg. \\ 
Blow horn when frustrated (MDSI) & 1.89 & 1.24 & 1.90 & 1.22 & 0.11 & 0.74 & 0.01 & neg. \\ 
Drive assertively (MDSI) & 2.60 & 1.29 & 2.59 & 1.29 & 0.02 & 0.90 & -0.01 & neg. \\ 
Fear of criticism (DC) & 2.27 & 1.26 & 2.26 & 1.24 & 0.01 & 0.97 & -0.01 & neg. \\ 
CVSCALE Uncertainty Avoidance & 21.4 & 3.06 & 21.4 & 3.06 & 0.02 & 0.88 & -0.01 & neg. \\ 
Cost & 4.15 & 1.05 & 4.15 & 0.97 & 0.33 & 0.57 & -0.00 & neg. \\ 
I will cause traffic/anger (DC) & 2.34 & 1.31 & 2.34 & 1.28 & 0.03 & 0.86 & -0.00 & neg. \\ 
   \hline
 \multicolumn{9}{p{.95\linewidth}}{\raggedright \textit{MDSI = Multidimensional Driving Style Inventory~\cite{taubman2004multidimensional}; DC = Driving Cognitions Questionnaire~\cite{ehlers2007driving}; M = Mean, SD = Standard Deviation, H = H-statistic, P = P-value, D = Cohen's D}}
\end{longtable}
 \endgroup

\section{Acknowledgements}
We would like to thank Rohan Bhide, Saumitra Sapre, Chloe Lee, and Janzen Molina for their help disseminating the survey. A special thanks to David Danks for his advice on modeling methods and to the Discovery Way Foundation for their generous usage of lab space. Chapter 4, in full, has been accepted for publication using the title \textit{Predicting Trust In Autonomous Vehicles: Modeling Young Adult Psychosocial Traits, Risk-Benefit Attitudes, And Driving Factors With Machine Learning.}, in the Proceedings of the 2025 CHI Conference on Human Factors in Computing Systems (CHI ’25). Kaufman, Robert, Lee, Emi, Bedmutha, Manas, Kirsh, David, Weibel, Nadir. The dissertation author was the primary investigator and author of this paper. \cite{kaufman2024predicting}

\chapter{Conclusions and Future Outlook}
\newpage
I hope this dissertation serves as both an entertaining and instructive contribution to the autonomous vehicle, artificial intelligence, and cognitive science communities. By integrating theories of complex systems, joint action, and situational awareness, I have demonstrated how accounting for the dynamic interplay between components of the human-AV system can improve interactions. Across three distinct studies, I explored the intricate relationship between humans and autonomous vehicles, posing them as a team jointly acting to satisfy the success criteria for a particular goal. Building the situational awareness \textit{of a rider} via optimized communications and the situational awareness \textit{of the AV} via an understanding of how to meet diverse rider needs in diverse contexts can enable AVs to foster trust, promote learning, and calibrate reliance. By anticipating and tailoring communications that meet the varying informational needs of individuals, contexts, and goals, we can design AVs to support successful interactions with their riders. Ultimately, it is my hope that this work will inspire future researchers and designers to adopt a systems-level approach, enabling the development of more equitable, adaptable, and useful autonomous vehicles and AI systems.

\section{Contributions review: Towards better AV design}
In Chapter 2, I investigated the potential of an AI driving coach by examining how variations in information type and presentation modality influenced a novice's ability to learn how to drive a racecar. The effectiveness of the coach depended on how well explanations guided attention and facilitated knowledge transfer, with the most successful outcomes occurring when clear, unambiguous information aligned with the novice's learning process. This was particularly true when cognitive demands were lower, allowing for better absorption of information. Participants who placed greater trust in the AI coach demonstrated improved learning outcomes and were more adept at mastering the unintuitive aspects of racing, underscoring trust as a necessary precondition for effective learning. This study illustrated the importance of tailoring AV communications to meet the specific informational needs of users, emphasizing the interplay between user goals, cognitive characteristics, and communication strategies. An AV that has awareness of these facets of the human-AV system can support learning and knowledge transfer. In discussion, I offered several design implications for developing AI coaches and broader human-machine interfaces.

In Chapter 3, I investigated the impact of explanation errors on trust and reliance perceptions of autonomous vehicles. Comparing explanations at different levels of accuracy showed that errors had a detrimental impact on people's perceptions of AVs, even in cases where the AV's driving itself was of high quality. The results showed that people take information an AV provides (including a judgment on the AV's accuracy) and make meaning of it within the context of a common goal. In the case of safe transportation, this may come in the form of an implicit risk assessment, where mistakes matter more when they can have the potential to cause real world harm. Contextual factors like the difficulty of driving and perceived harm directly affected trust and reliance judgments, as did the expertise and prior trust of the participant. Collectively, these highlight the emergent quality of human-AV interaction within a broader system, underscoring the importance of understanding the interplay of system components in order to better support users. AVs that can account for different rider and contextual demands will be better suited for building trust and calibrating reliance -- especially in cases where errors may occur. I concluded this study with a discussion on how to design accurate, contextually adaptive, and personalized AV explanations.

In Chapter 4, I deepen our understanding of how personal traits and risk assessments impact trust formation, using machine learning to predict trust in young adults using a wide range of personal factors. Results show that risk and benefit perceptions are the most significant predictors of trust in AVs, overshadowing the influence of other personal characteristics. The findings of this study give further insight into how trust forms: where goals dictate success functions, and a person's willingness to rely on an AV is a function of how well those goals can be met relative to the risks associated with meeting them. This aligns with prior work in cost-benefit analyses and (ir)rational decision-making~\cite{wang2019designing}. I break these findings into their implications for designing trustworthy AV systems, emphasizing the importance of enhancing situational awareness -- providing users with a clear understanding of the AV's capabilities, along with the associated risks and benefits of reliance. A key insight is the potential for AVs to adjust their communication strategies based on riders' individual trait profiles, which is a critical step toward equitable system design and informed policy development. This study presents a replicable methodology and a foundational set of personal traits that can guide future work in tailoring AV communications to meet specific informational needs.

\vspace{1em}
\noindent \textbf{In sum, this dissertation offers:}
\begin{enumerate}[topsep=0pt, nolistsep]
    \item Direct, actionable design insights to improve human-AV (and human-AI) interactions.
    \item A strong case for personalized, context-aware explainable AI explanations, as well as a process model by which these can be achieved.
    \item A framework based in systems thinking and cognitive theory that can deepen our understanding of how to design and study human-AI collaboration.
    \item Evidence supporting theories of trust formation and reliance with AI-based systems.
\end{enumerate}
\vspace{1em}

\section{A method to scaffold future research}
This dissertation took an unorthodox approach -- rather than following a linear narrative focused on a single problem, my studies tackled diverse questions in multiple areas of the human-AV system, at various stages of the research and design process. By exploring different components of the human-AV interactive system, from foundational trust-building mechanisms to practical design interventions, I aimed to demonstrate the value of a multifaceted investigation. This approach reflects the complexity of autonomous systems and the need for a broader, systems-oriented view that encompasses both technical and human-centered perspectives. In doing so, this work offers a flexible methodological framework that can guide future researchers and designers in studying and developing better AI, regardless of the specific phase of the process they are engaged in. Whether addressing early conceptual challenges or refining real-world applications, this dissertation encourages a holistic examination of the interplay between technology and its users, fostering more resilient and human-aligned AI-based systems.

\section{The future of AI-based systems}
The future of AI-based systems holds both immense promise and profound responsibility. As AI continues to permeate nearly every facet of human life, its potential to revolutionize industries, enhance societal well-being, and address global challenges is unparalleled. However, realizing this potential requires a concerted effort to design systems that are not only technically robust but also ethically sound, contextually adaptive, and deeply human-centered.

A core insight from this dissertation is that successful AI systems must be designed as dynamic participants in complex, interdependent ecosystems. In the case of autonomous vehicles, the interplay between human traits, system capabilities, and situational demands underscores the necessity of personalized and context-aware designs. The same principles extend beyond AVs to other domains where AI interfaces with humans: healthcare, education, environmental management, social governance, and more. In these areas, AI’s ability to tailor its actions and communications to meet diverse human needs can profoundly reshape outcomes, fostering trust, collaboration, and mutual understanding.

To achieve this vision, AI must evolve beyond a one-size-fits-all approach. Systems must learn to adapt dynamically to varying user goals, cognitive capacities, and cultural contexts. For instance, in healthcare, an AI assistant aiding in diagnostics must calibrate its explanations based on the expertise level of the practitioner and the stakes of the situation. In education, AI tutors should align their teaching strategies with individual learning styles, cognitive attributes, and motivations. These examples illustrate a broader imperative: AI systems must be capable of situational awareness that parallels human adaptability, enabling them to serve as true collaborators rather than mere tools.

Furthermore, the ethical and societal implications of AI cannot be overlooked. As this dissertation highlights, transparency and accountability are critical in building trust and preventing harm, particularly in safety-critical domains like autonomous driving. Designers and policymakers must grapple with questions of equity and accessibility to ensure AI benefits are distributed fairly. This involves addressing biases in AI systems, protecting user privacy, and fostering inclusivity across diverse populations.

Looking ahead, the integration of AI into human society presents opportunities to redefine what collaboration means in the 21st century. Imagine systems that do more than execute commands -- that actively anticipate needs, mediate interactions, and empower individuals to achieve their goals in ways previously unimaginable. Autonomous vehicles, for example, could transcend their role as mere transportation tools, becoming partners in enriching human experiences, from offering scenic routes to providing context-aware educational or cultural insights during a journey.

To realize this potential, a multidisciplinary approach is essential. Engineers, cognitive scientists, ethicists, policymakers, and many more must work together to address the technical and societal complexities of AI deployment. This dissertation provides a framework for advancing this collaboration by emphasizing the critical role of context-sensitive, transparent, and personalized designs. Yet, it also serves as a call to action for future research to explore unresolved challenges, such as managing competing human and system priorities, mitigating unforeseen risks, and ensuring long-term sustainability.

The future of AI is not predetermined, but a shared responsibility shaped by the choices we make today. By embedding principles of adaptability, equity, and human-centered design into every stage of AI development, we can ensure that these systems become catalysts for human flourishing rather than sources of division or harm. The work presented in this dissertation represents a small but meaningful step toward this vision, offering insights and methodologies that can inspire a broader transformation in how we design, deploy, and live with AI. Ultimately, I envision a future in which AI-based systems can support diverse needs, so that all people -- not just a privileged few -- can benefit from this remarkable technology.

\backmatter
\bibliographystyle{plainnat} % Or whatever style you want like plainnat
\bibliography{main}

\end{document}